\documentclass[useAMS,usenatbib,usegraphicx]{mn2e}

\usepackage{natbib}
\usepackage{appendix}
\usepackage{amsmath}
\usepackage{amssymb}
\usepackage{longtable}
\usepackage{graphicx}
\usepackage{verbatim}
\usepackage{lscape}	
\newcommand{\kms}{\ensuremath{\mathrm{km\,s}^{-1}}}
\newcommand{\asec}{\ensuremath{^{\prime \prime}}}
\usepackage{epsfig,color}
\usepackage{multicol}

\usepackage[T1]{fontenc} 
\usepackage{aecompl}

\newcommand\aap{{A\&A}}
\newcommand\aaps{{A\&AS}}
\newcommand\aj{{AJ}}
\newcommand\apj{{ApJ}}
\newcommand\apjl{{ApJL}}
\newcommand\apjs{{ApJS}}
\newcommand\mnras{{MNRAS}}
\newcommand\pasp{{PASP}}
\newcommand\nat{{Nature}}
\newcommand\physrep{{Phys. Rep.}}
\newcommand\prd{{Phys. Rev. D}}

\title[The Spitzer South Pole Telescope Deep Field Survey: Linking galaxies and haloes at z=1.5]
{The Spitzer South Pole Telescope Deep Field Survey: Linking galaxies and haloes at z=1.5}

\author[Martinez-Manso et al.]{Jesus Martinez-Manso$^{1}$
\thanks{E-mail: j.martinez.manso@gmail.com},  Anthony H. Gonzalez$^{1}$, Matthew L. N. Ashby$^{2}$,
\newauthor S. A. Stanford$^{3}$, Mark Brodwin$^{4}$, Gilbert P. Holder$^{5}$ and Daniel Stern$^{6}$\\
$^{1}$Department of Astronomy, University of Florida, Gainesville FL 32611, USA\\
$^{2}$Harvard-Smithsonian Center for Astrophysics, 60 Garden St., Cambridge MA 02138, USA\\
$^{3}$Department of Physics, University of California, One Shields Avenue, Davis CA 95616, USA\\
$^{4}$Department of Physics and Astronomy, University of Missouri, Kansas City MO 64110, USA\\
$^{5}$Department of Physics, McGill University, Montreal, Quebec, H3A 2T8, Canada\\
$^{6}$Jet Propulsion Laboratory, California Institute of Technology, 4800 Oak Grove Drive, Mail Stop 169-221, Pasadena CA 91109, USA}

\begin{document}


\pagerange{\pageref{firstpage}--\pageref{lastpage}} \pubyear{2014}

\maketitle

\label{firstpage}

\begin{abstract}
We present an analysis of the clustering of high-redshift galaxies in the recently completed 94 deg$^2$ {\it Spitzer}-SPT Deep Field survey. Applying flux and color cuts to the mid-infrared photometry efficiently selects galaxies at $z\sim1.5$ in the stellar mass range $10^{10}-10^{11}M_\odot$, making this sample the largest used so far to study such a distant population. We measure the angular correlation function in different flux-limited samples at scales $>6\asec$ (corresponding to physical distances $>0.05$ Mpc) and thereby map the one- and two-halo contributions to the clustering. We fit halo occupation distributions and determine how the central galaxy's stellar mass and satellite occupation depend on the halo mass. We measure a prominent peak in the stellar-to-halo mass ratio at a halo mass of $\log(M_{\rm halo} / M_\odot) = 12.44\pm0.08$, 4.5 times higher than the $z=0$ value. This supports the idea of an evolving mass threshold above which star formation is quenched. We estimate the large-scale bias in the range $b_g=2-4$ and the satellite fraction to be $f_\mathrm{sat}\sim0.2$, showing a clear evolution compared to $z=0$. We also find that, above a given stellar mass limit, the fraction of galaxies that are in similar mass pairs is higher at $z=1.5$ than at $z=0$. In addition, we measure that this fraction mildly increases with the stellar mass limit at $z=1.5$, which is the opposite of the behavior seen at low-redshift.
\end{abstract}

\begin{keywords}
cosmology: observations --- galaxies: evolution --- galaxies: high-redshift --- galaxies: halos --- large scale structure of universe
\end{keywords}

\section{Introduction}
Many observational studies have measured dark matter halo masses in order to find correlations with the properties of the galaxies they host. Various works have utilized gravitational lensing of background objects \citep{mandelbaum06, gavazzi07,bolton08,auger10,cacciato09,cacciato13,velander11}, virial temperatures derived from X-rays \citep{lin03,lin&mohr04,peterson06,hansen09} and dynamics of satellites \citep{more09, more11}. These methods have achieved high accuracy, but are also observationally expensive to carry out on large samples and for small haloes, which limits the statistical strength and range of application. 
A less direct but more comprehensive method of linking galaxies to haloes is abundance matching \citep{conroy06,vale&ostriker06,moster10,guo10,behroozi10,moster10,moster13}, which uses the merger trees from $N$-body dark matter simulations as input and assumes that the halo mass is the main determinant of galaxy luminosity and stellar mass. The basic idea is to cumulatively match observed galaxy luminosity functions and halo mass functions by placing progressively less luminous galaxies in less massive haloes. By design, this method reproduces the luminosity (or stellar mass) function, and is able to predict the clustering of galaxies in many cases \citep{conroy06,conroy09,moster10}. \\
\indent Direct measurements of galaxy clustering are another powerful way to connect galaxies with the underlying dark matter distribution. As a function of physical separation $r$, clustering is commonly measured in the form of the two-point spatial correlation function $\xi(r)$ (SCF; Peebles 1980). The relation between the distributions of galaxies and dark matter can be parametrized through the galaxy bias $b_g$ \citep{kaiser84,coles93,fry93, mo&white96, kauffmann97, ST99,tinker05, tinker10}, which is given by the scaling between the SCFs of these two fields:
\begin{equation}
\xi_g(r,z) = \xi_{m}(r,z) \, b_g^2(r,z).   \label{bias}
\end{equation}
The SCF of dark matter depends on the cosmology, and can be prescribed analytically given those parameters \citep{eisenstein99, smith03}. Thus, the bias of a galaxy sample is directly determined by its SCF. In general, the bias depends on the spatial scale and redshift \citep{fry96, moscardini98, tinker05, moster10}, since galaxies and dark matter do not evolve in the exact same manner in time or space. The measurement of $b_g(r,z)$ can therefore reveal a precise description of the connection between galaxies and dark matter. \\
\indent Many studies up to intermediate redshifts ($z<1$) have investigated galaxy clustering with samples selected in different ways \citep{phleps06,zheng07,coil08,blake08,brown08,mccracken08,meneux08,meneux09,simon09,ross10,foucaud10,abbas10,zehavi11,matsuoka11,wake11,
jullo12,leauthaud12,hartley13,mostek13,delatorre13,donoso14}. The most common conclusion is that clustering strength is correlated with luminosity, red color and morphology (towards early-type). Galaxies on the extreme of these properties are highly biased and therefore they live in massive haloes. \\
\indent These conclusions can be obtained just by analyzing the overall amplitude of the bias. However, the precise form of this observable as a function of spatial separation contains more information about the inner structure of the haloes. The halo occupation distribution (HOD) is a simple parametric framework to accurately model the bias \citep{ma&fry00,seljak00,peacock00,scoccimarro01, cooray02, berlind02,berlind03, kravtsov04,zheng05}. It considers galaxies to be either centrals or satellites, and the number of these that a halo can host is fully determined by the halo mass.\\
\indent One of the advantages of the HOD framework is that its parameters have a clear physical meaning, and thus when fitting the clustering one can gain a deeper insight into the connection between the galaxies and their host haloes. For example, the HOD framework can directly relate the average stellar mass of the central galaxies to a particular halo mass. As shown in many studies at $z=0-1$, the ratio of these masses is highest around a halo mass of $\sim 10^{12}M_\odot$ \citep{zheng07,yang12,zehavi12,leauthaud12,reddick13,behroozi13,moster13,wang13}. This implies that there is a characteristic halo mass where galaxy formation has been more efficient. The qualitative explanation for this is that at low halo masses the gravitational potential is not deep enough to halt the expulsion of gas due to stellar winds \citep{benson03}, while high-mass haloes have heated up the intra-halo medium by gravitational heating and AGN feedback \citep{croton06,bower06,vandevoort11a} so that infalling gas gets heavily shocked and cannot easily cool and condense \citep{birnboim&dekel03,dekel&birnboim06,keres05,keres09}. These two trends can be reduced to a comparison between dynamical and gas cooling times in haloes, such that $\tau_\mathrm{dyn}>>\tau_\mathrm{cool}$ for low masses and $\tau_\mathrm{dyn}<<\tau_\mathrm{cool}$ for high-masses. A possible consequence is that the peak halo mass $M_\mathrm{peak}$ is related to a characteristic quenching mass $M_q$ that sets $\tau_\mathrm{dyn} \sim \tau_\mathrm{cool}$ \citep{neistein06} and marks a transition between star forming and quenched haloes. Indeed, massive red galaxies with little star formation have been shown to live in massive haloes \citep{coil08, zehavi11}, supporting the idea of the red sequence of galaxies arising when they become quenched \citep{croton06, bower06}. This blue/red dichotomy is present in the nearby Universe \citep{kauffmann03,kauffmann04,baldry04}, and starts its build-up around $z\sim2$ \citep{bell04,cooper06,muzzin13a,wang13}. Thus, when haloes become large enough, they quench their star formation. A consequence of this is that the most massive galaxies today have no significant ongoing star formation. This effect has been called archeological downsizing \citep{cowie96,juneau05,conroy09}, and is also inferred from the lack of evolution in the massive end of the stellar mass function \citep{perez-gonzalez08,marchesini09,muzzin13a}. HOD models have shown that the stellar-to-halo mass ratio (SHMR) evolves in the sense that the peak moves to lower halo masses with increasing time, at least since $z\sim1$ \citep{coupon12,leauthaud12}. This trend has been predicted to persist up to $z=2$ by extensions of HOD that use conditional stellar mass functions \citep{yang03,yang12,wang13} and abundance matching studies \citep{moster13,behroozi13}. A possible mechanism for this would involve evolution in $M_q$, which is supported by the idea that the universal gas fraction drops with time and therefore star formation becomes more difficult with time at fixed halo mass \citep{vandevoort11a,vandevoort11b}. However, this is still a matter of debate \citep{conroy&ostriker08,tinker10b}. For instance, \citet{leauthaud12} present evidence in favor of this evolution being set by quenching below a critical galaxy-halo mass ratio instead of a critical halo mass. Such a mechanism would also shift the SHMR toward lower masses with time.  \\
\indent We have described the basic processes that can determine $M_\mathrm{peak}$, based on the comparison of $\tau_\mathrm{dyn}$ and $\tau_\mathrm{cool}$ as a function of halo mass. This basic model can be extended to include modes of galactic outflows, which are then directly constrained by the observed slope of the SHMR. The stellar mass growth of a galaxy is heavily regulated by the expulsion of gas, which could be mainly sourced by supernovae feedback \citep{murray05}. The stellar mass loss rate, $\dot{M}_\star$, can be broken down in two contributions: pressure-supported energy injection (energy-driven winds) and coherent momentum transfer (momentum-driven winds). The energy and momentum deposition rates, $\dot{E}$ and $\dot{P}$, can be related from first principles to the mass via a proxy of the kinetic velocity field, $\sigma_W$: $\dot{M}_\star \propto \dot{E}/\sigma_W^2$ and $\dot{M}_\star \propto \dot{P}/\sigma_W$. This suggests that galaxies with low velocity fields, and therefore low masses, may have their outflows dominated by energy-driven winds \citep{dutton09}. Thus, a larger contribution from this type of wind would result in a steeper low mass slope of the SHMR. \\
\indent At high masses (and high $\sigma_W$), these arguments would point to a dominance of momentum-driven winds. However, the winds in this regime are also sourced by radiative AGN feedback, which is expected to have a strong contribution \citep{vogelsberger13}. In addition, a large merger rate between central galaxies will result in a flattening of the SHMR \citep{leauthaud12}. With all these processes at play, the high-mass slope is less straightforward to interpret than the low-mass one, but it can still offer important constraints on this combination of mechanisms.\\
\indent Galaxy clustering combined with HOD modeling provide particularly solid measurements of the SHMR whenever the selection of galaxies spans the relevant range of stellar masses. At $z \gtrsim 1.5$, such measurements have proven to be very difficult given the lack of large volume-limited samples. \citet{wake11} use the 0.25 deg$^2$ NEWFIRM survey \citep{vandokkum09}, but the low number statistics made it difficult to map the turnover of the SHMR. In this study, we use a 94 deg$^2$ mid-infrared survey to select galaxies with stellar masses ranging from $10^{10}-10^{11}M_\odot$ and fit an HOD model to the angular correlation function. We present the most robust measurement to date of the peak of the SHMR at $z=1.5$. \\
\indent In addition, the HOD yields particularly strong constraints on the satellite population of a given galaxy sample. We determine what fraction of the galaxies are satellites, and how the abundance of these depends on the halo mass. Moreover, we measure a proxy for the occurrence of galaxy pairs of similar mass, and find that it mildly decreases toward high luminosities. Although we do not achieve a robust detection, this represents the opposite trend to what is seen at low-redshift. The processes that produce this relationship are strongly tied to the accretion and merger events between galaxies and haloes, as well as the quenching of star formation in satellites.  \\
\indent The paper is organized as follows. In Section \ref{s_datasets}, we describe all datasets that are used. In Section \ref{s_ctrlsamples} we describe how we adapt redshift and stellar mass distributions from a reference optical + mid-IR survey. In Section \ref{s_2pt} we define the two-point clustering statistic and the method used to compute it. In Section \ref{s_placing}, we describe the model that links galaxies to haloes. In Section \ref{s_fits}, we explain the fitting procedure of the HOD to the observed clustering. In Sections \ref{s_shmr}, \ref{s_satellites} and \ref{s_bg} we discuss the results obtained regarding the SHMR, the satellite galaxies and the large-scale bias, respectively. We end with a short summary in Section \ref{s_summary}. For the reader that is only interested in the results, we recommend reading Sections \ref{s_shmr} and beyond.\\
\indent Additionally, we include several appendices where many of the details are covered. Appendix \ref{s_photsims} presents a calibration of systematic effects in the photometry. Appendix \ref{a_EvsC} compares the results obtained from using different reference catalogs to draw redshift and stellar mass distributions. Appendix \ref{s_ic} calculates the systematic offset in the clustering amplitude due to the geometry of the survey. Appendix \ref{a_halo} presents the formalism of the halo model. Appendix \ref{nosc} investigates the removal of low-redshift sources from the sample using optical data. Appendix \ref{a_tests} explores different choices of free parameters used in the HOD fits to the clustering. \\
\indent Throughout this paper we use the following cosmology:  $\Omega_m=0.27$, $\Omega_\Lambda=0.73$ and $H_0=70$ \kms Mpc$^{-1}$. All magnitudes are in the Vega system and masses are in units of $M_\odot$.

\section{datasets} \label{s_datasets}
Our main dataset is the {\it Spitzer} South Pole Telescope Deep-Field Survey (SSDF; Ashby et al. 2013b), a 93.8 deg$^2$ photometric survey using the IRAC 3.6 and 4.5$\mu$m bands (hereafter [3.6] and [4.5]). The mosaics have a nominal integration time of 120 seconds. We used Source Extractor (SExtractor; Bertin \& Arnouts 1996) in dual image mode, detecting galaxies in [4.5] and extracting the flux from fixed 4\asec\ apertures in both IRAC channels. These aperture fluxes were then corrected to total fluxes using growth curves from isolated point sources found in the mosaics. A detailed description of the survey and a public photometric catalog are presented in \citet{ashby13b}. However, here we use a deeper private catalog and account for faint-end photometric bias and detection completeness (see Appendix A). We determine the 5$\sigma$ limit in [4.5] to be 18.19 mag, in agreement with \citet{ashby13b}.  \\
\indent We use the near-infrared 2MASS Point Source Catalog \citep{skrutskie06} to identify and remove sources brighter than $K_s(AB)=12$ mag, most of which are likely to be stars. In addition, we visually inspected some of these sources in the IRAC mosaics and determined an empirical relation between their $K_s$-band magnitude and the maximum radius where their 4.5$\mu$m flux caused a clear suppresion in the detection of nearby sources. This relation was then applied to the rest of the $K_s$-selected sample and the resulting radii were used to mask all SSDF sources enclosed within from the main catalog. For reference, the radii corresponding to $K_s(AB)=8$ and $K_s(AB)=12$ sources were 41\asec\ and 8.4\asec\, respectively. We also masked out low coverage gaps in the survey, yielding a final effective area of 88.8 deg$^2$.\\
\indent 
Finally, in order to better understand the redshift distribution of our IRAC-selected sample in the SSDF, we use public catalogs in two other regions of the sky: the COSMOS-UltraVista field (hereafter COSMOS, Muzzin et al. 2013) and
the Extended Groth Strip (hereafter EGS, Barro et al. 2011a,b). These two surveys have publically accessible IRAC photometry, photometric redshifts, and stellar masses. In the following Section we describe how we used these catalogs to infer the redshift and mass distributions of SSDF samples.

\section{Control Sample}\label{s_ctrlsamples}
This study requires knowledge of the redshift and stellar mass distribution of the SSDF galaxy sample. However, our main dataset is too limited to obtain reliable values for these observables. Therefore, the strategy is to import this information from a reference survey that contains optical data and IRAC photometry with a higher accuracy. We consider the catalogs from COSMOS and EGS, which include photometric redshifts and stellar masses. We will adopt COSMOS as the fiducial dataset because it is larger and has better statistics, and in Appendix \ref{a_EvsC} we show how our results do not change significantly when using EGS instead. The reference catalog is degraded to become a control sample whose photometric errors match those of the SSDF. Then, applying the same IRAC selection in both SSDF and the control sample allows us to match the derived distributions of redshift and mass. A brief description of the COSMOS photometry can be found in Appendix \ref{a_EvsC}. \\

\begin{figure}
\includegraphics[trim=2mm 0mm 05mm 0mm,clip=True,width=\columnwidth]{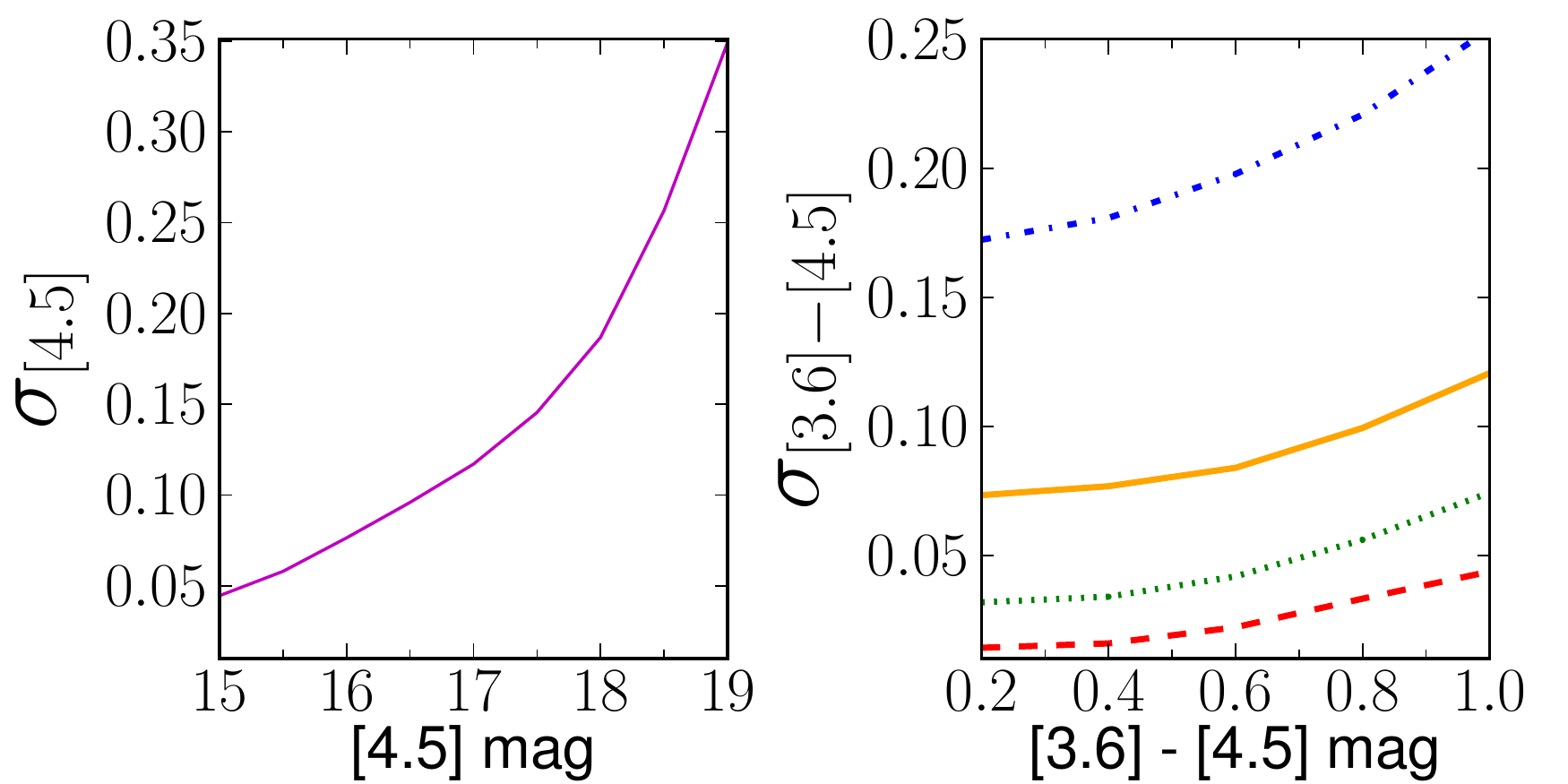}
\caption{\label{ch2_color_stds} Photometric scatter of SSDF sources derived from the simulations in Appendix \ref{s_photsims}. \textbf{Left:} Standard deviation in [4.5] magnitudes. There is higher scatter for fainter sources. \textbf{Right:} Standard deviation in [3.6]-[4.5] color (4\asec\ diameter aperture). Dashed, dotted, solid and dash-dotted represent fixed [4.5] input magnitudes of 15, 16, 17 and 18, respectively. Larger colors imply fainter [3.6] magnitudes, which is reflected as a mild increase in the scatter. }
\end{figure}

\indent For every source in the reference catalog, we have $[4.5]$ magnitudes, $[3.6] - [4.5]$ colors, photometric redshifts and stellar masses. The goal is to infer the SSDF distributions of these parameters by degrading the reference photometry, which is done using the SSDF photometric errors. We calculate the scatter in SSDF magnitudes and colors as a function of these same variables, using the results from the photometric simulations described in Appendix \ref{s_photsims}. These scatter profiles are shown in Figure \ref{ch2_color_stds}. At fixed [4.5] magnitude, the scatter in color increases for larger colors since these imply fainter [3.6] magnitudes. In the case of the reference sample, since it is 2 magnitudes deeper than SSDF (see Appendix \ref{a_EvsC}), we can safely consider its photometric scatter as negligible in comparison.\\
\indent The degradation of the reference catalog into a control sample consists of transforming the specific values (e.g., magnitude) of each source in the catalog into Gaussian probability density functions (PDFs). These PDFs are defined in the parameter space of apparent magnitude [4.5] ($\mathcal{M}$), $[3.6] - [4.5]$ color ($\mathcal{C}$) and photometric redshift ($z_\mathrm{phot}$): $\mathcal{P}\left( \mathcal{M} ,\, \mathcal{C} ,\, z_\mathrm{phot}  \right)$. The centroids are given by $\vec{\mu}_i= \left( \mathcal{M}^i ,\, \mathcal{C}^i ,\, z_\mathrm{phot}^i \right)$, which correspond to the parameter vectors of the sources in the reference catalog. The standard deviations are $\vec{\sigma}_i= \left( \sigma_{\mathcal{M}}^i,\, \sigma_{\mathcal{C}}^i ,\, \sigma_{z_\mathrm{phot}}^i \right)$. The first two components in $\vec{\sigma}$ are the functions $\sigma_{\mathcal{M}}(\mathcal{M})$ and $\sigma_{\mathcal{C}}(\mathcal{M},\mathcal{C})$, which are shown by the curves in Figure \ref{ch2_color_stds}. The redshift component does not have a counterpart in the SSDF catalog, but we apply a variable redshift smoothing kernel equivalent to 100 comoving Mpc, in order to filter the effect of large-scale structure. This amounts to $\sigma_{z_\mathrm{phot}}=0.02-0.1$ within our redshift range. However, we find that this redshift filtering has a minimal effect in the results, varying the $z=1.5$ clustering amplitude and galaxy number density at the $\sim1\%$ level.  \\

\subsection{Main redshift distribution}\label{ss_phi}
\begin{figure}
\includegraphics[trim=20mm 0mm 20mm 0mm,clip=True,width=\columnwidth]{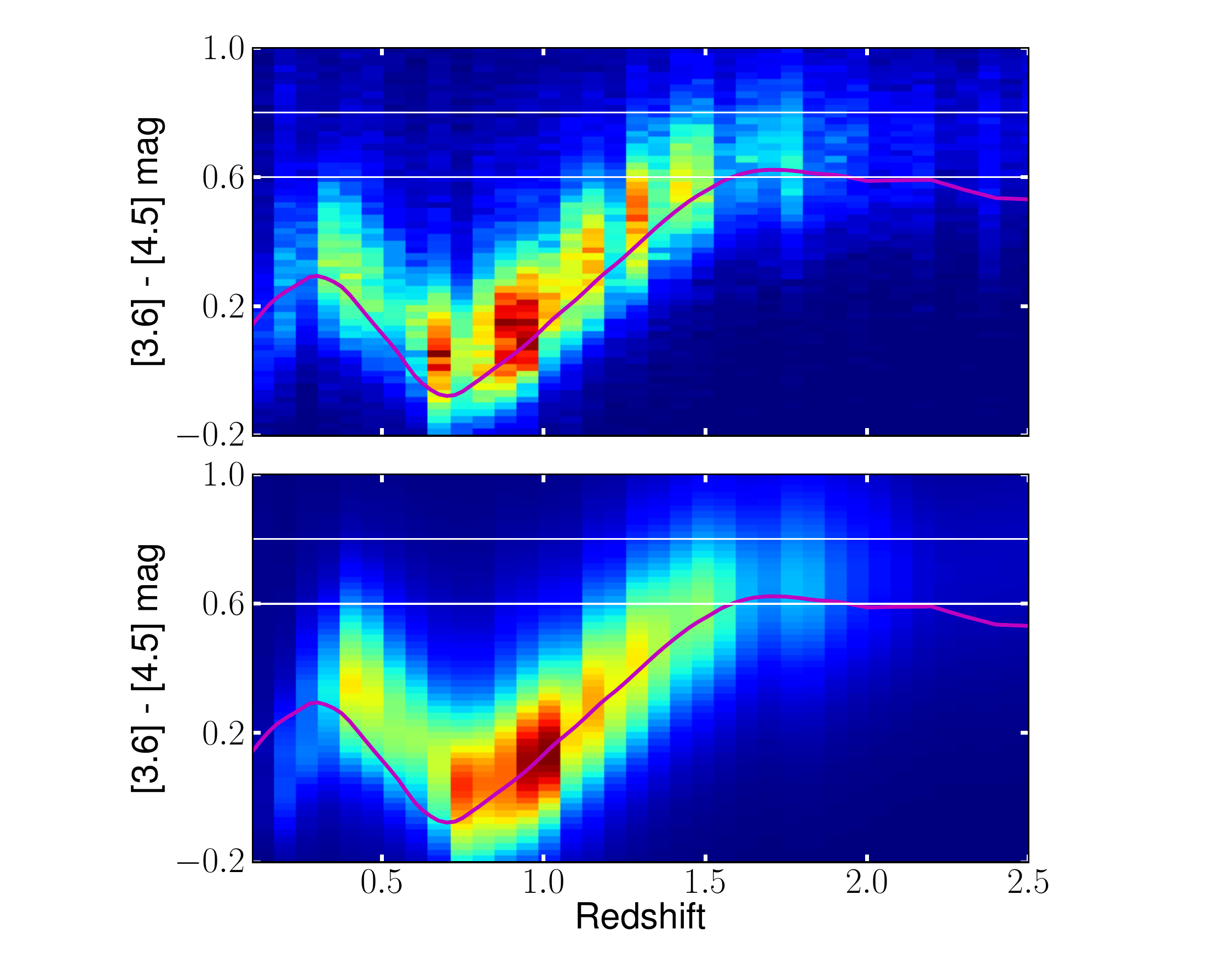}
\caption{\label{z_color} [3.6]-[4.5] color vs. redshift for galaxies with $15<\mathrm{[4.5]}<18.6$, based on the COSMOS reference catalog. The horizontal white lines indicate the color selection that we apply to our SSDF samples. The purple curve is the evolutionary track of a galaxy formed at $z_f=3.5$ using the BC03 model with a Chabrier IMF, shown for comparison. \textbf{Top:} Distributions of the raw reference catalog. \textbf{Bottom:} Reference catalog degraded to match the SSDF photometric properties, derived from Equation \ref{e_K}.   }
\end{figure}

If we consider galaxies with apparent magnitudes within some bracket $\Delta \mathcal{M}$, we can compute the distribution in color and redshift space $\mathcal{K}(\mathcal{C},z)$ of the control sample:

\begin{equation}
\mathcal{K}(\mathcal{C},z) = \frac{1}{N_\mathrm{ref}}\sum\limits_{j=1}^{N_\mathrm{ref}} \int\limits_{\Delta \mathcal{M}} dm\,\, \mathcal{P}( m ,\, \mathcal{C} ,\, z ; \,\vec{\mu}_j ,\, \vec{\sigma}_j ).  \label{e_K}
\end{equation}
Here, we have marginalized each individual PDF over $\Delta \mathcal{M}$ and summed them in the resulting space of $(\mathcal{C} ,\, z_\mathrm{phot})$, using the reference catalog (subscript ``ref''). A similar procedure to derive full redshift distributions based on the Bayesian combination of individual redshift likelihood functions was performed by \citet{brodwin06a,brodwin06b}. The normalization of $\mathcal{K}(\mathcal{C},z)$ is the total number of sources in the reference catalog, $N_\mathrm{ref}$. Figure \ref{z_color} shows the application of Equation (\ref{e_K}) for $\Delta \mathcal{M} \to 15<\mathrm{[4.5]}<18.6$, which are the limits for our full SSDF sample (see Section \ref{ss_subsamples}). The top panel correspond to the color versus redshift distribution from the raw reference catalog. The lower panel shows the control sample, which is how SSDF sources are expected to be distributed. For comparison, we have also plotted a galaxy evolutionary track for a single stellar population with solar metallicity and formation redshift of $z_f=3.5$, computed using the \citet{bc03} models with \citet{chabrier03} IMF.\\
\indent There is a clear correlation between color and redshift at $z>0.6$. This occurs because going from $z=0.6$ to $z=2$, the IRAC bands map the galaxy spectrum across the stellar bump at rest-frame $H$-band. This results in a monotonic change in observed color within $z=0.6-2$. An insightful description of this phenomenology can be found in \citet{muzzin13b}. We can take advantage of this effect to select galaxies in redshift using a color cut. A lower color threshold needs to be high enough to reject $z<0.3$ galaxies (see Figure \ref{z_color}), while also keeping a number of higher redshift sources that is large enough to measure a robust clustering signal. An upper threshold is also necessary, since very red colors $\mathrm{[3.6]}-\mathrm{[4.5]}\sim 1$ are characteristic of active galactic nuclei \citep{stern05}. The best compromise is a color cut of $0.6<\mathrm{[3.6]}-\mathrm{[4.5]}<0.8$, as shown in Figure \ref{z_color}. \\
\indent With a given $\Delta \mathcal{C}$, we can derive the redshift distribution of sources:

\begin{equation}
\phi(z) = \frac{1}{N_\mathrm{ref}}\sum\limits_{j=1}^{N_\mathrm{ref}} \int\limits_{\Delta \mathcal{M}}  \int\limits_{\Delta \mathcal{C}} dm\,dc\,\, \mathcal{P}( m ,\, c ,\, z ; \,\vec{\mu}_j ,\, \vec{\sigma}_j ).  \label{e_phi}
\end{equation}

Note that $\int \phi(z;\Delta \mathcal{M}, \Delta \mathcal{C}) dz$ is equal to 1 only when $\Delta \mathcal{M}$ and $\Delta \mathcal{C}$ represent the full ranges spanned by the reference sources. We denote such distribution as $\phi_\mathrm{full}(z)$, while the one corresponding to the color selection $\Delta \mathcal{C} \to 0.6<\mathrm{[3.6]}-\mathrm{[4.5]}<0.8$ is denoted as $\phi_\mathrm{cut}(z)$. \\
We are assuming that this color cut selects a representative sample of the $z>1$ galaxy population. However, it is important to check whether such a selection is biased toward young or old galaxies. It has been shown that older galaxies exhibit a higher clustering amplitude than their younger counterparts \citep[and references therein]{skibba14}. In our case, we are tracing a part of the spectral energy distribution that is much less sensitive to star formation history. To illustrate this point, we use the EzGal package \citep{mancone12} to compare the $\mathrm{[3.6]}-\mathrm{[4.5]}$ colors for passive and star-forming galaxies using assorted stellar population models \citep{bc03,m05,conroy&gunn09}. For the passive galaxies we assume a single burst model with formation redshift $z_f=3.5$. For the star-forming galaxies we run models with exponentially declining star formation, using $\tau=1$\,Gyr and an initial formation redshift $z_f=3.5$. The difference in $\mathrm{[3.6]}-\mathrm{[4.5]}$ is $\lesssim 0.05$. For the galaxy sample used in our analysis, this difference is comparable to or smaller than the photometric errors (see Fig. \ref{ch2_color_stds}). Thus, the associated systematic bias due to the color selection will be small compared to the photometric errors and intrinsic scatter in galaxy colors, and can be neglected for the current analysis.

\subsection{Definition of subsamples}\label{ss_subsamples}
\begin{figure}
\includegraphics[trim=15mm 0mm 0mm 0mm,clip=True,width=10cm]{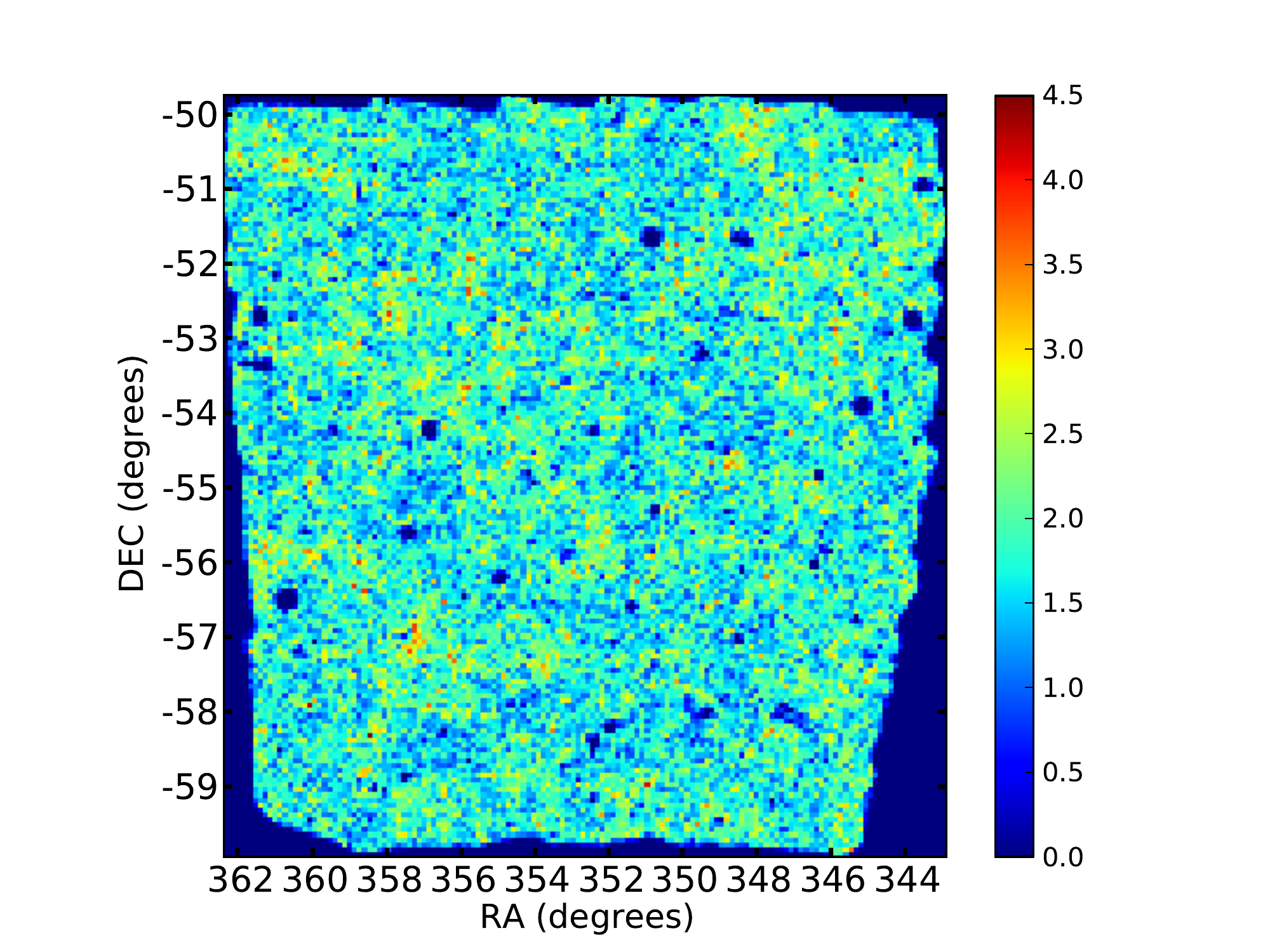}
\caption{\label{temp_field} Density map of the SSDF field for galaxies in our selected sample: $0.6<\mathrm{[3.6]-[4.5]}<0.8$ and $15<\mathrm{[4.5]}<18.6$. This corresponds to a redshift selection around $z\sim1.5$. Units are galaxies per square arcminute. Masking has been applied to bright stars and low coverage gaps, yielding a final size of 88.8 square degrees.  }
\end{figure}

\begin{figure}
\includegraphics[trim=5mm 0mm 10mm 0mm,clip=True,width=\columnwidth]{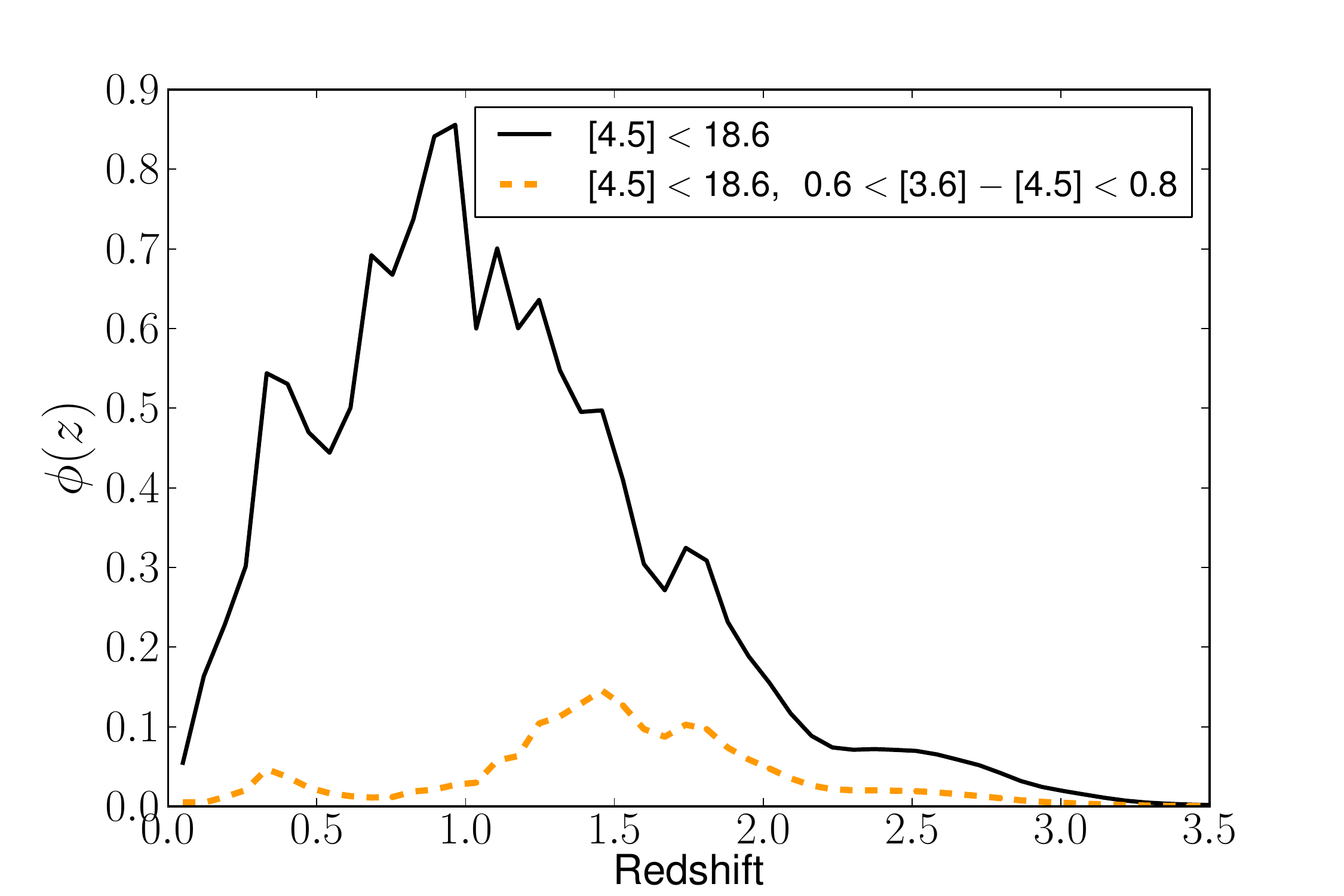}
\caption{\label{dndz1D} Redshift distribution of the COSMOS-based control sample using the faintest selection (15$<$[4.5]$<$18.6, $k=13$). The dashed lines represent the additional color cut selection. The color cut imposes a selection around $z\sim1.5$, although it only keeps about one third of the total number counts at that redshift.  }
\end{figure}

Our main science sample of SSDF galaxies is determined by the apparent magnitude and color cuts of $15<\mathrm{[4.5]}<18.6$ and $0.6<\mathrm{[3.6]}-\mathrm{[4.5]}<0.8$. The first of these cuts imposes an upper magnitude limit at the 80\% completeness level (see Appendix \ref{s_completeness}), and the second is tuned to select galaxies at high-redshift while avoiding AGN. A density map of this selection can be seen in Figure \ref{temp_field}, representing a slice of the Universe at $z\sim1.5$. \\
\indent We further split the main sample into 13 subsamples, with faint limits over the range $[4.5]=16.2-18.6$ in steps of 0.2 mag. The bright limit is $[4.5]=15$ in all of them. We do this instead of a selection within differential magnitude bins because the halo occupation framework presented in Section \ref{s_placing} requires cumulative samples in order to link halo masses and galaxy masses. We note that this approach carries the drawback of producing a correlation between the different samples. This correlation is strong between neighboring samples, but not dominant otherwise. Due to the steep variation of the stellar mass function, any given sample is mostly comprised by galaxies close to its low-mass threshold, making their clustering less sensitive to the most massive population (see Matsuoka et al. 2011).  \\
\indent The photometric scatter increases for fainter samples. Thus, we calculate the redshift distribution (see Equation \ref{e_phi}) for each sample, obtaining sets of $\phi_\mathrm{full}^k$, $\phi_\mathrm{cut}^k$, where $k=1-13$ is the sample index (going from brightest to faintest). We can also define the number density completeness as $f_\mathrm{N}^k(z) = \phi_\mathrm{cut}^k(z) / \phi_\mathrm{full}^k(z)$, which determines the fraction of galaxies as a function of redshift that the color cut retains. Figure \ref{dndz1D} shows a comparison of $\phi_\mathrm{cut} ,\, \phi_\mathrm{full}$ for $k=13$ (the largest sample). At the peak of the color-cut distribution we have that $f_\mathrm{N}^k\sim$0.3, and we will use this factor to scale up and correct the number density (see below). Figure \ref{dndz1D_set} shows $\phi_\mathrm{cut}^k$ for the smallest and largest samples (i.e., brighter and fainter thresholds, $k$=1,13), where each curve is shown normalized to 1. We also derive cosmic variance errors using the prescriptions from \citet{moster11}, which are based on analytical predictions of dark matter structure given a particular survey geometry (see also Brodwin et al. 2006a). The peak in these redshift distributions is consistently around $z=1.5$ in all samples. In general, the samples consist of a $z\geq1$ population that has approximately the same absolute luminosity and stellar mass (see Section \ref{ss_stellarmasses}), plus a $z\sim0.3$ contribution of ``contaminant'' galaxies that are intrinsically much less luminous. These contaminants represent 12\% (37\%) of all galaxies in our full (brightest) sample. When setting a brighter flux threshold, the high-redshift population becomes less dominant since these galaxies are closer to the turnover of the luminosity function. The consequence of this is the clear trend where brighter samples have a stronger low-redshift bump. The contribution of the latter to the clustering is modeled in the following Sections. Alternatively, we show in Appendix \ref{nosc} that our results remain unchanged if instead we employ shallow optical data to remove most of the low-redshift sources. \\ 
\indent With these redshift distributions, we can calculate the spatial number density of observed galaxies at the pivot redshift $z_p \equiv 1.5$. Here, we use the SSDF survey area and the effective number of observed sources in the subsamples, $N_\mathrm{obs}^k$. This number is derived by summing the inverse of the completeness value for all galaxies, using the relation from Figure \ref{completeness}. Then, the true number of galaxies within $z_p \pm \delta z/2$ can be written as

\begin{equation}
N_\mathrm{true}^k = \frac{N_\mathrm{obs}^k}{f_N^k(z_p)} \frac{\phi_\mathrm{cut}^k(z_p)}{\int \phi_\mathrm{cut}^k(z^\prime)dz^\prime} \delta z.
\end{equation}
The sampled volume reads as
\begin{equation}
V  = \frac{dV(z_p)}{dz}\delta z = \frac{c \, \Omega \, \chi^2(z_p)}{H(z_p)}\delta z,
\end{equation}
where $\chi(z)$ is the comoving radial distance, $H(z)$ is the Hubble function, $c$ is the speed of light and $\Omega=0.0271$ steradians is the solid angle subtended by the survey. Hence, the number density at $z_p$ results in
\begin{equation}
n_g^k  = \frac{N_\mathrm{true}^k}{V}. \label{ng_obs}
\end{equation}
Note that this quantity is the result of combining the SSDF observed number counts (via $N^k_\mathrm{obs}$) and the color fractions of the control sample. Table 1 shows the values of these number densities for all samples.

\begin{figure}
\includegraphics[trim=5mm 0mm 15mm 0mm,clip=True,width=\columnwidth]{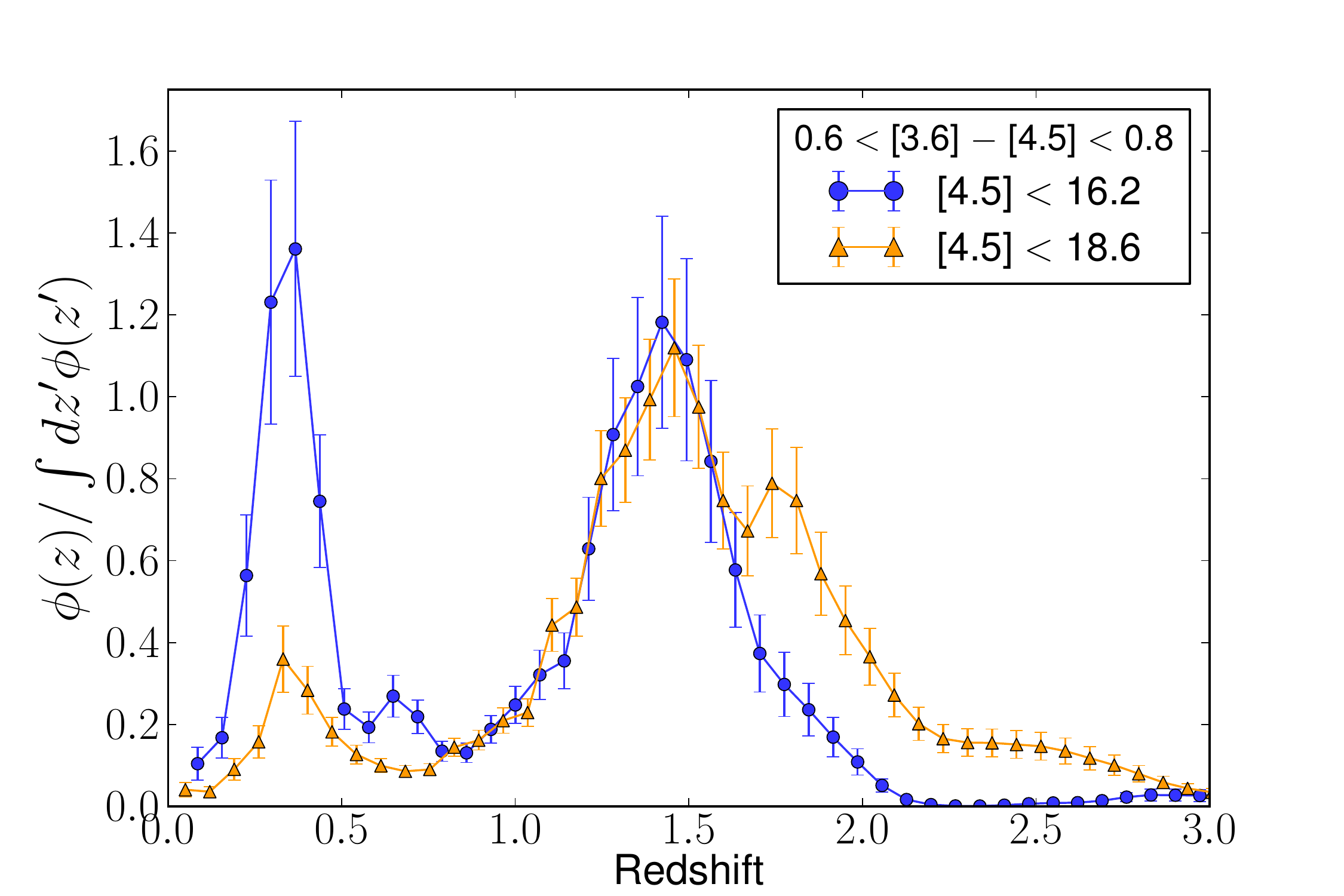}
\caption{\label{dndz1D_set} 
Normalized redshift distributions of the COSMOS-based control sample using the faintest (15$<$[4.5]$<$18.6, orange triangles) and brightest (15$<$[4.5]$<$16.2, blue circles) flux thresholds with the color cut. Brighter sample thresholds induce a higher contribution of low-redshift sources (see text). }
\end{figure}

\begin{figure}
\includegraphics[trim=0mm 0mm 10mm 0mm,clip=True,width=\columnwidth]{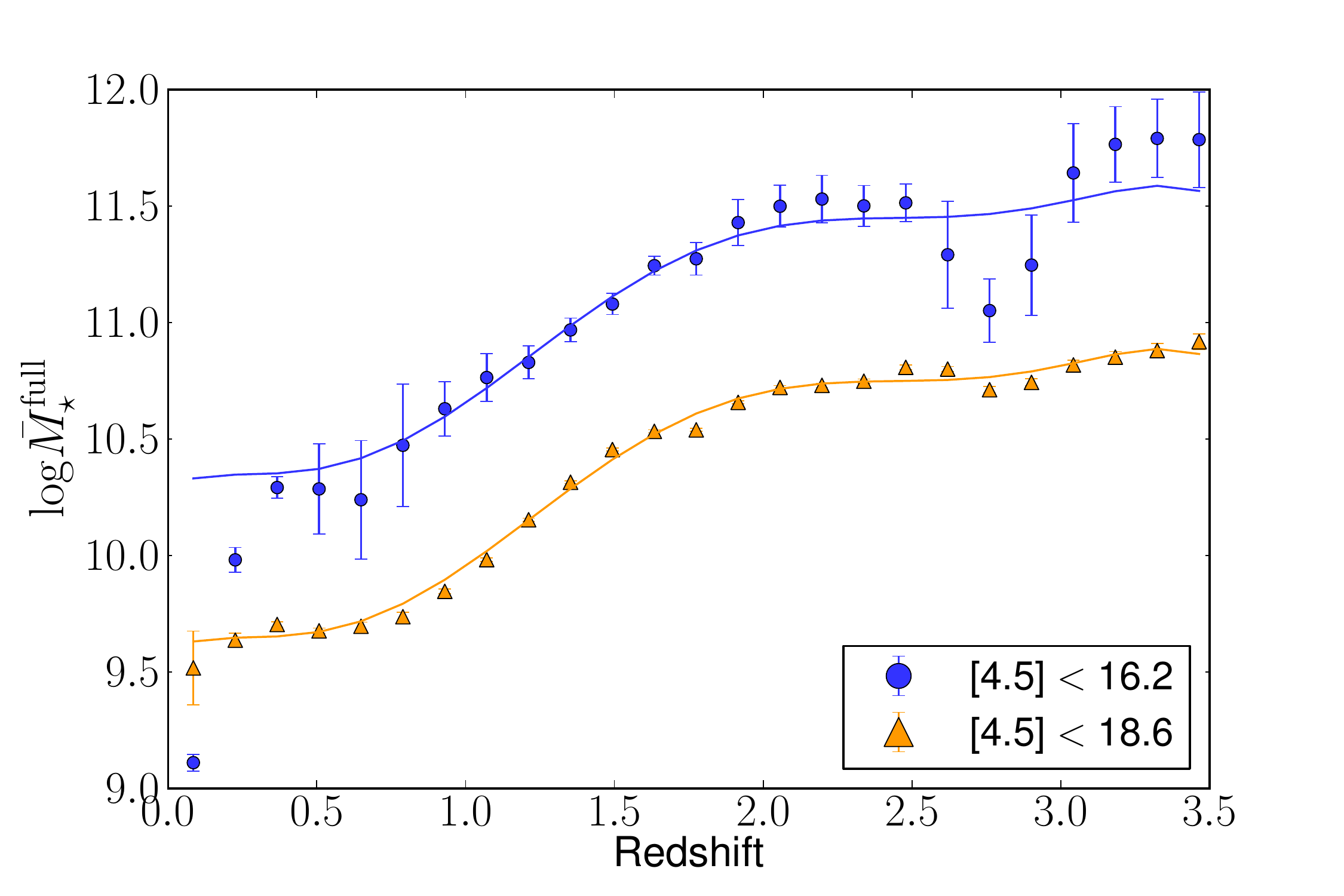}
\caption{\label{medmass_z} Redshift evolution of the median stellar mass in our brightest and faintest samples. The lower solid curve is a polynomial fit to the points from the latter. Those corresponding to the bright sample are noisier, so we offset the lower solid curve to match them. This is physically motivated by the approximation that mass scales linearly with flux.  }
\end{figure}

\begin{figure}
\includegraphics[trim=0mm 0mm 15mm 0mm,clip=True,width=\columnwidth]{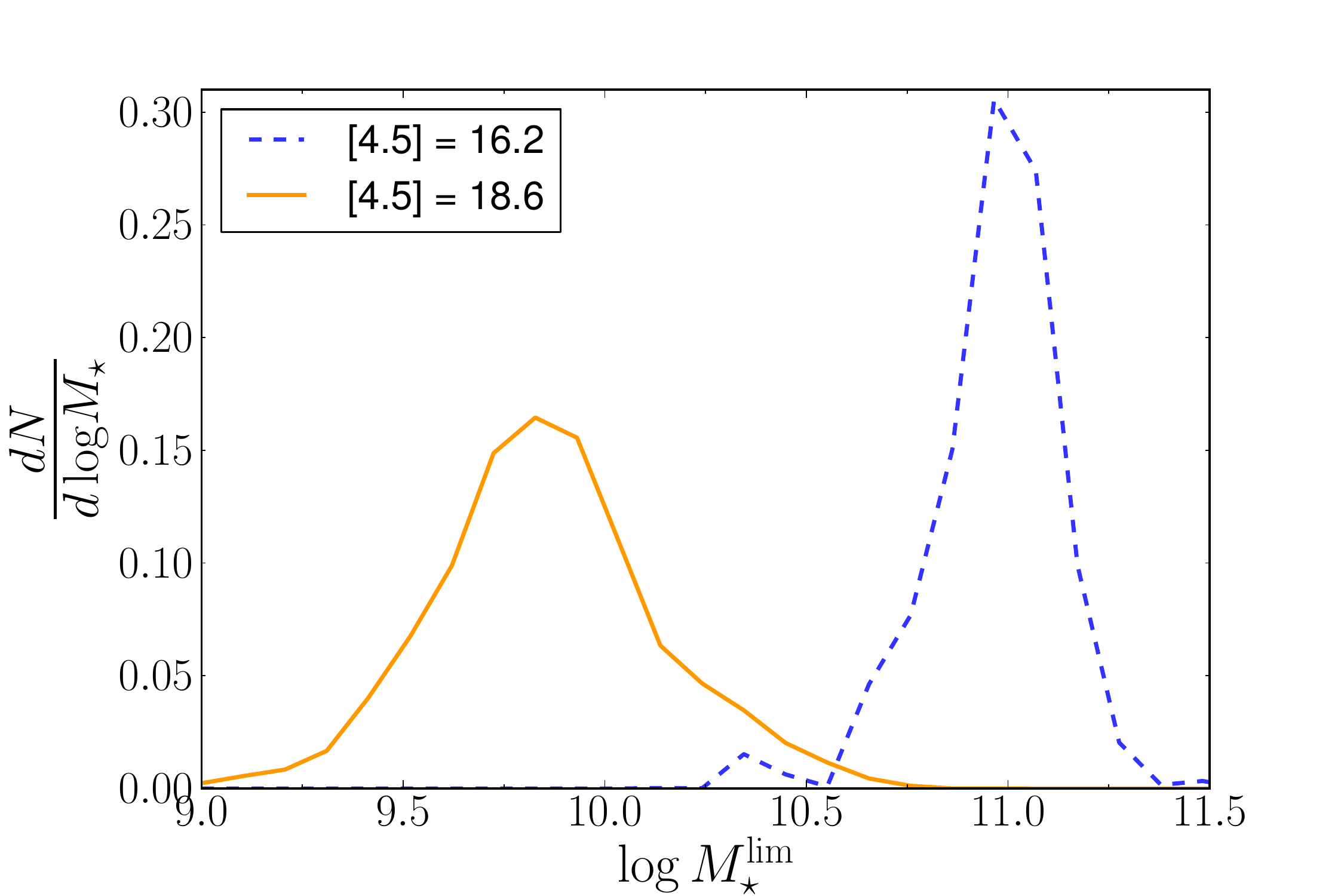}
\caption{\label{mass_hist} Normalized distributions of the $z=1.5$ stellar mass at the brightest and faintest magnitude limits of our samples.    }
\end{figure}

\subsection{Stellar masses}\label{ss_stellarmasses}

The stellar masses in the reference catalog are also retrieved to construct our control sample. We use those based on Bruzual \& Charlot (2003, hereafter BC03) stellar grids, \citet{chabrier03} IMF and \citet{calzetti00} dust extinction. Unless otherwise noted, all stellar masses are given under these prescriptions.  \\
\indent For the purposes of this paper, we need to calculate stellar masses for two different selections of galaxies. One is the median mass of all galaxies within each sample, derived at every redshift bin, $\bar{M}_\star^\mathrm{full}$. This mass will be used to derive a redshift scaling of the galaxy bias in Section \ref{ss_redshift_scaling}. The other, $\bar{M}_\star^\mathrm{lim}$, is the median mass of the galaxies at the pivot redshift ($z_p=1.5$) and at the magnitude limit of each sample. This is the stellar mass that will be linked in Section \ref{s_shmr} to a particular halo mass.\\

\indent Ideally, in order to calculate $\bar{M}_\star^\mathrm{full}$ we would compute the median mass of those galaxies within the given selection range in the parameter space of redshift, magnitude and color. However, classifying galaxies on whether they fall in that range is not straightforward, since each galaxy is represented by an extended probability distribution in the parameter space. A more insightful approach is to compute the probability that a galaxy's true parameter vector falls within the specified range, which reads as
\begin{equation}
\varrho_j(z)  = \int\limits_{\Delta \mathcal{M}}  \int\limits_{\Delta \mathcal{C}} dm\,dc\,\, \mathcal{P}( m ,\, c ,\, z ; \,\vec{\mu}_j ,\, \vec{\sigma}_j ),
\end{equation}
where $0<\varrho^j<1$ and $j$ is an index that identifies every galaxy in the reference catalog. Thus, we can calculate $\bar{M}_\star^\mathrm{full}$ as the weighted median stellar mass over all galaxies in the reference catalog, where the set of $\varrho^j$ act as weight coefficients to the individual stellar masses $M_\star^j$. By definition, the weighted median mass represents the mass value where the weighted integral of the mass distribution above and below that value is the same, and we write it in the following condensed form

\begin{equation}
\bar{M}_\star^\mathrm{full}  =\mathrm{Median}\left[ \vec{M_\star};\,\mathrm{weights}= \vec{\varrho}\right] , \label{mall}
\end{equation}
where we have omitted the implicit dependence on $z$. Conveniently, the weighted distribution of masses follows closely a log-normal distribution. Thus, Equation \ref{mall} returns almost the same value as the weighted mean of $\mathrm{log}\vec{M}_\star$, which allows us to adopt the standard deviation from the latter distribution:

\begin{equation}
\Sigma_M  =\frac{ \sum \left(\mathrm{log}M_\star^j - \mathrm{log}\bar{M}_\star^\mathrm{full} \right)^2 \varrho_j}{\sum \varrho_j} ,
\end{equation}
which is typically $\sim0.2$ dex. Even though this scatter is rather large, these log-mass distributions are single-peaked and approximately symmetric, so their mean value is well-defined and physically meaningful. We use the scatter to estimate the error in the mean as
\begin{equation}
\bar{\Sigma}_M  =\frac{\Sigma_M}{\sqrt {\mathcal{N}_\mathrm{ind}}} \label{meanerrs}
\end{equation}
with 
\begin{equation}
\mathcal{N}_\mathrm{ind}=\frac{\left( \sum \varrho_j\right) ^2 }{\sum {\varrho_j}^2}.
\end{equation}
Here, $\mathcal{N}_\mathrm{ind}$ represents the effective number of independent elements in the ensemble. This number is proportional to the sum of contributing weights and inversely proportional to their scatter. It equals the total number of elements in the reference catalog in the limit of $\varrho_j\rightarrow1$.\\
\indent Figure \ref{medmass_z} shows $\bar{M}_\star^\mathrm{full}(z)$ for all samples. The errors are from Equation (\ref{meanerrs}) and the solid curve is a 5th order polynomial fit to the points of the faintest sample. 
We use that curve plus an offset to fit the data from the rest of the samples, since it becomes noisier at brighter limits. Here, we take advantage of the fact that stellar mass scales with flux approximately in a linear manner. It is clear from the figure that the mass is tightly correlated with the redshift of observation within $z=0-1.5$. Beyond that, the relation flattens out significantly. The reason for this is that at $z\geq1.5$, the [4.5] band samples the rising spectral slope of the stellar bump \citep{muzzin13b}. This offsets the k-correction in a way that galaxies of a certain intrinsic near-infrared luminosity have a similar apparent [4.5] magnitude across a range of redshift. A consequence of this is that any [4.5] limited sample becomes roughly stellar mass limited at $z>1.5$ (see also Figure 14 in Barro et al. 2011b). Nonetheless, we do not attempt to take advantage of this effect by averaging stellar masses at high-redshift. The modeling in this work is based on well-defined median masses as a function of redshift, independent of the form of that redshift dependence. However, the flattening of this curve does benefit our study to some extent. Since there is an inherent uncertainty in how well represented the SSDF data is with the control sample, it is convenient that the stellar masses are naturally more constrained than a case where they had a strong redshift dependence.\\
\indent We can calculate $\bar{M}_\star^\mathrm{lim}$ at $z_p$ in an analogous way, considering a selection within $\Delta \mathcal{C}$. The corresponding weights are
\begin{equation}
\kappa_j(\mathcal{M})  =  \int\limits_{\Delta \mathcal{C}} dc\,\, \mathcal{P}( \mathcal{M} ,\, c ;z_p, \,\vec{\mu}_j ,\, \vec{\sigma}_j ), \label{weights}
\end{equation}
and replacing $\varrho_j$ with $\kappa_j$ in Equations (7-10) gives us the median mass $\bar{M}_\star^\mathrm{lim}$, along with its error. An example of the stellar mass histograms that are obtained using these weights can be seen in Figure \ref{mass_hist}. They are shown for the brightest and faintest magnitude limits of our samples. These distributions are symmetric and have a well-defined mean. The scatter is generally large, with values around $\sim$0.2 dex. However, the error in the logarithmic mean (see Equation (\ref{meanerrs})) is typically quite small $\sim$0.03 dex. The values of these limiting masses are displayed in Table 1. Thus, we have calculated the median mass of all galaxies in our samples as a function of redshift, and the median mass of galaxies around the pivot redshift at each sample magnitude limit.

\begin{table*}
\begin{center}
\caption{\label{t_samples} Sample properties and HOD fits. \textbf{Column 1:} Upper limiting magnitude for each sample. The lower limit is fixed at 15 mag. \textbf{Column 2:} Total number of observed SSDF sources, corrected for completeness. \textbf{Column 3:} Estimated fraction between observed and true number of sources at $z_p$. \textbf{Column 4:} Number density at $z_p$, corrected by $f_N$. Units are $10^{-4}\mathrm{Mpc}^{-3}$. Errors are derived by the method of \citet{moster11}. \textbf{Columns 5-6:} Median stellar masses of $z=z_p$ galaxies at and above the flux limit of the sample, respectively. \textbf{Columns 7-11:} Parameters from the HOD fits. $M_1^\prime$ are best-fit values, the rest are derived parameters. The reduced chi-squared is given by $\chi^2_\nu= \chi^2 /(28-1)$. Since $\alpha=1$, one can directly compute $M_1 = M_1^\prime + M_0$, with $\mathrm{log}M_0=0.76\,\mathrm{log}M_1^\prime+2.3$.}
  \begin{tabular}{ccccccccccc}
  \hline
[4.5] limit & $N_\mathrm{obs}$ & $f_N$ & $n_g$ & $\mathrm{log}\bar{M}^\mathrm{lim}_{\star}$ & $\mathrm{log}\bar{M}^\mathrm{full}_{\star}$
 & $\mathrm{log}M_1^\prime$ & $\mathrm{log}M_\mathrm{min}$ & $b_g$ & $f_\mathrm{sat}$ & $\chi^2_\nu$  \\

 \hline
                                              
16.2 & 17713 & 0.43 & 0.4$\pm$0.1 & 10.93$\pm$0.02 & 11.05$\pm$0.02 & 14.28$\pm$0.09 & 13.17$\pm$0.05 &
 3.95$\pm$0.13 & 0.06$\pm$0.01 & 0.33\\ [0.07cm]                                                         
16.4 & 29385 & 0.42 & 0.7$\pm$0.2 & 10.86$\pm$0.02 & 10.99$\pm$0.02 & 14.03$\pm$0.11 & 13.00$\pm$0.06 &
 3.57$\pm$0.12 & 0.09$\pm$0.01 & 0.74\\ [0.07cm]                                                         
16.6 & 47242 & 0.40 & 1.3$\pm$0.3 & 10.80$\pm$0.01 & 10.91$\pm$0.02 & 13.80$\pm$0.09 & 12.84$\pm$0.05 &
 3.28$\pm$0.08 & 0.12$\pm$0.01 & 1.27\\ [0.07cm]
16.8 & 72506 & 0.40 & 2.2$\pm$0.4 & 10.72$\pm$0.01 & 10.83$\pm$0.01 & 13.62$\pm$0.09 & 12.70$\pm$0.05 &
 3.04$\pm$0.07 & 0.15$\pm$0.02 & 1.79\\ [0.07cm]
17.0 & 105801 & 0.40 & 3.2$\pm$0.6 & 10.63$\pm$0.01 & 10.74$\pm$0.01 & 13.51$\pm$0.08 & 12.57$\pm$0.04
& 2.85$\pm$0.05 & 0.15$\pm$0.01 & 1.93\\ [0.07cm]
17.2 & 146773 & 0.40 & 4.5$\pm$0.8 & 10.54$\pm$0.01 & 10.65$\pm$0.01 & 13.39$\pm$0.09 & 12.47$\pm$0.05
& 2.72$\pm$0.06 & 0.16$\pm$0.02 & 1.76\\ [0.07cm]
17.4 & 195346 & 0.39 & 6.0$\pm$1.0 & 10.46$\pm$0.01 & 10.57$\pm$0.01 & 13.28$\pm$0.08 & 12.38$\pm$0.05
& 2.62$\pm$0.04 & 0.18$\pm$0.01 & 1.22\\ [0.07cm]
17.6 & 249444 & 0.38 & 7.6$\pm$1.3 & 10.36$\pm$0.01 & 10.48$\pm$0.01 & 13.20$\pm$0.08 & 12.30$\pm$0.04
& 2.52$\pm$0.04 & 0.18$\pm$0.01 & 1.04\\ [0.07cm]
17.8 & 308064 & 0.37 & 9.3$\pm$1.5 & 10.26$\pm$0.01 & 10.37$\pm$0.01 & 13.12$\pm$0.10 & 12.24$\pm$0.06
& 2.46$\pm$0.04 & 0.20$\pm$0.02 & 0.90\\ [0.07cm]
18.0 & 370735 & 0.35 & 11.1$\pm$1.8 & 10.16$\pm$0.01 & 10.27$\pm$0.01 & 13.06$\pm$0.09 & 12.17$\pm$0.05
& 2.40$\pm$0.04 & 0.20$\pm$0.02 & 1.03\\ [0.07cm]
18.2 & 435672 & 0.34 & 13.0$\pm$2.1 & 10.06$\pm$0.01 & 10.18$\pm$0.01 & 13.00$\pm$0.07 & 12.12$\pm$0.05
& 2.35$\pm$0.03 & 0.21$\pm$0.01 & 0.95\\ [0.07cm]
18.4 & 503212 & 0.32 & 15.0$\pm$2.3 & 9.97$\pm$0.01 & 10.08$\pm$0.01 & 12.95$\pm$0.07 & 12.07$\pm$0.05
& 2.30$\pm$0.03 & 0.22$\pm$0.01 & 0.88\\ [0.07cm]
18.6 & 575131 & 0.30 & 17.2$\pm$2.6 & 9.87$\pm$0.01 & 9.99$\pm$0.01 & 12.89$\pm$0.07 & 12.03$\pm$0.05
& 2.27$\pm$0.03 & 0.23$\pm$0.01 & 0.56\\

\hline
\end{tabular}
\end{center}
\end{table*}

\section{two-point clustering}\label{s_2pt}

Given a population of galaxies in a three-dimensional space, one can define the joint probability of finding two such objects in volume elements $\delta V_1$, $\delta V_2$ separated by a distance $r$ \citep{peebles80, phillips78}:
\begin{equation}
\delta P(r) = \bar{\mathcal{N}}^2(1+\xi_g(r))\delta V_1 \delta V_2 . \label{scf}
\end{equation}
Here, $\bar{\mathcal{N}}$ is the density of galaxies and $\xi_g$ is the SCF, which quantifies the clustering strength of the field as a function of $r$. The SCF can also be interpreted as the differential probability of finding two objects separated by a given distance, with respect to the case of a random distribution. \\
\indent The SCF for a galaxy population can be directly computed if the individual distances (redshifts) to those galaxies are known. However, in our case we are limited to individual sky positions and the ensemble redshift distribution. Therefore, we are interested in the angular correlation function (ACF), which is the projection of the SCF onto the 2D sphere. Analogously to the SCF, the ACF represents the differential probability with respect to a random distribution of finding two galaxies separated by a particular angle. The ACF is related to the SCF through the Limber projection \citep{limber53}, which integrates the SCF along the line of sight using the normalized redshift distribution $\phi(z)$ as a weight kernel \citep{phillips78, coupon12}:

\begin{equation}
\omega(\theta) = \frac{2}{c} \int_0^\infty dz H(z) \phi^2(z) \int_0^\infty dy \,\, \xi_g (r=\sqrt{y^2+D_c^2(z)\theta^2}), \label{limber}
\end{equation}
where $D_c(z)$ is the radial comoving distance, $H(z)$ is the Hubble function, $c$ is the speed of light and $\theta$ is the angular separation given in radians.  \\
\indent In order to measure $\omega (\theta)$ we use the estimator presented in \citet{hamilton93}, which counts the number of galaxy pairs with respect to those of a random sample distributed in the same geometry:

\begin{equation}
\hat{\omega}(\theta)=\frac{\mathrm{RR}(\theta) \mathrm{GG}(\theta) }{\mathrm{GR}^2(\theta)} - 1, \label{ham1}
\end{equation}
where GG, GR and RR are total number of galaxy-galaxy, galaxy-random and random-random pairs separated by an angle $\theta$. We have also tested the estimator from \citet{landy&szalay93}, which returns results that are practically indistinguishable from those using Equation \ref{ham1}.   \\
\indent In order to account for the completeness shown in Figure \ref{completeness}, we make a small generalization of Equation \ref{ham1}. Instead of counting all pairs with values of 1, we use a weighted scheme where each pair of sources $\alpha,\, \beta$ is counted as a product of weights $\upsilon_\alpha \upsilon_\beta$. Random sources have $\upsilon=1$, and galaxies have weights equivalent to the inverse of the completeness value at its apparent magnitude. We counts pairs by brute force in discrete angular bins using the graphics processing unit (GPU) on a desktop computer. We have developed our own code, which yields computation times of the order of 1000 times faster than using a CPU-based run with 16 cores. Our code is written in PyCUDA\footnote{documen.tician.de/pycuda/}, which is a Python wrapper of the CUDA, the programming language that interfaces with the device. \\
\indent The estimator in Equation \ref{ham1} implicitly assumes that the average galaxy density of the survey is the same as the all-sky value. However, since the survey is a small fraction of the sky, its density is higher (structures cluster more toward smaller scales) and this results in a systematic suppression of $\hat{\omega}(\theta)$. We correct for this effect, even though it is not significant for our results. Details can be found in Appendix \ref{s_ic}. The values of the corrected ACF for all samples are displayed in Table \ref{acf_values}.

\subsection{Error estimation}\label{s_errors}

We estimate errors with the jackknife technique, which uses the observed data and is very effective in recovering the covariance of $\hat{\omega}(\theta)$ between different scales. First, the entire sample is divided into $N_\mathrm{jack}=64$ spatial regions of equal size. Then, the correlation is run $N_\mathrm{jack}$ times, each one excluding one of those regions from the sample. The value of the estimator is the average $\bar{\omega}(\theta)$ of those iterations and the covariance between angular bins is given by \citep{scranton02}

\begin{equation}
C_{jk} = \frac{N-1}{N}\sum\limits_{i=0}^N \left[ \hat{\omega}_i(\theta_j)-\bar{\omega}(\theta_j) \right]\left[ \hat{\omega}_i(\theta_k)-\bar{\omega}(\theta_k) \right]. \label{Cij}
\end{equation}
We also compare the jackknife errors with those obtained from mock simulations, which are described in Appendix A. We find that both sets of errors have a good agreement, with differences around 20\%. Although our mock simulations only cover large-scales, the systematic differences between mock and jackknife errors are not expected to vary significantly across different scales for a projected statistic like $\omega(\theta)$ \citep{norberg09}.

\section{placing galaxies in haloes}\label{s_placing}

The galaxy bias $b_g$ (see Equation \ref{bias}) encodes all the information that can be extracted from the two-point galaxy distribution, given a particular cosmology. Thus, our aim is to construct a precise model of $b_g$ and adjust the resulting correlation function to match the observed clustering of galaxies. The main idea behind this model is to assume a halo distribution and place galaxies in haloes according to a set of simple rules, as explained below.

\subsection{The Halo Occupation Distribution}\label{ss_hod}
The distribution of dark matter haloes under the CDM paradigm has been well-studied both phenomenologically and through simulations \citep{ma&fry00, cooray02, berlind02}, leading to a halo model where the halo mass function, the bias $b_h$ and the halo density profile are determined by the halo mass. The Halo Occupation Distribution (HOD) is a statistical framework that has been developed to link the halo model with the distribution of galaxies \citep{berlind02, cooray02, kravtsov04}. The HOD is mainly described with the probability $P(N|M)$ that a halo of a given virial mass $M$ hosts $N$ galaxies; one central and $N-1$ satellites distributed according to a NFW profile. All galaxies are linked to some halo, and the occupation is independent of their formation history and environment \citep{zentner05}. This assumption is generally valid, since the induced changes in the galaxy bias due to environment are expected to be only at the $\sim$5\% level \citep{croton07,zu08}, while the overall uncertainties in galaxy clustering studies are typically larger. For our work in particular, the main source of error arises from the uncertainty in the shape of the redshift distribution, which is explored in Appendix \ref{a_EvsC} by comparing results from the use of COSMOS and EGS as reference catalogs. The variations in galaxy bias are around 10\%\, and they do not alter qualitatively any of the final conclusions. Therefore, given that the environmental effects in the galaxy bias are expected to be smaller, we consider them negligible for the current purposes.\\
\indent The average distribution of central galaxies as a function of halo mass can written as \citep{zheng05, zheng07}: 

\begin{equation}
N_c(M)  = \frac{1}{2}\left[ 1+\mathrm{erf}\left( \frac{\mathrm{log}M-\mathrm{log}M_\mathrm{min}}{\sigma_{\mathrm{log}M}}   \right) \right].    \label{e_nc}
\end{equation}
This implies that $N_c(M_\mathrm{min})=0.5$. Thus, $M_\mathrm{min}$ sets a step-like transition where half of the haloes above this mass will host a central galaxy, and this transition is smoothed by the scatter $\sigma_{\mathrm{log}M}$. The number of satellites galaxies is drawn from a Poisson distribution with mean

\begin{equation}
N_s(M)  = N_c(M) \left( \frac{M-M_0}{M_1^\prime} \right) ^\alpha,    \label{e_ns}
\end{equation}
and are assumed to follow a NFW \citep{nfw} density profile from the halo center. The factor $N_c(M)$ accounts for the constraint that only haloes with a central galaxy may host satellites. Equation \ref{e_ns} represents a power law, where $\alpha$ sets the steepness, $M_1^\prime$ defines the typical mass scale for this distribution being close to unity and $M_0$ represents the mass below which the power-law is cut off. In addition, one can derive the characteristic mass where a halo hosts exactly one satellite on average, $M_1$, by imposing $N_s(M_1)\equiv 1$ and noting that generally $N_c(M\approx M_1^\prime)=1$. In the case where $M_0=0$ it reduces simply to $M_1 = M_1^\prime$, and when $\alpha=1$ then $M_1 = M_1^\prime+M_0$. The occupation distribution of the total number of galaxies in a halo can be expressed as the sum of the central and satellite terms:
\begin{equation}
N(M)   = N_c(M) + N_s(M) .    \label{}
\end{equation}
Other HOD derived quantities are the effective galaxy bias
\begin{equation}
b_g^\mathrm{eff} = \frac{1}{n_g}\int dM \frac{dn(M)}{dM}N(M)b_h(M), \label{e_bg}
\end{equation}
and the fraction of satellite galaxies
\begin{equation}
f_\mathrm{sat} = \frac{1}{n_g}\int dM \frac{dn(M)}{dM} N_s(M).\label{e_fsat}
\end{equation}
Further details about the halo model used here can be found in the Appendix \ref{a_halo}.

\subsection{Redshift scaling} \label{ss_redshift_scaling}

\begin{figure}
\includegraphics[trim=10mm 0mm 5mm 0mm,clip=True,width=\columnwidth]{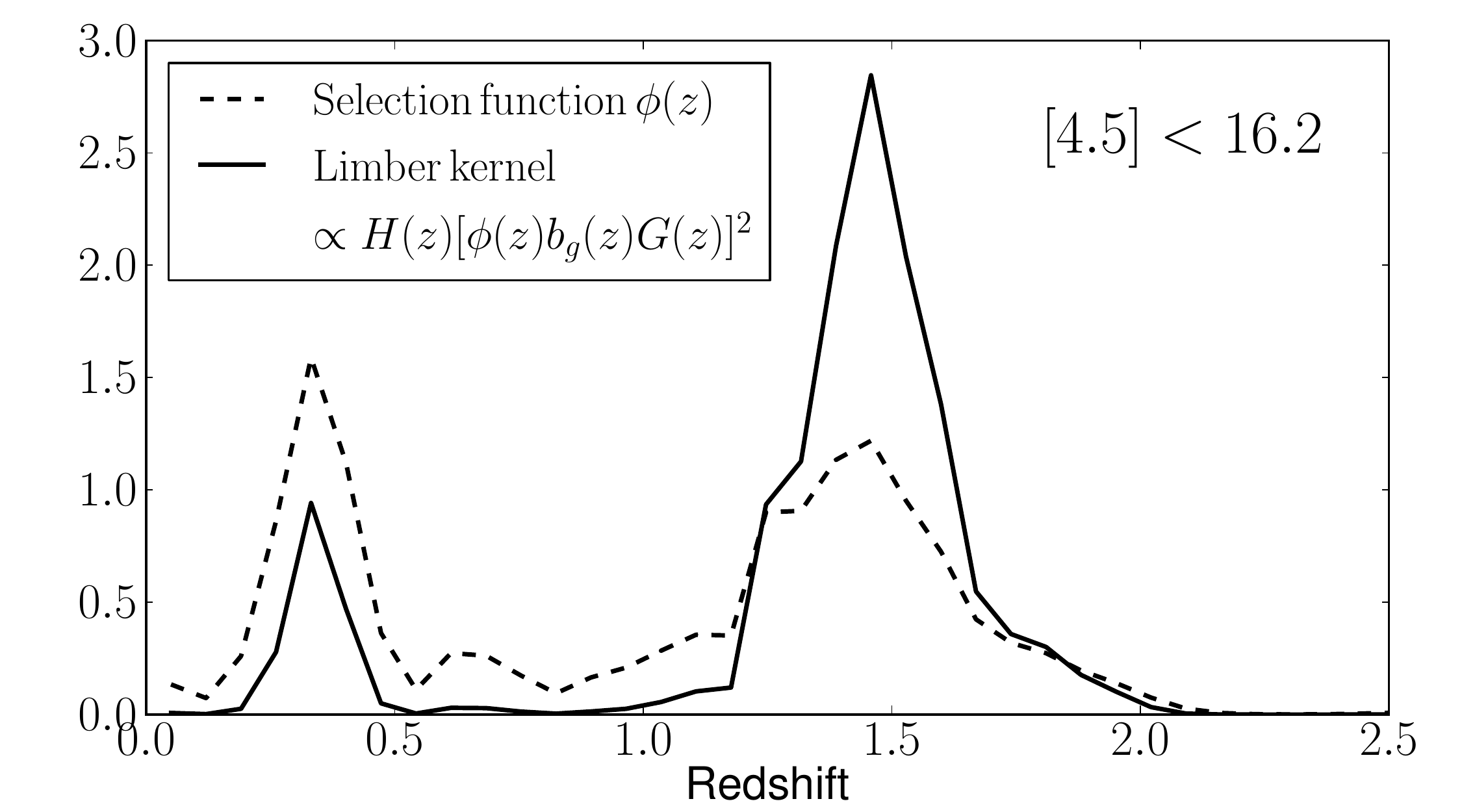}
\caption{\label{dndz_limber} Comparison of the normalized redshift distribution $\phi(z)$ of the brighest sample (15$<$[4.5]$<$16.2), which has the most prominent low-redshift bump, and the corresponding Limber kernel. The y-scaling is arbitrary in either curve. In the Limber projection, the bias function boosts the contribution of high-$z$ galaxies, since they are also more massive. This effect minimizes the contribution of the low-redshift bump to the ACF.  }
\end{figure}

We aim to fit the halo model at the pivot redshift $z_p=1.5$. However, the galaxies in our samples have redshift distributions that are too broad to be neglected or averaged over (see Fig. \ref{dndz1D_set}). Thus, our approach is to produce $\xi_{g}^p \equiv \xi_{g}(z=z_p)$, and scale it using a simple prescription to generate $\xi_{g}$ at all other redshifts. We then use Equation \ref{limber} to make a redshift projection of $\xi_{g}(z)$ onto $\omega(\theta)$. \\
\indent The scaling we apply is based on how the large-scale clustering (represented by the two-halo term $\xi_g^{2h}$, see Equation \ref{1_2halo}) varies with redshift. This change in amplitude is driven by the growth factor $G(z)$ of the dark matter and the galaxy bias $b_g(z)$. Hence, we can write 
\begin{equation}
\xi_{g}(z) =  \frac{G^2(z)}{G^2(z_p)} \frac{b_g^2(z)}{{b^\mathrm{eff}_g}^2}  \xi_{g}^p, \label{xi_z}
\end{equation}
where $b^\mathrm{eff}_g$ and $\xi_{g}^p$ are set at $z_p$ by construction. Here, we have made the approximation that the entire correlation function can be scaled with a single factor. However, the relative amplitude of the one and two-halo terms is known to evolve \citep{conroy06,watson11}, in the sense that typically the one-halo term is more prominent at higher redshift. We have tested how $\omega(\theta)$ would change if we allow for some differential redshift scaling between the one and two-halo terms of $\xi_{g}$. This was done applying a linearly redshift-dependent factor to the one-halo term, in addition to the general scaling from Equation \ref{xi_z}. In this way, below and above $z_p$ the one-halo becomes reduced and boosted, respectively. We find that this has a very little effect on the resulting ACF. This is because the redshift distributions of our galaxies are more or less symmetric, so that the relative scaling of the one-halo above and below $z_p$ is almost canceled when these contributions are summed together. In reality, this relative scaling might have a more complex dependence on redshift, but we believe that the linear representation we considered here is adequate given the symmetric and peaked forms of our redshift distributions. Thus, we find that our model is not sensitive to the particular evolution of the one-halo term and do not incorporate it in the determination of our results.  \\
\indent Our general approach is to calculate the bias as a function of the evolving median stellar mass, $b_g(z) = b_g(\bar{M}_\star^\mathrm{full}(z))$. For this purpose, we make use of the galaxy bias as a function of stellar mass and redshift presented in Moster et al. (2010, hereafter M10), $b_g^\mathrm{M10}(M_\star, z)$. However, we do not use their bias values directly since we need to enforce that $b_g(z_p)=b^\mathrm{eff}_g$, i.e., the bias function has to match the HOD bias at the redshift of the fit. Our bias function is normalized to hold that constraint, but the scaling at other redshifts is adopted from M10 (for a given stellar mass). To accomplish this, first we define the stellar mass $M^\prime_\star$ where $b_g^\mathrm{M10}(M^\prime_\star, z_p)= b^\mathrm{eff}_g$. Ideally, $M^\prime_\star$ would be equal to the median mass of the sample from Section \ref{ss_stellarmasses}, $\bar{M}_\star^\mathrm{full}(z_p)$, but they differ. This is not surprising, since the modeling in M10 is based on abundance matching, which is different from our clustering approach and can potentially yield differing values of the bias. In addition, some variations are expected given the differences in models and codes used to derive stellar masses in M10 and our reference sample.  However, the M10 masses by themselves are not relevant to us, and they simply represent a quantity or label that links brighter populations of galaxies with a larger bias. Therefore, it is sufficient to assume \emph{a priori} that all these masses hold a monotonic relationship with sample luminosity, which has been proven correct \emph{a posteriori}. In other words, $M^\prime_\star$ does scale monotonically with $\bar{M}_\star^\mathrm{full}$ across all samples. So, for a given sample, what we calculate is the offset $\Delta \mathrm{log}M_\star =  \mathrm{log}M^\prime_\star-\mathrm{log}\bar{M}_\star^\mathrm{full}$ at $z_p$. In this way, we are able to ``convert'' our stellar masses into M10 masses. Our bias function then becomes 
\begin{equation}
b_g(z)=b_g^\mathrm{M10}(\mathrm{log}\bar{M}_\star^\mathrm{full}(z) + \Delta \mathrm{log}M_\star, z). \label{e_bm10}
\end{equation}
In Section \ref{ss_stellarmasses} we calculated $\bar{M}_\star^\mathrm{full}(z)$ for all samples. For the brightest ones, the data becomes a bit noisy due to the low number density (see Figure \ref{medmass_z}). This mass distribution is consistent with having a constant shape and varying it by some normalization that scales with the magnitude limit of the sample. Thus, we adopt the functional form of the largest sample $\bar{M}_\star^\mathrm{full}(z;\,k=13)$, which is given by the polynomial fit shown in Figure \ref{medmass_z}. The normalization of this function does not need to be taken into account, since it will be implicitly incorporated in $\Delta \mathrm{log}M_\star$.\\
\indent The stellar mass dependence of the bias has an important effect on $\omega(\theta)$. Because the stellar mass of our samples is larger at high-redshift (about 10 times larger at $z=1.5$ than at $z=0.5$; see Figure \ref{medmass_z}), the bias will place a stronger weight there than at low-redshift. This helps to minimize the contribution of the undesired low-redshift bump at $z\sim0.3$. Additionally, there are other functions weighing in the Limber projection, which is proportional to $ H(z)[\phi(z) b_g(z) G(z)]^2$, as inferred from Equations \ref{limber} and \ref{xi_z}. Figure \ref{dndz_limber} shows the comparison between the redshift distribution $\phi(z)$ and the full Limber kernel. It is shown for the brightest sample because it is the one with the highest fraction of low-redshift contaminants. In the end, the low-redshift contribution to the clustering is minimized due to the decrease in $\bar{M}_\star^\mathrm{full}(z)$, which suppresses $b_g(z)$. This effect is convenient for our analysis, since it makes the clustering properties of our samples highly representative of the $z\sim1.5$ Universe. Moreover, in Appendix \ref{nosc} we investigate how the final results are impacted by the use of available optical data in the SSDF field to remove low-redshift sources. We find that the changes in the results are negligible compared to keeping these sources and modeling their weak contribution to the clustering, as done in this Section.\\
\indent A possible concern at this point is that the results from M10 would be ``built-in'' to ours through the coupling with Equation \ref{e_bm10}. However, the normalization of the bias is set by our own data, and it is the redshift modulation that we incorporate from these authors. In addition, we have explored variations of $b_g^\mathrm{M10}(M_\star,z)$ and determined that our model is not very sensitive to such changes. As seen in Figure \ref{dndz_limber}, the redshift modulation plays a role in weighing galaxies at $z=0.3$ vs $z=1.5$. Basically, any function that down-weights the low-redshift bump will do so in a manner that it becomes quickly subdominant. We just need a function that reflects approximately the variation of the bias with stellar mass and redshift, which is precisely what is provided by scalings from M10. Our model does not strongly depend on the detailed form of this function, and we have verified that our results are not pre-set in a significant way by those in M10.

\section{HOD Model fits}\label{s_fits}
\begin{figure}
\includegraphics[trim=0mm 0mm 10mm 0mm,clip=True,width=\columnwidth]{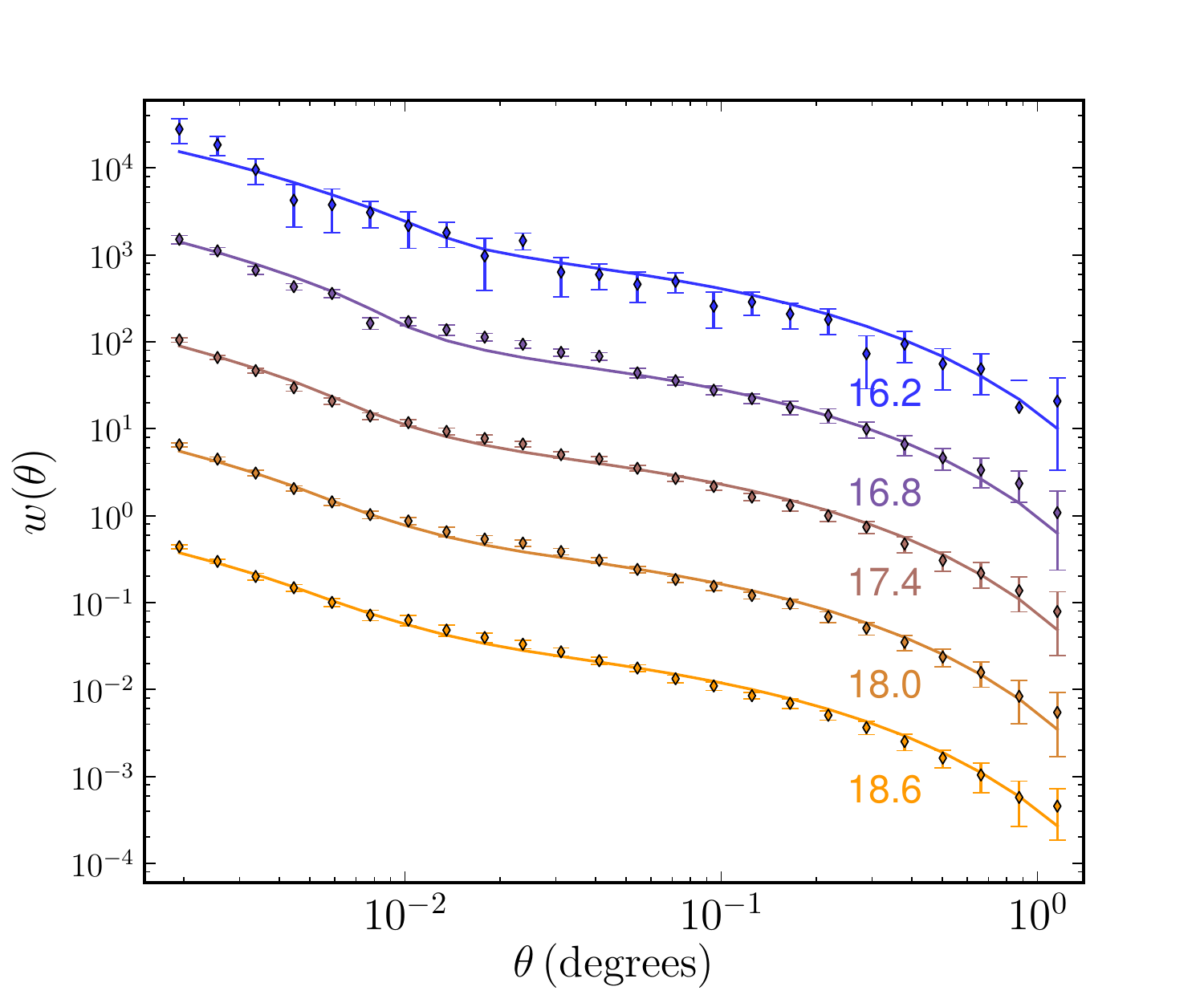}
\caption{\label{w_fits} Observed angular correlation function (points with error bars) of the samples $15<[4.5]<\{16.2,16.8,17.4,18.0,18.6\}$. The solid curves correspond to the best model fits. An extra decade has been added between consecutive curves for easier visualization. The error bars are drawn from the diagonal elements of the covariance matrix (Equation \ref{Cij}). In general, neighboring points are positively correlated, while those far apart are anti-correlated. This is an inherent property of the ACF estimation \citep{norberg01,scranton02}, but can be effectively taken into account via a metric of the form of Equation \ref{chi2}.   }
\end{figure}

The fitting procedure is based on maximizing the likelihood of the model given the observable $\mathcal{L}(${\footnotesize mod}$|${\footnotesize obs}$)=e^{-\chi^2}$, with
\begin{equation}
\chi^2 = \sum\limits_{i=0}^N \sum\limits_{j=0}^N \left[ \omega_m(\theta_j)-\bar{\omega}(\theta_j) \right] C_{ij}^{-1}  \left[ \omega_m (\theta_k)-\bar{\omega}(\theta_k) \right]. \label{chi2}
\end{equation}
Here, $\omega_m$ and $\bar{\omega}$ are the model and observed ACFs, $C_{ij}$ is the covariance matrix from Equation \ref{Cij} and $N=28$ is the number of angular bins.
The halo occupation model we consider has a total of 5 parameters: $M_\mathrm{min}$, $M_1^\prime$, $M_0$, $\alpha$ and $\sigma_{\mathrm{log}M}$. Even though the signal-to-noise of our ACFs is very good ($11-31\sigma$ with respect to the null hypothesis), the fact that it is the result of projecting the SCF across a wide redshift distribution reduces our constraining power on the HOD model. Thus, to avoid over-ftting the data, we choose to fix a number of parameters. We have run sets of Monte Carlo Markov chains to explore the sensitivity of the model to different choices of constraints. To evaluate this sensitivity, we use the Akaike criterion \citep{akaike74}, which states that an extra free parameter is justified only when the new best-fit $\chi^2$ is reduced by an amount larger than 2. For either $\sigma_{\mathrm{log}M}$ and $M_0$, this criterion is not fullfilled. Thus, we follow \citet{conroy06} and set $\mathrm{log}M_0 =0.76\,\mathrm{log}M_1 +2.3 $. We also fix $\sigma_{\mathrm{log}M}=0.2$, following a number of studies that support typical values $>0.15$ \citep{more09,behroozi10, more11,wake11,moster13,reddick13, behroozi13}. In the case of $\alpha$, we have that $\Delta \chi^2 \approx 3$, which would mildly favor setting it free. However, this parameter has an intrinsic degeneracy with $M_1^\prime$ and when left free to float, the best fit values show a significant stochastic component in their behavior with respect to sample luminosity. It cannot be constrained as well as $M_1$, and thus we decide to fix it to a common choice in the literature that is also supported by simulations, $\alpha=1$ \citep{kravtsov04,zentner05,tinker05,zheng05,zehavi11,wake11,leauthaud12}. None of the final conclusions in this work change whether or not we allow $\alpha$ to vary freely. Additionally, Equation $\ref{ng}$ fixes $M_\mathrm{min}$ through the observed galaxy number density, leaving $M_1^\prime$ as the only parameter left in the fit. In Appendix \ref{a_tests} we comment on how the resulting HOD model changes if we leave nearly all parameters free in the fit.\\
\indent Obtaining the best-fit value of $M_1^\prime$ is straightforward. The error in the fit can be estimated from the width of the likelihood distribution, but it does not account for departures arising from cosmic variance. To account for that, we perform a set of 100 random realizations of the redshift distribution and number density at $z_p$, which we call $\phi^\mathrm{rd}_\mathrm{cut}(z)$ and $n_g^\mathrm{rd}$. We find the best HOD fit $M_1^\prime$ each time, along with the corresponding derived parameters. Each redshift $j$-bin in $\phi^\mathrm{rd}_{\mathrm{cut},\,j}$ is drawn from a normal distribution with mean $\phi_{\mathrm{cut},\,j}$ and standard deviation as in Section \ref{ss_phi} (see Figure \ref{dndz1D_set}). The value for $n_g^\mathrm{rd}$ is produced in a similar manner; using a normal distribution with mean and standard deviation equal to the value and error of $n_g$ in Table 1. The scatter in all parameters from the random realizations is clearly dominant over that arising from the width of the $M_1^\prime$ likelihood in the fiducial fit, especially due to the variations in $n_g$. We can therefore approximate the final errors as those from the random realizations.  \\
\indent Figure \ref{w_fits} shows the observed ACF and the model fits for a few samples. The values and errors for all relevant parameters are given in Table 2.

\section{The stellar-to-halo mass ratio}\label{s_shmr}

Haloes of masses equal to $M_\mathrm{min}$ host on average 0.5 central galaxies with luminosities greater than the sample threshold (Equation \ref{e_nc}). \citet{zheng07} showed analytically that central galaxies living in these particular haloes have a median luminosity corresponding to the limit of the sample. This links halo masses with luminosities, albeit with some scatter $\gtrsim0.15$ dex \citep{zehavi11, coupon12}. However, we are interested in the connection with stellar masses, which also have a well-defined mean and scatter at fixed luminosity (see Section \ref{ss_stellarmasses}). Hence, we can link $M_\mathrm{min}$ to $M^\mathrm{lim}_\star$, with a scatter ($\sigma_{\mathrm{log}M}$) that ought to be close to the quadratic sum of the scatters from the luminosity-$M_\mathrm{min}$ and luminosity-$M^\mathrm{lim}_\star$ relations. We measure the latter to be around 0.2 dex, and the former is expected to be similar. Thus, the fact that we fix $\sigma_{\mathrm{log}M}=0.2$ might seem an underestimation. However, as we will discuss in Appendix \ref{a_tests}, an unconstrained HOD fit does not prefer larger values for this parameter. Also, the final results do not change significantly by increasing it to larger values as 0.4 dex. We thus retain our choice and proceed. \\
\indent The values of $\bar{M}^\mathrm{lim}_\star$ and $M_\mathrm{min}$ for our samples can be found in Table \ref{t_samples}. Their ratio yields the SHMR, which is plotted as a function of halo mass in Figure \ref{f_shmr} for our different sets of stellar masses. The vertical error bars are a combination of the halo mass uncertainty and the error in the median stellar masses (equation \ref{meanerrs}), i.e., it does not represent the scatter in stellar mass at fixed halo mass. It is interesting that the error bars do not get notably bigger for brighter samples, even though the ACFs of those are much noisier and the stellar mass errors are indeed larger. The reason is that $M_\mathrm{min}$ becomes progressively less sensitive to the HOD fit at higher luminosities. The fit is based on $M_1^\prime$, which falls close to the steep drop of the halo mass function (equation \ref{e_mf}) in the bright samples and makes the overall HOD model be weakly affected by the satellite occupation  (e.g., Equation \ref{ng}). Thus, the error contribution from $M_1^\prime$ is minimized, and that from $n_g$ and $M^\mathrm{lim}_\star$ increases, keeping the total error roughly constant across the different samples.  \\

\subsection{Comparison to other results at $z=1.5$}
\begin{figure*}
\begin{center}
\includegraphics[width=1.8\columnwidth]{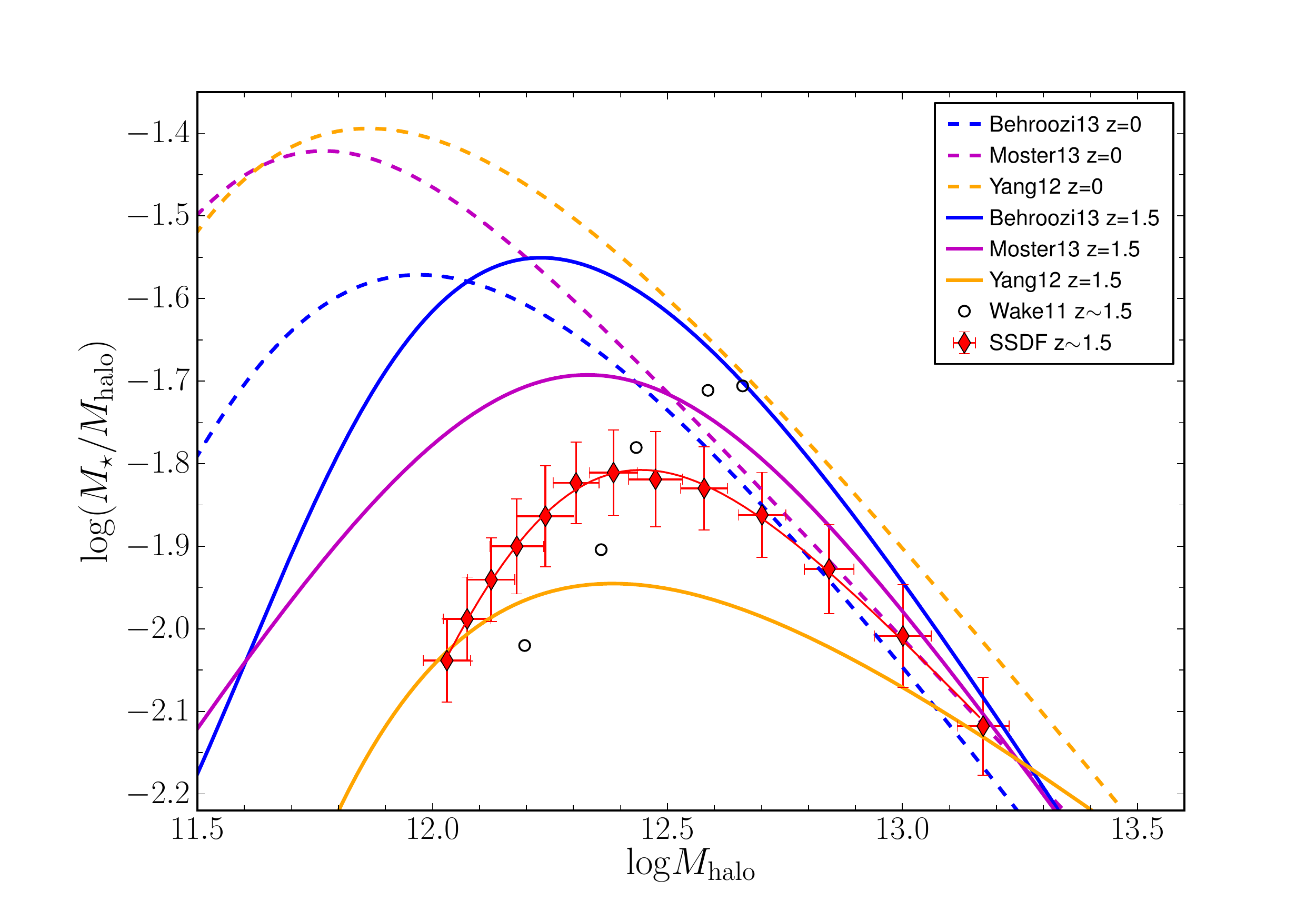}
\end{center}
\vspace{-2ex}
\caption{\label{f_shmr} Stellar-to-halo mass ratio from our study and predictions from other authors. Dashed and solid lines are predictions at $z=0$ and $z=1.5$, respectively. Our points are plotted as $\mathrm{log}(\bar{M}_\star^\mathrm{lim}/M_\mathrm{min})$ versus $\mathrm{log}M_\mathrm{min}$. The error bars are strongly correlated between neighboring points, since our galaxy samples are defined in cumulative magnitude bins. We fit the parametrizations from those authors to our data, robustly measuring a maximum at $\mathrm{log}M_\mathrm{peak}=12.44\pm0.08$. This characteristic mass scale is $\sim 4$ times larger than what is found at $z=0$. The M13 fit is shown as the thin red curve. We also include data from \citet{wake11} as empty circles, where their M05-based stellar masses have been increased by 50\% to approximately match the BC03 masses used by other authors.    }
\end{figure*}

\begin{figure}
\includegraphics[trim=5mm 0mm 5mm 0mm,clip=True,width=\columnwidth]{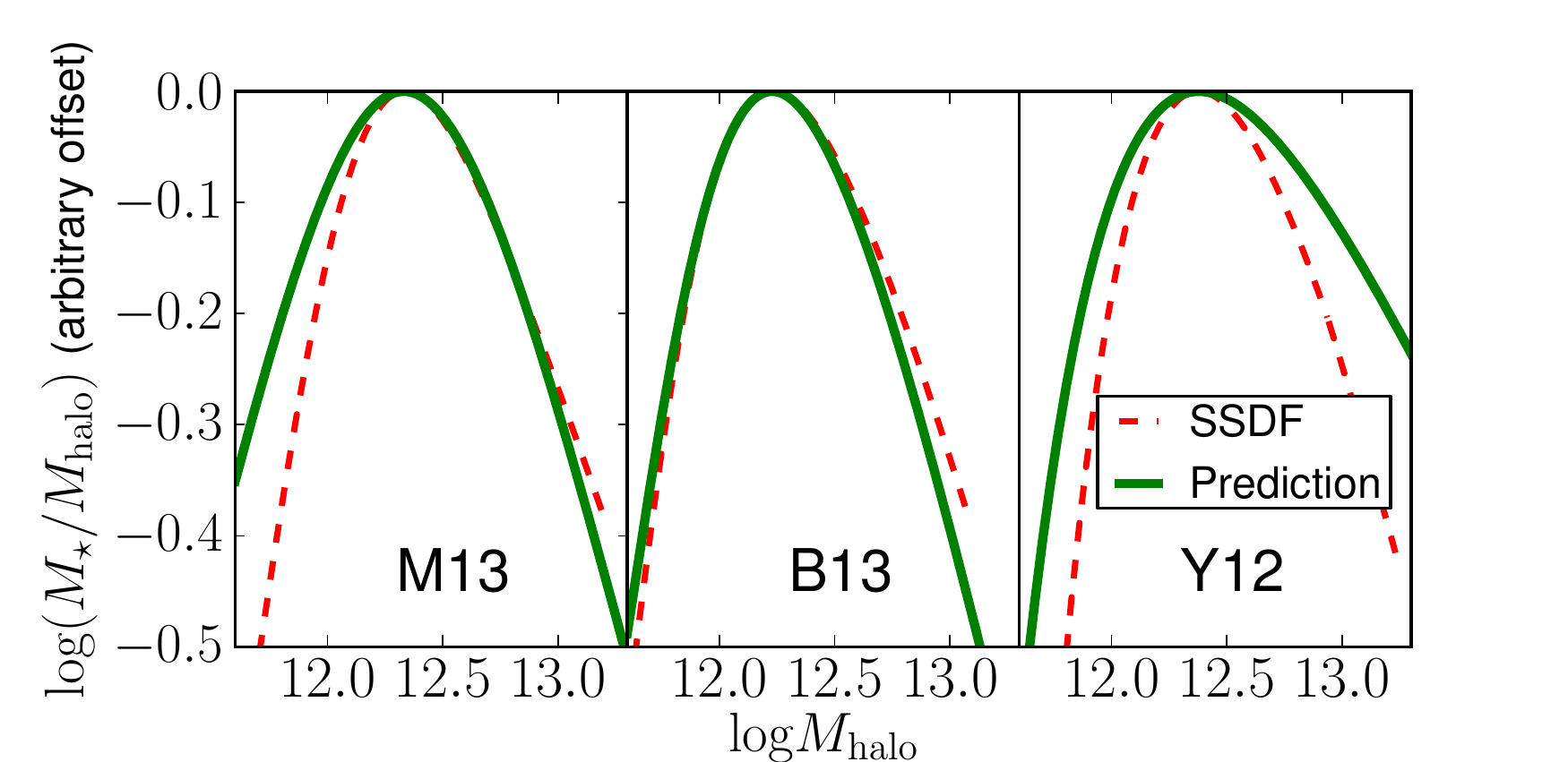}
\caption{\label{f_shmr2} Comparison of the high and low mass slopes between model predictions \{M13,B13,Y12\} at $z=1.5$ and the fits of their parametric models to our data. In each panel, we have offset all curves to the same peak value and shifted our curves in mass to match the peak position of the prediction. This has been done to help the eye in comparing the slopes at either side of the peak. Our data shows a moderate discrepancy compared to the predictions.   }
\end{figure}

There are several studies that have tried to constrain the SHMR at $z>1$, based on abundance matching \citep{moster13,behroozi13}, HOD modeling \citep{zheng07,wake11, coupon12} and extensions using conditional luminosity functions \citep{yang12, wang13}. Some of these works also provide their own parametric form for the SHMR as a function of halo mass, and we will use three of them to fit our points. These are the forms from \citet{yang12}, \citet{moster13} and \citet{behroozi13} (hereafter Y12, M13, B13), which read:

\begin{equation}
\mathcal{S}^Y(m) = M_0^Y \left(\frac{m}{M_p^Y}\right)^{\alpha^Y+\beta^Y} \left(1+\frac{m}{M_p^Y}\right)^{-\beta^Y},  \label{shmr_y}
\end{equation}

\begin{equation}
\mathcal{S}^M(m) = 2 N^M \left[  \left(\frac{m}{M_p^M}\right)^{-\beta^M} + \left(\frac{m}{M_p^M}\right)^{\gamma^M}    \right]^{-1}    \label{shmr_m}
\end{equation}
and
\begin{equation}
\mathrm{log}\mathcal{S}^B(m) =\mathrm{log}(\epsilon^B M_p^B) + f\left(\mathrm{log}\frac{m}{M_p^B}\right)-f(0)- \mathrm{log}(m)  \label{shmr_b}
\end{equation}
with
\begin{equation}
f(x) = -\mathrm{log}(1+10^{\alpha^B x}) + \frac{\delta^B [\mathrm{log}(1+\mathrm{exp}(x))]^{\gamma^B}}{1+\mathrm{exp}(10^{-x})}.
\end{equation}
The superscript labels $\{Y,M,B \}$ refer to the author names. The position of the peak is mostly modulated by the pivot mass $M_p^{ Y,M,B}$. In Y12 and M13, the low mass logarithmic slopes are set by $\beta^M$, $\alpha^Y+\beta^Y-1$ and the high-mass slopes by $\gamma^M$,  $\alpha^Y-1$, respectively. In the case of B13, the link between the slopes and the parameters is less straightforward, but the low and high-mass regimes are mostly modulated by $\alpha^B$ and $\delta^B$. $\gamma^B$ tunes the high-mass behavior of $\mathrm{log}\mathcal{S}^B(m)$ going from logarithmic at $\gamma^B=0$ to power-law at $\gamma^B=1$ (see Behroozi et al. 2013). We set all of these parameters free when perform orthogonal regression fits of $S^{ Y,M,B}$ to our measurements of the SHMR. However, we do enforce $\gamma^B\leq1$ and $\beta^Y\leq100$ (see Yang et al. 2012), which are limits by definition. \\
\indent These authors mainly use measures of the stellar mass functions at different redshifts to build a redshift evolution model of the SHMR. They provide explicit redshift dependence for all parameters in Equations \ref{shmr_m}-\ref{shmr_b}. Thus, we use them to compare our measurements to the predicted SHMR of these authors at the redshift of our survey. This is shown in Figure \ref{f_shmr}, where we plot their predictions at low and high-redshift. In addition, the specific parameter values of the $z=1.5$ curves, for both the predictions and the fits to our data, are displayed in Table \ref{t_shmr}. The normalization values $N^M$, $\mathrm{log}\epsilon^B$ and $\mathrm{log}M_0^Y$ are also fitted, although we disregard any interpretation of them because there are important systematic uncertainties in the stellar masses between different authors. A thorough examination of these to allow a meaningful comparison is beyond the scope of this paper. For the current purposes, we simply assume that the differences in stellar masses are due to a simple logarithmic offset. This assumption holds well when comparing different sets of masses in the COSMOS and EGS catalogs. In addition, M13 and B13 use stellar masses based on BC03 and Chabrier IMF, which matches our fiducial choice of masses. Y12 use masses produced with the \citet{pegase} models and Kroupa IMF, but we still do not expect a significant deviation from a constant offset when compared to our masses \citep{barro11b}. We have checked this based on the masses from this particular model that are also available in the EGS control catalog. We also note that we use the SHMR in Y12 that is based on fits ``CSMF/SMF1'', where only stellar mass functions are utilized.  \\
\indent We limit the comparison between all measurements to the centroid position and slopes of the SHMR. There are some discrepancies when comparing our results to the predictions from the other authors, but these are not dramatic (see below). The centroid of the SHMR is computed as the actual peak position in the parametric relations, and the fits of these models to our data show $\mathrm{log}M_\mathrm{peak}=12.44\pm0.08$ (see Table \ref{t_shmr}). We did not compute confidence intervals for the $M_\mathrm{peak}$ predictions because it requires knowledge of the explicit covariance between their parameters fits. Our value is larger than these predictions; based only on our errors, it lies 0.8$\sigma$ above Y12, 1.2$\sigma$ above M13 and 2.7$\sigma$ above B13. Because their errors are not being taken into account, these offsets should not be treated as absolute levels of inconsistency with respect to our study.    \\
\indent For the slopes, there are also some slight discrepancies. To better visualize this comparison, we have plotted in Figure \ref{f_shmr2} the prediction and the fits to our data for each parametric model. All curves in each panel are scaled in the x-axis to match the peak of the prediction, and scaled in the y-axis to set all peak heights to zero. The idea is to fix the peak position (in both axes) of all curves to better compare the slopes on either side. In this case, the \emph{slopes} are the approximate power-law index at either side of the peak, and is not necessarily linked to a parameter in a unique manner (except for M13, where the slopes are independently controlled by $\beta^M$, $\gamma^M$). In comparison to M13, our low mass slopes are steeper (higher $\beta^M$) and the high-mass slopes are shallower (lower $\gamma^M$) than their predictions. With respect to B13, the low mass slopes are in agreement but our high-mass slopes are shallower. In the case of Y12, our slopes are steeper at both low and high-mass.\\
\indent As explained in Section 1, the low-mass slope can be directly related to the importance of energy- versus momentum-driven winds. In general, we find a steeper low-mass slope than the predictions, which favors energy-driven winds. At high-masses, the interpretation of the slope is less clear, since AGN feedback and galaxy mergers should also have an important contribution.  \\
\indent Another important study to compare our measurements with is \citet{wake11}. It is based on HOD modeling of $\sim10^{10} M_\odot$ stellar mass limited galaxies at $z\sim 1.5$, which makes it similar to our work. These authors had the advantage of using data with accurate photometric redshifts and stellar masses, but also the drawback of sampling a small region of the sky (NEWFIRM survey, 0.25 deg$^2$). They had very few galaxies around $10^{11} M_\odot$ and therefore it was not possible to map the full peak of the SHMR. Their data is shown in Figure \ref{f_shmr}, where we have scaled the stellar masses by 50\% to roughly transform them from the M05 model to BC03. These authors performed a parametric fit and found a peak at $\mathrm{log}M_\mathrm{peak}=12.63$ (an estimated uncertainty was not provided), which lies 2.5$\sigma$ above our result of $12.44\pm0.08$.

\begin{table}
\begin{center}
\caption{\label{t_shmr} Best-fit SHMR parameters. \textbf{Columns 1,2:} SHMR functions (see equations \ref{shmr_y}-\ref{shmr_b}) and parameters that describe them. \textbf{Column 3:} Prediction of the parameter values at $z=1.5$, derived by these authors using data from luminosity and stellar mass functions at different redshifts. \textbf{Column 4:} Parameter values derived from fitting these functions to our clustering data at $z=1.5$.}
  \begin{tabular}{cccc}
  \hline \hline
Function & Parameter & Prediction $z=1.5$ & SSDF Fit  \\
 \hline \hline
 & $N^M$ & 0.020$\pm$0.007 & 0.0139$\pm$0.0003 \\ [0.1cm]
$\mathcal{S}^M$ & log$M_p^M$ & 12.31$\pm$0.32 & 12.25$\pm$0.02 \\ [0.1cm]
 (Moster & $\beta^M$ & 0.88$\pm$0.20 & 1.64$\pm$0.09\\ [0.1cm]
 et al. 2013)& $\gamma^M$ & 0.81$\pm$0.12 & 0.60$\pm$0.02 \\ [0.1cm]
 & log$M_\mathrm{peak}$ & 12.33 & 12.44$\pm$0.08 \\ [0.1cm]
\hline \\[-0.2cm]
 & log$\epsilon^B$ & -1.70$\pm$0.16 & -2.03$\pm$0.17 \\ [0.1cm]
 $\mathcal{S}^B$& log$M_p^B$ & 11.88$\pm$0.13 & 12.03$\pm$0.15 \\ [0.1cm]
 (Behroozi & $\alpha^B$ & -1.64$\pm$0.09 & -2.10$\pm$0.16 \\ [0.1cm]
 et al. 2013) & $\gamma^B$ & 0.12$\pm$0.25 & 0.32$\pm$0.26 \\[0.1cm]
 & $\delta^B$ & 2.65$\pm$0.90 & 3.31$\pm$1.15\\[0.1cm]
 & log$M_\mathrm{peak}$ & 12.23 & 12.44$\pm$0.07\\ [0.1cm]
\hline \\[-0.2cm]
 & log$M_0^Y$ & 9.57$\pm$0.31 & 10.65$\pm$0.13 \\ [0.1cm]
 $\mathcal{S}^Y$& log$M_p^Y$ & 10.48$\pm$0.22 & 10.37$\pm$0.30 \\ [0.1cm]
 (Yang & $\alpha^Y$ & 0.56$\pm$0.11 & 0.16$\pm$0.04 \\[0.1cm]
 et al. 2012) & $\beta^Y$ & 35$\pm$30 & 100\\[0.1cm]
 & log$M_\mathrm{peak}$ & 12.38 & 12.44$\pm$0.06 \\
\hline
\end{tabular}
\end{center}
\end{table}

\subsection{Evolution with redshift}
\begin{figure}
\includegraphics[trim=0mm 0mm 5mm 0mm,clip=True,width=\columnwidth]{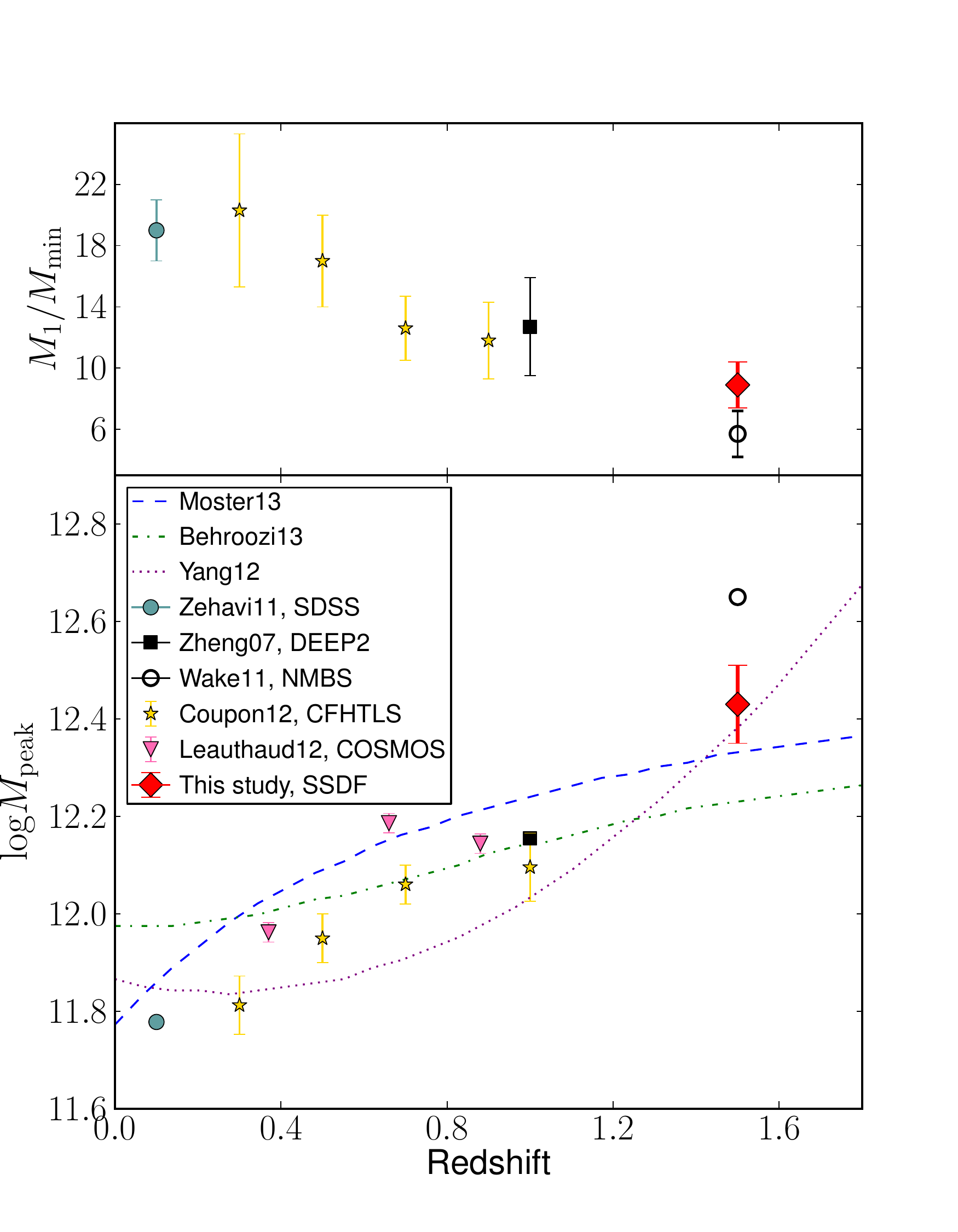}
\caption{\label{f_mpeak}\textbf{Top:} Evolution in the $M_1/M_\mathrm{min}$ ratio for samples with density $n_g=10^{-3}\mathrm{Mpc}^{-3}$, collected from different HOD studies. A decline in this ratio with redshift is measured consistently and agrees with results from $N$-body simulations. The basic interpretation is that at high-redshift there is a larger rate of halo infall, which increases the fraction of similar mass galaxies and reduces $M_1/M_\mathrm{min}$ (see text). \textbf{Bottom:} Evolution in the peak halo mass of the SHMR. Our results show that $M_\mathrm{peak}$ has decreased by a factor of 4.5 through $z=1.5-0$. In combination with other HOD measurements (points), the evolution seems to be monotonic. The curves show predictions from CLF and abundance matching studies.     }
\end{figure}

At this point, we can compare our result for the peak in the SHMR with other studies at different epochs and trace its evolution with redshift (Figure \ref{f_mpeak}). We include HOD results from \citet{zehavi11}, \citet{zheng07}, \citet{leauthaud12}, \citet{coupon12} and \citet{wake11}, as well as predictions from M13, B13 and Y12. As mentioned earlier, our peak lies above the predictions and below the value inferred by \citet{wake11}. Looking at the trend with values at other epochs, the peak mass seems to have evolved in a monotonic and quasi-linear way with redshift. Our data supports a change of log$M_\mathrm{peak}=12.44\rightarrow 11.8$ through $z=1.5\rightarrow0$. This means that the halo mass scale that is most efficient at forming and accreting stars to the central galaxy has decreased by a factor of 4.5 during this redshift range. Thus, the downsizing trend of galaxies has continued steadily during the last 10 Gyrs. Low mass galaxies have grown faster than their haloes, while the opposite trend happened for high-mass galaxies. 

\section{Satellite galaxies}\label{s_satellites}

\subsection{Satellite fraction}
The satellite fractions for all of our samples are displayed on Table \ref{t_samples} and plotted in the lower panel of Figure \ref{f_combined}. Note that the satellites making up this fraction are above the sample flux limit, i.e., $f_\mathrm{sat}$ does not refer to the total fraction of satellites that a central galaxy at the flux limit has. The satellite fraction clearly decreases towards the brighter end, which is a manifestation of the drop in the halo mass function. Based upon the model we use, $M_1$ is the scale that sets the occupation number of satellites in a halo of a given mass, and the number of such haloes is given by the mass function. If $M_1$ approaches the cutoff scale of the mass function, then the satellite contribution to the total density will be reduced compared to that of central galaxies. This effect is seen in most studies \citep{zheng07,zehavi11,wake11,coupon12,tinker13}. \\
\indent Our faint-limit value is $f_\mathrm{sat}\sim 0.2$ (see Figure \ref{f_combined}). Compared to the results of $f_\mathrm{sat}\sim 0.3$ obtained at $z=0$ by \citet{zehavi11}, our value is suggestive of a mild increase in the satellite fraction with cosmic time. A similar conclusion was also reached by \citet{coupon12} based on their comprehensive study of samples at $0<z<1$, and such evolution is predicted by some simulations (e.g., Wetzel et al. 2009, 2013). We caution, however, that the sample in \citep{zehavi11} extends to fainter absolute magnitudes that our data set ($M^\star+2.4$ versus $M^\star+1.2$), and so the evidence from this comparison is suggestive rather than conclusive.

\subsection{The $M_1/M_\mathrm{min}$ relation}
\begin{figure}
\includegraphics[trim=0mm 0mm 5mm 0mm,clip=True,width=\columnwidth]{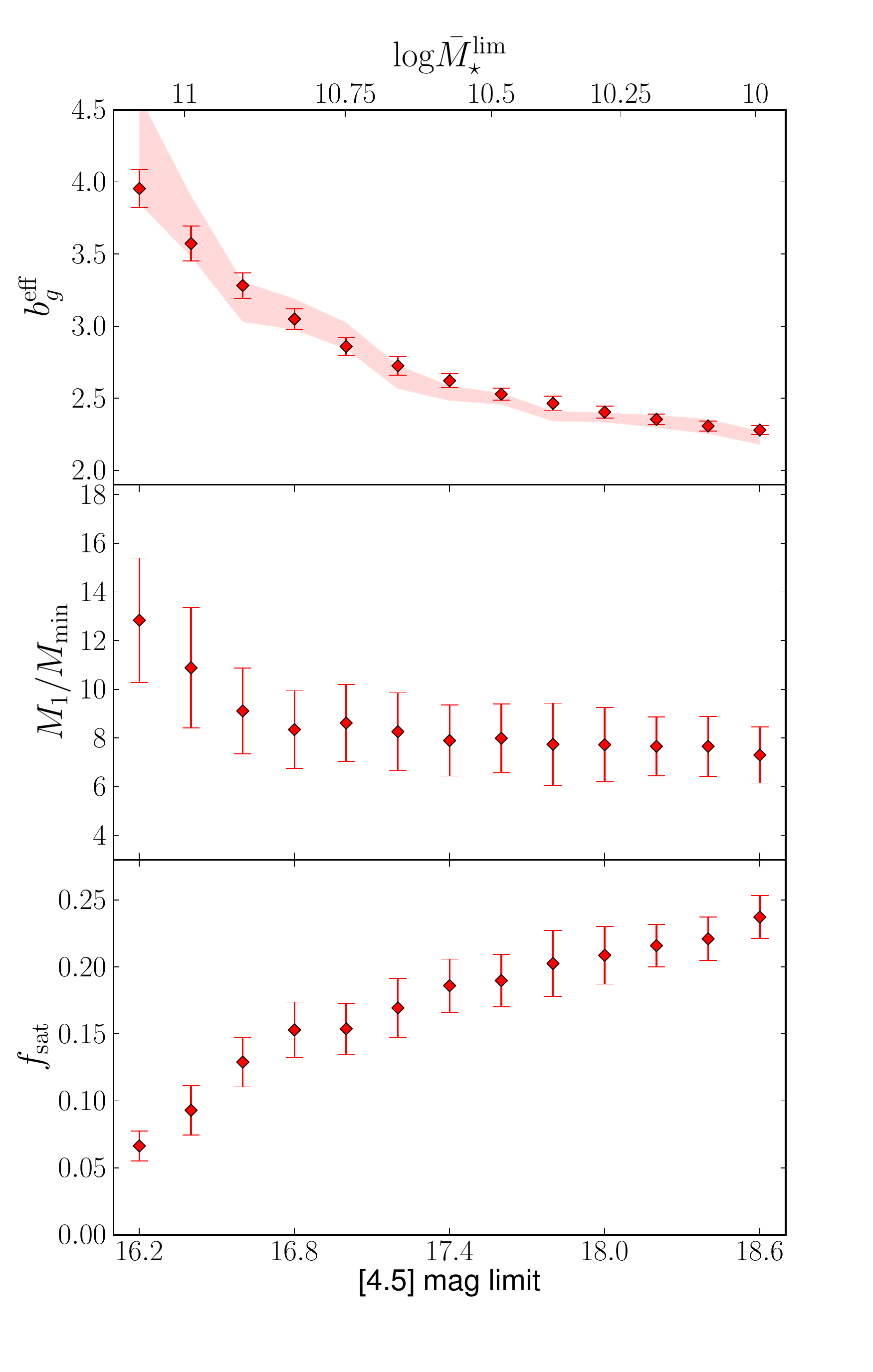}
\caption{\label{f_combined} Results from the HOD fits. Each point denotes a sample defined by a limiting apparent magnitude threshold, which is associated with the median stellar mass $\bar{M}_\star^\mathrm{lim}$. In the top panel, the shaded region represents the $\pm 1\sigma$ interval of direct large-scale bias fits. These are consistent with the HOD bias.   }
\end{figure}

\begin{figure}
\includegraphics[trim=3mm 0mm 5mm 0mm,clip=True,width=\columnwidth]{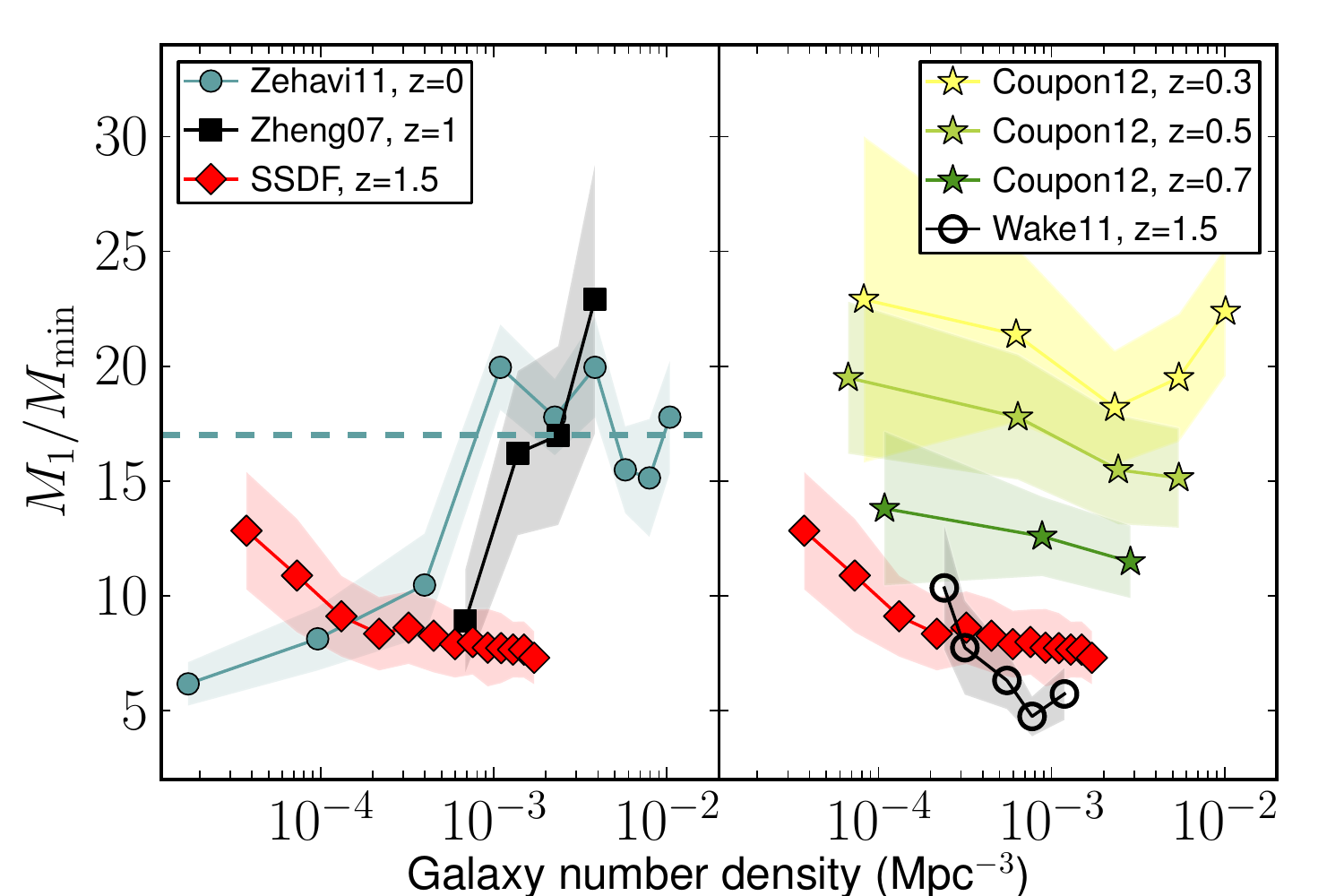}
\caption{\label{f_ratio_ng} Ratio between $M_1$ and $M_\mathrm{min}$ as a function of cumulative galaxy number density. We compare our data (red points) with other studies, grouping them in those showing a decrease (left panel) or an increase (right panel) with number density. The dashed line indicates the ratio of 17 presented in \citet{zehavi11} as the typical value for low-redshift galaxies.     }
\end{figure}

A deeper insight into the relationship between haloes and their satellites is given by the $M_1/M_\mathrm{min}$ ratio. As mentioned in previous Sections, haloes typically become occupied by a central galaxy at $M_\mathrm{min}$ and gain an additional satellite at $M_1$. Thus, at fixed $M_\mathrm{min}$, lowering $M_1$ would directly increase the overall satellite fraction. However, $M_1/M_\mathrm{min}$ holds further clues in relation to the galaxies that occupy these haloes. Because of the decline in the halo mass function towards the massive end, most of the haloes are small and have masses around $M_\mathrm{min}$. These will typically host galaxies that are also small, with stellar mass close to the sample limit $\bar{M}_\star^\mathrm{lim}$. The satellites considered have masses that are also near this limit and living in haloes near $M_1$, where the central galaxy can have a mass much larger than $\bar{M}_\star^\mathrm{lim}$. However, if $M_1$ approaches $M_\mathrm{min}$, then its central galaxy will have a mass closer to $\bar{M}_\star^\mathrm{lim}$. In the case of $M_1/M_\mathrm{min}\gtrsim1$, the satellite will have a stellar mass around $\bar{M}_\star^\mathrm{lim}$ and the central will be slightly more massive than that. Thus, when this ratio is smaller, there is an increased fraction of centrals that have a satellite of similar stellar mass. \\
\indent In the local Universe, $M_1/M_\mathrm{min}\approx 17$ \citep{zehavi11,beutler13}. On the other hand, we measure $M_1/M_\mathrm{min}\approx 9$ at $z=1.5$. A decline of this ratio with redshift had been predicted by simulations \citep{kravtsov04,zentner05} and measured by abundance matching \citep{conroy06} and other HOD studies. Ratios at different redshifts and fixed number density $n_g=10^{-3}\,\mathrm{Mpc}^{-3}$ are shown in the top panel of Figure \ref{f_mpeak}, where it can be seen that there is a general increase towards later times. The reason why both $f_\mathrm{sat}$ and $M_1/M_\mathrm{min}$ are higher at low-redshift is due to the evolution of the mass function. At low-redshift, there are very large haloes that can host multiple satellites, which helps increase the average satellite fraction. However, the fraction of galaxies that are satellites with masses close to the limit of the sample is still larger at high-redshift. This happens because the halo infall timescale is lower, which enhances their accretion rate onto other structures and reduces the gap between $M_1$ and $M_\mathrm{min}$ (see below and Conroy et al. 2006).  \\
\indent However, the most interesting result we derive is the trend of $M_1/M_\mathrm{min}$ with sample luminosity. The middle panel of Figure \ref{f_combined} shows indications of a slight rise at high luminosities, which is not obviously expected. At $z\lesssim1$, this ratio has been observed to be constant or decrease with increasing luminosity at fixed redshift \citep{zheng07,blake08,abbas10,zehavi11,matsuoka11,leauthaud12,beutler13}. Simulations also predict that the accretion rate is larger for more massive haloes at all times \citep{zentner05,wetzel09,mcbride09,fakhouri08,fakhouri10}, which would lower $M_1/M_\mathrm{min}$ at the bright end. We measure the opposite trend at $z=1.5$. Interestingly, \citet{wake11}, the only other study at this redshift that measured HOD for stellar mass selected samples, also obtained a slight increase of this ratio with sample mass. However, those authors did not explore this effect in depth. The results from \citet{coupon12} also hint a similar trend at $z<1$ in haloes of mass $<10^{13}M_\odot$, although they are consistent with a constant ratio. \\
\indent In order to better compare the results from a few different authors, we plot $M_1/M_\mathrm{min}$ as a function of cumulative number density in Figure \ref{f_ratio_ng}. For visual clarity, we show in the left (right) panel those results that follow an increasing (decreasing) trend with density, along with our data. A caveat in this comparison is that, in reality, the number density of a given population does not remain constant through redshift. However, the data from \citet{zheng07} and \citet{coupon12} do not follow the same trends, even though they sample similar redshifts. Thus, from the observational side, the $0<z<1$ data do not offer a consensus regarding the trend with luminosity. At $z=1.5$, our results and those from \citet{wake11} do support a minimal rise in $M_1/M_\mathrm{min}$ with luminosity. \\
\indent As shown in Figure \ref{f_ratio_ng}, two families of curves can be defined. Our results, \citet{wake11} and \citet{coupon12} show a similar shape, offset in the y-axis according to redshift. Meanwhile, \citet{zehavi11} and \citet{zheng07} are similar to one another. One possible effect leading to the disparity between the two sets of results may be the particular selection of galaxy samples. Those in \citet{zehavi11}, \citet{zheng07} and \citet{coupon12} are limited by absolute magnitude in the optical. \citet{wake11} selects directly in stellar mass, and we make a luminosity selection that is later matched to a stellar mass limited sample. Thus, we find no clear explanation for the existence of these two families of curves regarding sample selection. In addition, all these authors (including us) use a very similar form of the HOD, and variations in the chosen cosmology do not have such a strong impact. Regarding possible systematics effects in our modeling, we test different possibilities in Appendices \ref{a_tests} and \ref{nosc} and find nothing that would alter our conclusions. \\

\subsection{Physical mechanisms for a mass-dependent evolution }
The rise of $M_1/M_\mathrm{min}$ with luminosity is not clearly detected. However, \citet{zehavi11} and \citet{zheng07} very clearly measure the opposite behavior at redshifts $z=0$ and $z=1$, respectively, so that even if our data follows a flat trend at $z=1.5$, it would imply that evolution has taken place. Interestingly, there is no obvious mechanism that could be responsible for this change, and we speculate with some possibilities in what follows. The dynamical processes at play can be reduced to a competition between accretion and destruction of satellites. Regarding the former, big structures have recently assembled a larger fraction of their mass than smaller counterparts, at all times \citep{wechsler02,zentner05,fakhouri10}. In other words, the specific growth rate of haloes is an increasing function of mass. Regarding the latter, the dynamical time in bigger haloes is larger, contributing to a slower destruction of accreted satellites. These effects yield a larger number of recently accreted and undisrupted satellites in larger haloes, which would produce a decrease in $M_1/M_\mathrm{min}$ towards higher masses.\\
\indent So, what additional mechanism can reverse this trend at high-redshift? This mechanism could involve the ratio of destruction to accretion being larger at high-masses, which is possible if the dynamical timescale decreases considerably with mass. However, a caveat in these scenarios is that we are implicitly considering that galaxies are accreted or disrupted in the same way as haloes, which does not have to hold. What we are really tracking are galaxies, since $M_1/M_\mathrm{min}$ is inversely proportional to the occurrence of galaxy pairs with masses close to the sample limit. Thus, there could be a star formation dependent process that drives the trend we see with stellar mass. For example, \citet{wetzel13} show that the star formation in satellites fades at the same rate as the central galaxy for a few Gyrs after accretion, but then undergoes a rapid quenching period. They find that quenching timescale is shorter for more massive satellites. Thus, if the central galaxy outgrows the satellites in a way proportional to its own mass, this would produce a lower fraction of similar mass pairs and play in favor of our trend. In addition, such a mechanism would need to become milder toward low-redshift, so that the trend becomes inverted. 
This allows us to restate our previous question:  what physical process would more efficiently quench satellites in similar mass pairs and is more important at high redshift? We do not have a plausible answer for this question. \\

\section{Galaxy bias}\label{s_bg}
At small scales, the complex baryonic processes of galaxy formation break the homology between the spatial distribution of galaxies and dark matter. However, at large-scales, the gravitational effects of dark matter dominate the dynamics and the overdensity of some selection of galaxies is expected to match that of dark matter multiplied by a scaling factor, the galaxy bias $b_g$. In the HOD models, this quantity is described as a number-weighted average of the halo bias (see Equation \ref{e_bg}) and ideally would match the square-root of the ratio between the large-scale SCF of galaxies and dark matter (equation \ref{bias}).\\
\indent Our measurements of the effective galaxy bias are shown in the top panel of Figure \ref{f_combined}, where we plot against apparent magnitude threshold. Bright (massive) galaxies have a larger bias than faint (small) ones, a trend that has been determined in many other studies \citep{benoist96, norberg01, tegmark04, zehavi05, brodwin08, brown08, foucaud10,zehavi11, matsuoka11,coupon12, beutler13, jullo12, mostek13} and is expected because luminous galaxies reside on average in more massive haloes, which are more biased with respect to dark matter \citep{white87, kauffmann97}. In addition to $b_g^\mathrm{eff}$, we also fit the large-scale bias directly to our measured ACF at $\theta > 0.05$ deg ($\gtrsim 4$ comoving Mpc at $z=1.5$). This fit does not depend on the HOD, and is performed by scaling the dark matter SCF in a similar way to the procedure in Section \ref{ss_redshift_scaling}, but leaving the $z=1.5$ bias as a free parameter. The inputs from the galaxy population are the redshift distributions and the evolution in the median mass to modulate the bias across redshifts, but not the galaxy number density. The shaded regions in the top panel of Figure \ref{f_combined} represent the $\pm 1\sigma$ confidence intervals for the direct bias fit, which is in good agreement with the HOD values. Thus, we find that the HOD modeling of our data makes a good description of the large-scale bias. \\
\indent Nonetheless, we note that this description is not perfect. In Appendix \ref{a_tests} we comment on how a HOD fit with all parameters allowed to vary freely makes $\sigma_{\mathrm{log}M}$ float down to unphysical values $\simeq0$, trying to maintain a high bias that otherwise would yield a smaller value due small shifts in the fitted $M_\mathrm{min}$ and $n_g$. The overall HOD is not very sensitive to $\sigma_{\mathrm{log}M}$ and therefore this is not a significant problem. However, it points towards the amplitude of our observed ACFs being slightly too large to be perfectly reproduced by the combination of halo bias and halo mass functions.

\subsection{Comparison to \citet{wake11}} \label{s_wake}
We find a slight bias excess in our data, a result that has been noted to a larger extent in other HOD studies of stellar mass limited samples at $1<z<2$. \citet{matsuoka11} and \citet{wake11} find that their ACFs are too strong to be reproduced by a halo model with the observed density of galaxies. Those fits to the clustering plus density were compared to fits to the clustering only, where the number density was not fixed to the observed value. The latter fit was able to reproduce the ACFs, but with a bias about 50\% higher than the clustering plus density fit in their most massive and distant samples. The $z=1.5$ samples of \citet{wake11} are directly comparable to our study, since they are defined by lower stellar mass limits. In the lower panel of Figure \ref{f_bg_comp} we show our standard HOD bias measurements as a function of stellar mass and the bias results from those authors. To make the comparison more direct, we plot our results for the Maraston (2005; hereafter M05) evolutionary models with the Kroupa (2001) IMF. Here we comment on the two types of HOD fits in \citet{wake11}, and how they compare to our results: 
\begin{itemize}
\item \textbf{Clustering only}: \citet{wake11} find the effective bias from this fit to be the closest to a direct measurement of the large-scale bias. However, these values are high compared to our findings. The 0.2 dex offset relative to our work could be due to a difference in stellar mass estimates (Figure \ref{f_bg_comp}). Given that both studies employ very similar stellar mass models (M05 stellar grids, Kroupa IMF, and Calzetti 2000 extinction), this possibility seems unlikely. Another explanation would be sample variance due to the small size of the survey in \citet{wake11}, which could lead to an excess in the clustering signal. Our survey is almost 400 times larger in area and therefore significantly less impacted by this effect. 
\item \textbf{Clustering + density}: \citet{wake11} find biases from this fit that are also not fully consistent with ours. Their bias is larger (smaller) than our values in the low (high) mass end. However, the observed number densities are discrepant in the opposite way, and the stellar mass at which the densities and biases match is roughly the same, $\mathrm{log}M^\mathrm{lim}_\star \sim 10.6$. In any HOD model, the effective bias is anticorrelated with number density if the rest of the parameters are held fixed. Thus, if \citet{wake11} and our study had the same observed densities, the HOD bias from both surveys could perhaps be in full agreement.
\end{itemize}

Thus, we speculate that the high clustering amplitude in \citet{wake11} might be predominantly a consequence of cosmic variance (as also suggested by those authors), and their clustering + density HOD fits would be consistent with ours if the observed $z\sim1.5$ comoving number densities were the same. 

\subsection{Comparison to other studies}
\begin{figure}
\includegraphics[trim=0mm 0mm 5mm 0mm,clip=True,width=\columnwidth]{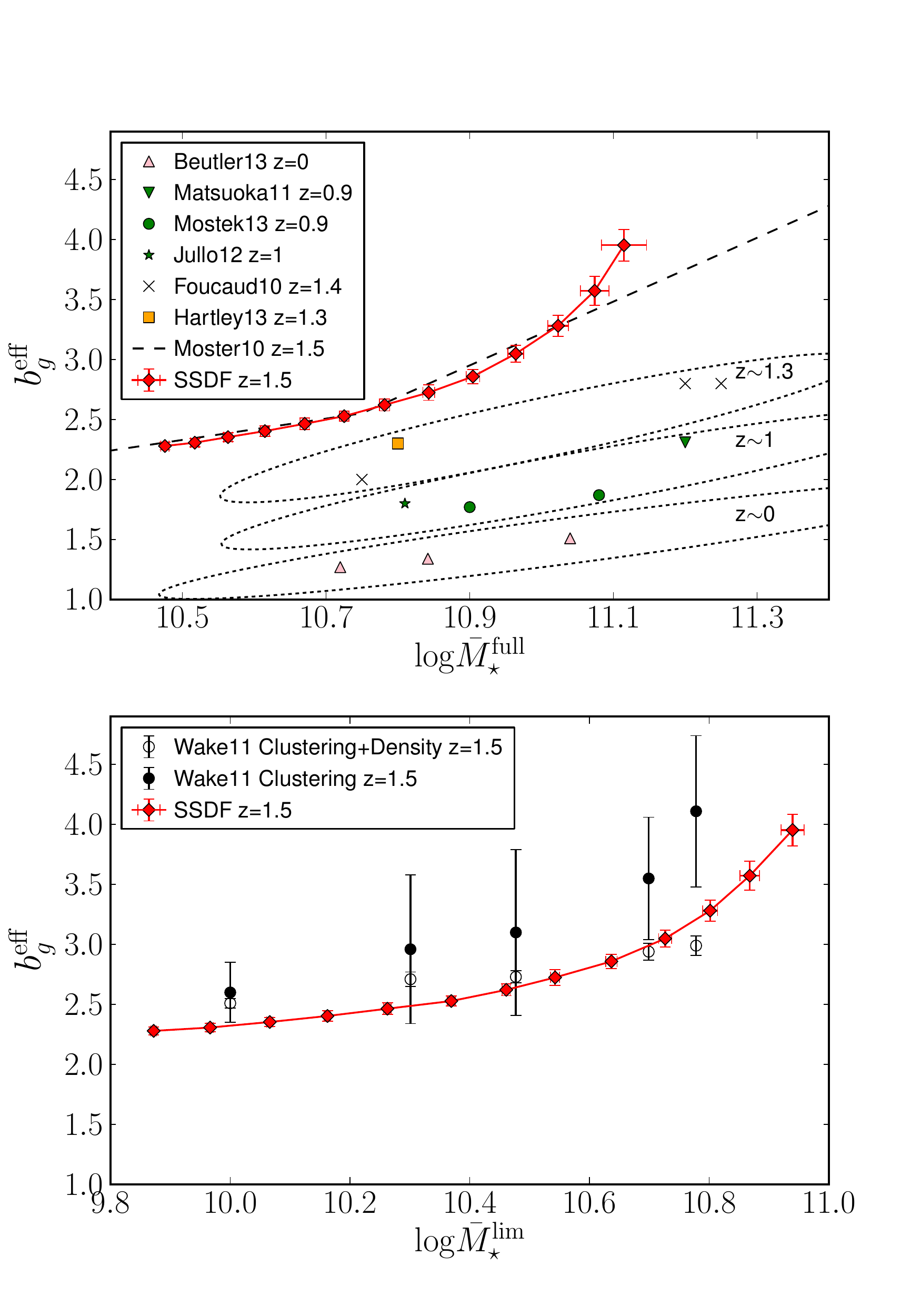}
\caption{\label{f_bg_comp} \textbf{Top:} Comparison of our HOD bias with other values from the literature, as a function of stellar mass. There are some differences in the way masses from the other studies are defined, but in general they represent the median stellar mass of a given sample. For this reason, we show $\bar{M}_\star^\mathrm{full}$ instead of $\bar{M}_\star^\mathrm{lim}$. We use BC03/Chabrier IMF stellar masses, as do most of the other authors. We note the increase of bias at fixed stellar mass as a function of redshift. \textbf{Bottom:} Comparison of our HOD bias with \citet{wake11} as a function of stellar mass limit. For more direct comparison between the results, here we model our
galaxies with M05/Kroupa IMF. The clustering-only fits of these authors yield considerably larger bias values than ours, but this might be due to sample variance in their survey (see text).   }
\end{figure}

We also compare our bias results with other measurements at different redshifts based on stellar masses in the top panel of Figure \ref{f_bg_comp}. These studies include M10, \citet{foucaud10}, \citet{matsuoka11}, \citet{jullo12}, \citet{hartley13}, \citet{mostek13} and \citet{beutler13}. The comparisons are less straightforward than with \citet{wake11}, since there are some differences with the stellar mass models used by each author. In addition, the selection is not always done with stellar mass lower limits, but in stellar mass ranges. Thus, we choose to plot the bias against the median stellar mass of the full galaxy samples. We show our results with BC03 and Chabrier IMF masses, since this is the most common choice among the other authors. \\
\indent At fixed stellar mass, the bias increases with increasing redshift. This result has also been shown in most studies that use multi-redshift data \citep{ross10,foucaud10,moster10,matsuoka11,wake11,jullo12,hartley13}. Such behavior is expected from analytical derivations \citep{fry96,moscardini98} and can be qualitatively understood if we assume that most of the galaxies are formed around a particular redshift and at high density peaks in the dark matter distribution. Such galaxies would then be initally very biased, but with time their spatial distribution would relax to match that of dark matter. Thus, the bias is generally expected to evolve toward lower values.\\
\indent M10 do not use clustering measurements, but an abundance-matching technique based on the stellar mass functions. They provide predictions of the galaxy bias for several redshifts and stellar mass ranges. We have plotted an interpolation of these values in the top panel of Figure \ref{f_bg_comp}, taking the middle point of their mass ranges as the effective median mass. Overall, there is a good agreement with our results. 

\subsection{Bias of central galaxies}
\indent Ideally, one would like to predict the bias of an individual galaxy based on its stellar mass. However, this is not possible because there is some intrinsic scatter (represented by the HOD fudge parameter $\sigma_{\mathrm{log}M}$ ) related to other physical processes that might also intervene, such as environment or assembly history. In addition, there can be ensemble scatter, which arises if the bias and mass are drawn from population averages. This is what we have done so far in this work, establishing a connection between the effective bias and the median mass of a given sample (see Table \ref{t_samples}), which are moments of broad mass distributions. Thus, we wish to reduce the amount of ensemble scatter in the bias-stellar mass mapping, which can be done straightforwardly by considering the bias of central galaxies. Basically, we exploit the connection outlined in Section \ref{s_shmr}, where central galaxies with stellar mass $\bar{M}_\star^\mathrm{lim}$ typically occupy haloes of mass $M_\mathrm{min}$. Therefore, the bias of such galaxies can be computed as $b_\mathrm{c}(\bar{M}_\star^\mathrm{lim})=b_h(M_\mathrm{min})$. Here, there is no averaging over halo masses and the stellar mass distribution is less broad than that of the full galaxy samples. We fit a 4th order polynomial to our results and thus provide a functional form of the bias of central galaxies as a function of the stellar mass logarithm $m$:

\begin{equation}
\begin{split}
b_{c}(m) = 1.6+ p_1 (m-9.8)+ p_2 (m-9.8)^2+ \\
p_3(m-9.8)^3+p_4(m-9.8)^4, \label{e_bc}
\end{split}
\end{equation}
\begin{equation}
\vec{p}=\left[ 0.22\pm0.07, 1.38\pm0.38,-2.79\pm0.62,2.23\pm0.31 \right], \nonumber
\end{equation}
which is valid over the range $9.8<m<11$.

\section{Summary}\label{s_summary}
We use a recently completed {\it Spitzer}-IRAC survey over 94 deg$^2$ to study the relation between dark matter and galaxies through their angular two-point clustering. Our data allows us to select galaxies at $z\sim1.5$ with stellar masses in the range $10^{10}-10^{11}M_\odot$. In order to derive stellar mass and redshift distributions, we employ the optical+MIR data from the COSMOS field \citep{muzzin13b} as a reference catalog, adapting it to the photometry of our survey. Then, we develop a statistical method that links sources between the SSDF and the reference catalog by matching their IRAC photometry, accounting for the relative photometric errors in both datasets. We are able to infer with high confidence the distribution of stellar mass and redshift in the SSDF for a particular IRAC selection. IRAC magnitudes and colors are well correlated with these quantities for galaxies in the range of $1<z<2$.  \\
\indent The angular correlation functions are fit with an HOD model, which offers physical insight into the relationship between dark matter haloes and the galaxies they host, both centrals and satellites. Our main results are:

\begin{itemize}
\item We fully map the stellar-to-halo mass ratio across its peak, which lies in the middle of the mass range we probe. The halo mass at the peak is found to be $\mathrm{log}M_\mathrm{peak}=12.44\pm0.08$. This is 4.5 times higher than what is found at $z\sim0$, supporting the trend of ``archeological downsizing'' since $z=1.5$. An evolving quenching mass scale $M_q$ related to $M_\mathrm{peak}$ could be responsible for this effect.

\item We compare our SHMR curves with the predictions from other authors at $z=1.5$. Our results show a higher $M_\mathrm{peak}$ than \citet{yang12}, \citet{moster13} and \citet{behroozi13}. The low- and high-mass slopes of the relation are more consistent with \citet{moster13} and \citet{behroozi13} than \citet{yang12}. In particular, we measure a slightly steeper low-mass slope than these predictions, which could support a large contribution from energy-driven winds in low-mass galaxies. 

\item The effective bias of galaxies is in the range 2-4 for galaxies of stellar mass $10^{10}-10^{11}M_\odot$, respectively. This is in good agreement with an HOD-independent fit of the large-scale bias. When compared to low-redshift studies, we find that at fixed stellar mass the bias decreases with time, in agreement with expectations from theory. We also provide a fitted form of the bias of central galaxies as a function of stellar mass. This relation suffers less from ensemble scatter than one that uses the sample average of the bias, $b_g^\mathrm{eff}$.

\item The satellite fraction is $\sim0.2$ for galaxies of stellar mass $M_\star\sim10^{10}M_{\odot}$ and decreases toward the high-mass end. In comparison to the higher fractions $\sim0.3$ measured at low-redshift (Zehavi et al. 2011; with the caveat that the two studies extend to different absolute magnitudes), this agrees with the hierarchical CDM scenario, where with time there are bigger virialized structures that can host multiple satellites.

\item We find mild evidence of an increase of $M_1/M_\mathrm{min}$ in more massive samples. This is at odds with what is generally found at lower redshifts (e.g., Zehavi et al. 2011, Zheng et al. 2007) and predicted by some simulations \citep{wechsler02,zentner05,wetzel09}. If true, this effect implies that at $z=1.5$ the overall fraction of $M_\star\sim10^{11}M_\odot$ galaxies in similar mass pairs is smaller than at lower masses. We do not find a clear reason for this trend.

\end{itemize}

Regarding possible systematic effects in our treatment, we stress that our results are robust. In general, we have not found that any of the choices we have made about the fitting parameters or overall HOD model would make a qualitative difference in our conclusions. Fixing a different number of parameters or allowing for an differential evolution between the one and two halo terms does not produce significant changes. This is in part due to the strong constraint placed by the observed galaxy number density, which is the main driver for setting $M_\mathrm{min}$ and the bias. In addition, if the prior on the density is dropped, the results become naturally more noisy but still consistent with our fiducial model. As explored in Appendix \ref{a_EvsC}, our results are also robust with the use of either COSMOS or EGS datasets as the reference catalog, even though the photometry and data products in those surveys were generated in very different ways. This fact strongly supports the robustness of our methods and conclusions.\\
\indent In the near future, deep optical catalogs in the SSDF field will be available from the Dark Energy Survey. Combining such data with the IRAC catalogs used in this study will yield an enormous boost to this kind of science. Accurate photometric redshifts and stellar masses for individual galaxies will enable a much cleaner selection. Such data will also allow HOD modeling through many redshift slices in the range of $0<z<2$, delivering a consistent and comprehensive description of the evolution in the halo-galaxy connection.\\
\indent Additionally, there are dark matter convergence maps on the SSDF field derived from CMB lensing with the South Pole Telescope \citep{carlstrom11}. Cross correlations in the SSDF field have already been performed by \citet{bleem12} and \citet{holder13} with an early IRAC galaxy catalog and {\it Herschel} data, respectively. These studies focused on $z\lesssim1$ sources and measured a positive signal, although no halo model was fitted. In Martinez-Manso et al. (in prep.), we use the accurately calibrated IRAC catalogs in this paper to explore the cross correlation of $z\sim1.5$ stellar mass-selected galaxies with dark matter maps. This study will offer a direct connection between these matter fields, and allow for an independent test of the halo model framework.\\

We would like to thank Jean Coupon, Andrew Wetzel, Andrew Zentner, Andrey Kravtsov and Matthieu B\'ethermin for their helpful comments on this work.

\clearpage
\begin{landscape}
\begin{table}
\begin{center}
\caption{\label{acf_values} Measured angular correlation for all our samples, which are denoted by their limiting [4.5] magnitude in the first row. These values have been corrected for the integral constraint.}
\tabcolsep=0.1cm
  \begin{tabular}{cccccccc}
  \hline \hline
$\theta$ (degrees) & 16.2 & 16.4 & 16.6 & 16.8 & 17.0 & 17.2 & 17.4  \\
 \hline \hline
0.0019 & 2.8$\pm$0.9$\times 10^{0}$ & 1.9$\pm$0.4$\times 10^{0}$ & 1.7$\pm$0.2$\times 10^{0}$ & 1.5$\pm$0.2$\times 10^{0}$ & 1.3$\pm$0.1$\times 10^{0}$ & 1.2$\pm$0.1$\times 10^{0}$ & 1.1$\pm$0.1$\times 10^{0}$\\ [0.07cm]                            
0.0025 & 1.8$\pm$0.5$\times 10^{0}$ & 1.6$\pm$0.3$\times 10^{0}$ & 1.2$\pm$0.1$\times 10^{0}$ & 1.1$\pm$0.1$\times 10^{0}$ & 8.6$\pm$0.6$\times 10^{-1}$ & 7.5$\pm$0.4$\times 10^{-1}$ & 6.5$\pm$0.3$\times 10^{-1}$\\ [0.07cm]                         
0.0033 & 9.5$\pm$3.0$\times 10^{-1}$ & 9.3$\pm$1.6$\times 10^{-1}$ & 7.5$\pm$0.9$\times 10^{-1}$ & 6.6$\pm$0.7$\times 10^{-1}$ & 5.8$\pm$0.5$\times 10^{-1}$ & 5.3$\pm$0.3$\times 10^{-1}$ & 4.6$\pm$0.2$\times 10^{-1}$\\ [0.07cm]                     
0.0044 & 4.2$\pm$2.1$\times 10^{-1}$ & 5.1$\pm$1.2$\times 10^{-1}$ & 5.0$\pm$0.6$\times 10^{-1}$ & 4.3$\pm$0.3$\times 10^{-1}$ & 3.6$\pm$0.3$\times 10^{-1}$ & 3.3$\pm$0.2$\times 10^{-1}$ & 2.9$\pm$0.2$\times 10^{-1}$\\ [0.07cm]                     
0.0058 & 3.7$\pm$1.9$\times 10^{-1}$ & 3.6$\pm$0.8$\times 10^{-1}$ & 3.7$\pm$0.5$\times 10^{-1}$ & 3.6$\pm$0.3$\times 10^{-1}$ & 3.0$\pm$0.2$\times 10^{-1}$ & 2.5$\pm$0.2$\times 10^{-1}$ & 2.0$\pm$0.1$\times 10^{-1}$\\ [0.07cm]                     
0.0077 & 3.0$\pm$1.0$\times 10^{-1}$ & 1.5$\pm$0.5$\times 10^{-1}$ & 1.7$\pm$0.3$\times 10^{-1}$ & 1.6$\pm$0.2$\times 10^{-1}$ & 1.7$\pm$0.1$\times 10^{-1}$ & 1.4$\pm$0.1$\times 10^{-1}$ & 1.4$\pm$0.1$\times 10^{-1}$\\ [0.07cm]
0.0102 & 2.1$\pm$0.9$\times 10^{-1}$ & 1.9$\pm$0.4$\times 10^{-1}$ & 2.0$\pm$0.2$\times 10^{-1}$ & 1.7$\pm$0.1$\times 10^{-1}$ & 1.4$\pm$0.1$\times 10^{-1}$ & 1.3$\pm$0.1$\times 10^{-1}$ & 1.1$\pm$0.1$\times 10^{-1}$\\ [0.07cm]
0.0135 & 1.8$\pm$0.5$\times 10^{-1}$ & 1.6$\pm$0.3$\times 10^{-1}$ & 1.4$\pm$0.2$\times 10^{-1}$ & 1.3$\pm$0.1$\times 10^{-1}$ & 1.1$\pm$0.1$\times 10^{-1}$ & 1.0$\pm$0.1$\times 10^{-1}$ & 9.3$\pm$0.8$\times 10^{-2}$\\ [0.07cm]
0.0178 & 9.7$\pm$5.7$\times 10^{-2}$ & 1.0$\pm$0.2$\times 10^{-1}$ & 1.1$\pm$0.1$\times 10^{-1}$ & 1.1$\pm$0.1$\times 10^{-1}$ & 9.9$\pm$0.8$\times 10^{-2}$ & 8.8$\pm$0.7$\times 10^{-2}$ & 7.7$\pm$0.7$\times 10^{-2}$\\ [0.07cm]
0.0235 & 1.4$\pm$0.3$\times 10^{-1}$ & 1.3$\pm$0.1$\times 10^{-1}$ & 1.0$\pm$0.1$\times 10^{-1}$ & 9.3$\pm$0.9$\times 10^{-2}$ & 7.9$\pm$0.7$\times 10^{-2}$ & 7.4$\pm$0.5$\times 10^{-2}$ & 6.6$\pm$0.5$\times 10^{-2}$\\ [0.07cm]
0.0311 & 6.3$\pm$3.0$\times 10^{-2}$ & 8.5$\pm$1.9$\times 10^{-2}$ & 7.6$\pm$1.0$\times 10^{-2}$ & 7.5$\pm$0.7$\times 10^{-2}$ & 6.3$\pm$0.5$\times 10^{-2}$ & 5.7$\pm$0.4$\times 10^{-2}$ & 5.0$\pm$0.4$\times 10^{-2}$\\ [0.07cm]
0.0411 & 5.9$\pm$1.9$\times 10^{-2}$ & 6.8$\pm$1.1$\times 10^{-2}$ & 7.2$\pm$0.7$\times 10^{-2}$ & 6.8$\pm$0.7$\times 10^{-2}$ & 5.7$\pm$0.4$\times 10^{-2}$ & 4.9$\pm$0.3$\times 10^{-2}$ & 4.4$\pm$0.2$\times 10^{-2}$\\ [0.07cm]
0.0543 & 4.5$\pm$1.7$\times 10^{-2}$ & 5.3$\pm$1.0$\times 10^{-2}$ & 4.7$\pm$0.7$\times 10^{-2}$ & 4.3$\pm$0.5$\times 10^{-2}$ & 4.2$\pm$0.3$\times 10^{-2}$ & 3.8$\pm$0.3$\times 10^{-2}$ & 3.5$\pm$0.2$\times 10^{-2}$\\ [0.07cm]
0.0717 & 4.9$\pm$1.2$\times 10^{-2}$ & 3.9$\pm$0.8$\times 10^{-2}$ & 3.6$\pm$0.5$\times 10^{-2}$ & 3.5$\pm$0.3$\times 10^{-2}$ & 3.2$\pm$0.2$\times 10^{-2}$ & 2.8$\pm$0.2$\times 10^{-2}$ & 2.6$\pm$0.1$\times 10^{-2}$\\ [0.07cm]
0.0947 & 2.5$\pm$1.1$\times 10^{-2}$ & 2.6$\pm$0.6$\times 10^{-2}$ & 2.8$\pm$0.5$\times 10^{-2}$ & 2.7$\pm$0.3$\times 10^{-2}$ & 2.5$\pm$0.2$\times 10^{-2}$ & 2.3$\pm$0.2$\times 10^{-2}$ & 2.1$\pm$0.1$\times 10^{-2}$\\ [0.07cm]
0.1250 & 2.8$\pm$0.8$\times 10^{-2}$ & 2.7$\pm$0.5$\times 10^{-2}$ & 2.1$\pm$0.3$\times 10^{-2}$ & 2.2$\pm$0.2$\times 10^{-2}$ & 1.9$\pm$0.2$\times 10^{-2}$ & 1.7$\pm$0.1$\times 10^{-2}$ & 1.6$\pm$0.1$\times 10^{-2}$\\ [0.07cm]
0.1651 & 2.0$\pm$0.6$\times 10^{-2}$ & 1.8$\pm$0.4$\times 10^{-2}$ & 1.8$\pm$0.3$\times 10^{-2}$ & 1.7$\pm$0.3$\times 10^{-2}$ & 1.5$\pm$0.2$\times 10^{-2}$ & 1.4$\pm$0.1$\times 10^{-2}$ & 1.3$\pm$0.1$\times 10^{-2}$\\ [0.07cm]
0.2180 & 1.8$\pm$0.5$\times 10^{-2}$ & 1.3$\pm$0.3$\times 10^{-2}$ & 1.4$\pm$0.3$\times 10^{-2}$ & 1.4$\pm$0.2$\times 10^{-2}$ & 1.2$\pm$0.1$\times 10^{-2}$ & 1.0$\pm$0.1$\times 10^{-2}$ & 9.9$\pm$1.3$\times 10^{-3}$\\ [0.07cm]
0.2879 & 7.2$\pm$4.4$\times 10^{-3}$ & 1.1$\pm$0.3$\times 10^{-2}$ & 1.0$\pm$0.2$\times 10^{-2}$ & 9.8$\pm$2.0$\times 10^{-3}$ & 9.2$\pm$1.7$\times 10^{-3}$ & 8.1$\pm$1.3$\times 10^{-3}$ & 7.3$\pm$1.1$\times 10^{-3}$\\ [0.07cm]
0.3802 & 9.4$\pm$3.6$\times 10^{-3}$ & 8.3$\pm$2.2$\times 10^{-3}$ & 7.4$\pm$2.0$\times 10^{-3}$ & 6.6$\pm$1.7$\times 10^{-3}$ & 5.7$\pm$1.4$\times 10^{-3}$ & 5.2$\pm$1.0$\times 10^{-3}$ & 4.7$\pm$0.9$\times 10^{-3}$\\ [0.07cm]
0.5021 & 5.5$\pm$2.7$\times 10^{-3}$ & 4.4$\pm$2.1$\times 10^{-3}$ & 5.1$\pm$1.5$\times 10^{-3}$ & 4.6$\pm$1.2$\times 10^{-3}$ & 3.7$\pm$0.9$\times 10^{-3}$ & 3.3$\pm$0.8$\times 10^{-3}$ & 3.0$\pm$0.7$\times 10^{-3}$\\ [0.07cm]
0.6630 & 4.8$\pm$2.4$\times 10^{-3}$ & 3.2$\pm$1.8$\times 10^{-3}$ & 3.1$\pm$1.3$\times 10^{-3}$ & 3.3$\pm$1.2$\times 10^{-3}$ & 2.7$\pm$1.0$\times 10^{-3}$ & 2.3$\pm$0.8$\times 10^{-3}$ & 2.1$\pm$0.7$\times 10^{-3}$\\ [0.07cm]
0.8755 & 1.7$\pm$1.8$\times 10^{-3}$ & 2.6$\pm$1.4$\times 10^{-3}$ & 2.5$\pm$1.0$\times 10^{-3}$ & 2.3$\pm$0.9$\times 10^{-3}$ & 1.4$\pm$0.8$\times 10^{-3}$ & 1.5$\pm$0.6$\times 10^{-3}$ & 1.3$\pm$0.5$\times 10^{-3}$\\ [0.07cm]
1.1561 & 2.0$\pm$1.7$\times 10^{-3}$ & 1.7$\pm$1.2$\times 10^{-3}$ & 1.6$\pm$0.9$\times 10^{-3}$ & 1.0$\pm$0.8$\times 10^{-3}$ & 8.5$\pm$6.7$\times 10^{-4}$ & 7.7$\pm$5.8$\times 10^{-4}$ & 7.8$\pm$5.4$\times 10^{-4}$\\ [0.07cm]
1.5266 & -5.1$\pm$13.0$\times 10^{-4}$ & 1.3$\pm$9.4$\times 10^{-4}$ & 1.0$\pm$7.2$\times 10^{-4}$ & -1.1$\pm$6.8$\times 10^{-4}$ & -0.2$\pm$6.1$\times 10^{-4}$ & -1.0$\pm$4.8$\times 10^{-4}$ & 1.0$\pm$4.5$\times 10^{-4}$\\ [0.07cm]
2.0158 & -7.4$\pm$9.5$\times 10^{-4}$ & 0.3$\pm$7.1$\times 10^{-4}$ & -2.9$\pm$5.9$\times 10^{-4}$ & -3.7$\pm$5.5$\times 10^{-4}$ & -5.4$\pm$4.5$\times 10^{-4}$ & -5.1$\pm$3.6$\times 10^{-4}$ & -4.6$\pm$2.9$\times 10^{-4}$\\ [0.07cm]
2.6618 & -4.0$\pm$10.0$\times 10^{-4}$ & 1.4$\pm$7.1$\times 10^{-4}$ & -2.7$\pm$6.0$\times 10^{-4}$ & -4.2$\pm$4.9$\times 10^{-4}$ & -2.6$\pm$4.3$\times 10^{-4}$ & -1.9$\pm$3.9$\times 10^{-4}$ & -3.0$\pm$3.3$\times 10^{-4}$\\ [0.07cm]
3.5148 & -9.4$\pm$9.3$\times 10^{-4}$ & -1.2$\pm$0.5$\times 10^{-3}$ & -1.0$\pm$0.4$\times 10^{-3}$ & -7.4$\pm$4.3$\times 10^{-4}$ & -4.7$\pm$3.3$\times 10^{-4}$ & -4.4$\pm$2.9$\times 10^{-4}$ & -4.2$\pm$2.5$\times 10^{-4}$\\ [0.07cm]

\hline
\end{tabular}
\end{center}
\end{table}
\end{landscape}

\clearpage
\begin{landscape}
\begin{table}
\begin{center}
\caption{\label{acf_values} Continuation of previous table.}
\tabcolsep=0.1cm
  \begin{tabular}{ccccccc}
  \hline \hline
$\theta$ (degrees) & 17.6 & 17.8 & 18.0 & 18.2 & 18.4 & 18.6  \\
 \hline \hline
0.0019 & 9.0$\pm$0.4$\times 10^{-1}$ & 7.5$\pm$0.3$\times 10^{-1}$ & 6.5$\pm$0.3$\times 10^{-1}$ & 5.7$\pm$0.2$\times 10^{-1}$ & 5.0$\pm$0.2$\times 10^{-1}$ & 4.3$\pm$0.2$\times 10^{-1}$\\ [0.07cm]                                                   
0.0025 & 5.7$\pm$0.3$\times 10^{-1}$ & 5.0$\pm$0.2$\times 10^{-1}$ & 4.4$\pm$0.2$\times 10^{-1}$ & 3.9$\pm$0.2$\times 10^{-1}$ & 3.4$\pm$0.2$\times 10^{-1}$ & 2.9$\pm$0.1$\times 10^{-1}$\\ [0.07cm]                                                   
0.0033 & 4.0$\pm$0.2$\times 10^{-1}$ & 3.5$\pm$0.2$\times 10^{-1}$ & 3.0$\pm$0.2$\times 10^{-1}$ & 2.5$\pm$0.2$\times 10^{-1}$ & 2.3$\pm$0.1$\times 10^{-1}$ & 1.9$\pm$0.1$\times 10^{-1}$\\ [0.07cm]                                                   
0.0044 & 2.6$\pm$0.1$\times 10^{-1}$ & 2.3$\pm$0.1$\times 10^{-1}$ & 2.0$\pm$0.1$\times 10^{-1}$ & 1.8$\pm$0.1$\times 10^{-1}$ & 1.6$\pm$0.1$\times 10^{-1}$ & 1.4$\pm$0.1$\times 10^{-1}$\\ [0.07cm]                                                   
0.0058 & 1.9$\pm$0.1$\times 10^{-1}$ & 1.6$\pm$0.1$\times 10^{-1}$ & 1.4$\pm$0.1$\times 10^{-1}$ & 1.2$\pm$0.1$\times 10^{-1}$ & 1.1$\pm$0.1$\times 10^{-1}$ & 1.0$\pm$0.1$\times 10^{-1}$\\ [0.07cm]                                                   
0.0077 & 1.2$\pm$0.1$\times 10^{-1}$ & 1.1$\pm$0.1$\times 10^{-1}$ & 1.0$\pm$0.1$\times 10^{-1}$ & 9.0$\pm$1.0$\times 10^{-2}$ & 8.0$\pm$0.9$\times 10^{-2}$ & 7.1$\pm$0.9$\times 10^{-2}$\\ [0.07cm]                                                   
0.0102 & 1.0$\pm$0.1$\times 10^{-1}$ & 9.4$\pm$0.9$\times 10^{-2}$ & 8.6$\pm$0.8$\times 10^{-2}$ & 7.5$\pm$0.8$\times 10^{-2}$ & 6.7$\pm$0.8$\times 10^{-2}$ & 6.2$\pm$0.8$\times 10^{-2}$\\ [0.07cm]
0.0135 & 8.1$\pm$0.8$\times 10^{-2}$ & 7.4$\pm$0.8$\times 10^{-2}$ & 6.5$\pm$0.7$\times 10^{-2}$ & 5.8$\pm$0.7$\times 10^{-2}$ & 5.3$\pm$0.7$\times 10^{-2}$ & 4.8$\pm$0.7$\times 10^{-2}$\\ [0.07cm]
0.0178 & 6.5$\pm$0.6$\times 10^{-2}$ & 5.9$\pm$0.5$\times 10^{-2}$ & 5.3$\pm$0.5$\times 10^{-2}$ & 5.0$\pm$0.5$\times 10^{-2}$ & 4.4$\pm$0.5$\times 10^{-2}$ & 3.9$\pm$0.5$\times 10^{-2}$\\ [0.07cm]
0.0235 & 5.8$\pm$0.4$\times 10^{-2}$ & 5.3$\pm$0.4$\times 10^{-2}$ & 4.8$\pm$0.4$\times 10^{-2}$ & 4.1$\pm$0.3$\times 10^{-2}$ & 3.7$\pm$0.3$\times 10^{-2}$ & 3.3$\pm$0.3$\times 10^{-2}$\\ [0.07cm]
0.0311 & 4.5$\pm$0.3$\times 10^{-2}$ & 4.1$\pm$0.3$\times 10^{-2}$ & 3.8$\pm$0.3$\times 10^{-2}$ & 3.4$\pm$0.3$\times 10^{-2}$ & 3.0$\pm$0.3$\times 10^{-2}$ & 2.7$\pm$0.2$\times 10^{-2}$\\ [0.07cm]
0.0411 & 4.1$\pm$0.2$\times 10^{-2}$ & 3.4$\pm$0.2$\times 10^{-2}$ & 3.0$\pm$0.2$\times 10^{-2}$ & 2.7$\pm$0.2$\times 10^{-2}$ & 2.4$\pm$0.2$\times 10^{-2}$ & 2.1$\pm$0.2$\times 10^{-2}$\\ [0.07cm]
0.0543 & 3.1$\pm$0.2$\times 10^{-2}$ & 2.7$\pm$0.2$\times 10^{-2}$ & 2.4$\pm$0.1$\times 10^{-2}$ & 2.1$\pm$0.1$\times 10^{-2}$ & 1.9$\pm$0.1$\times 10^{-2}$ & 1.7$\pm$0.1$\times 10^{-2}$\\ [0.07cm]
0.0717 & 2.4$\pm$0.1$\times 10^{-2}$ & 2.0$\pm$0.1$\times 10^{-2}$ & 1.8$\pm$0.1$\times 10^{-2}$ & 1.6$\pm$0.1$\times 10^{-2}$ & 1.5$\pm$0.1$\times 10^{-2}$ & 1.3$\pm$0.1$\times 10^{-2}$\\ [0.07cm]
0.0947 & 1.8$\pm$0.1$\times 10^{-2}$ & 1.7$\pm$0.1$\times 10^{-2}$ & 1.5$\pm$0.1$\times 10^{-2}$ & 1.3$\pm$0.1$\times 10^{-2}$ & 1.2$\pm$0.1$\times 10^{-2}$ & 1.0$\pm$0.1$\times 10^{-2}$\\ [0.07cm]
0.1250 & 1.5$\pm$0.1$\times 10^{-2}$ & 1.3$\pm$0.1$\times 10^{-2}$ & 1.2$\pm$0.1$\times 10^{-2}$ & 1.0$\pm$0.0$\times 10^{-2}$ & 9.5$\pm$0.8$\times 10^{-3}$ & 8.4$\pm$0.7$\times 10^{-3}$\\ [0.07cm]
0.1651 & 1.1$\pm$0.1$\times 10^{-2}$ & 1.0$\pm$0.1$\times 10^{-2}$ & 9.6$\pm$1.1$\times 10^{-3}$ & 8.7$\pm$1.1$\times 10^{-3}$ & 7.7$\pm$1.0$\times 10^{-3}$ & 6.9$\pm$0.9$\times 10^{-3}$\\ [0.07cm]
0.2180 & 8.7$\pm$1.2$\times 10^{-3}$ & 7.8$\pm$1.0$\times 10^{-3}$ & 6.8$\pm$0.9$\times 10^{-3}$ & 6.2$\pm$0.7$\times 10^{-3}$ & 5.7$\pm$0.7$\times 10^{-3}$ & 5.0$\pm$0.6$\times 10^{-3}$\\ [0.07cm]
0.2879 & 6.3$\pm$1.0$\times 10^{-3}$ & 5.6$\pm$0.9$\times 10^{-3}$ & 5.0$\pm$0.8$\times 10^{-3}$ & 4.6$\pm$0.7$\times 10^{-3}$ & 4.1$\pm$0.7$\times 10^{-3}$ & 3.6$\pm$0.6$\times 10^{-3}$\\ [0.07cm]
0.3802 & 4.1$\pm$0.8$\times 10^{-3}$ & 3.7$\pm$0.7$\times 10^{-3}$ & 3.4$\pm$0.7$\times 10^{-3}$ & 3.0$\pm$0.6$\times 10^{-3}$ & 2.8$\pm$0.6$\times 10^{-3}$ & 2.5$\pm$0.5$\times 10^{-3}$\\ [0.07cm]
0.5021 & 2.7$\pm$0.6$\times 10^{-3}$ & 2.4$\pm$0.5$\times 10^{-3}$ & 2.3$\pm$0.5$\times 10^{-3}$ & 2.0$\pm$0.4$\times 10^{-3}$ & 1.8$\pm$0.4$\times 10^{-3}$ & 1.6$\pm$0.3$\times 10^{-3}$\\ [0.07cm]
0.6630 & 1.8$\pm$0.6$\times 10^{-3}$ & 1.5$\pm$0.5$\times 10^{-3}$ & 1.5$\pm$0.5$\times 10^{-3}$ & 1.2$\pm$0.4$\times 10^{-3}$ & 1.1$\pm$0.4$\times 10^{-3}$ & 1.0$\pm$0.3$\times 10^{-3}$\\ [0.07cm]
0.8755 & 1.0$\pm$0.5$\times 10^{-3}$ & 8.8$\pm$4.7$\times 10^{-4}$ & 8.3$\pm$4.3$\times 10^{-4}$ & 7.6$\pm$3.8$\times 10^{-4}$ & 7.2$\pm$3.4$\times 10^{-4}$ & 5.7$\pm$3.0$\times 10^{-4}$\\ [0.07cm]
1.1561 & 6.5$\pm$4.5$\times 10^{-4}$ & 6.6$\pm$4.2$\times 10^{-4}$ & 5.4$\pm$3.7$\times 10^{-4}$ & 5.6$\pm$3.3$\times 10^{-4}$ & 5.4$\pm$3.0$\times 10^{-4}$ & 4.5$\pm$2.6$\times 10^{-4}$\\ [0.07cm]
1.5266 & 1.2$\pm$3.7$\times 10^{-4}$ & 1.8$\pm$3.2$\times 10^{-4}$ & 1.7$\pm$2.7$\times 10^{-4}$ & 1.9$\pm$2.6$\times 10^{-4}$ & 1.7$\pm$2.3$\times 10^{-4}$ & 1.2$\pm$2.1$\times 10^{-4}$\\ [0.07cm]
2.0158 & -3.3$\pm$2.6$\times 10^{-4}$ & -2.0$\pm$2.3$\times 10^{-4}$ & -1.4$\pm$2.0$\times 10^{-4}$ & -1.1$\pm$1.9$\times 10^{-4}$ & -0.7$\pm$1.7$\times 10^{-4}$ & -0.8$\pm$1.6$\times 10^{-4}$\\ [0.07cm]
2.6618 & -2.1$\pm$2.6$\times 10^{-4}$ & -2.4$\pm$2.3$\times 10^{-4}$ & -1.8$\pm$2.0$\times 10^{-4}$ & -1.7$\pm$1.8$\times 10^{-4}$ & -2.0$\pm$1.5$\times 10^{-4}$ & -1.8$\pm$1.4$\times 10^{-4}$\\ [0.07cm]
3.5148 & -3.3$\pm$2.0$\times 10^{-4}$ & -2.8$\pm$1.8$\times 10^{-4}$ & -2.7$\pm$1.5$\times 10^{-4}$ & -2.6$\pm$1.4$\times 10^{-4}$ & -2.4$\pm$1.3$\times 10^{-4}$ & -1.8$\pm$1.1$\times 10^{-4}$\\ 

\hline
\end{tabular}
\end{center}
\end{table}
\end{landscape}

\begin{appendix}

\section{Photometric simulations}\label{s_photsims}

Our selection criteria are based on aperture photometry drawn from the \citet{ashby13b} 
4.5\,$\mu$m-selected catalog, but we used sources fainter than their 5$\sigma$ sensitivity 
threshold. For this reason, we carried out an independent analysis of the SSDF source 
extraction in order to estimate the completeness and photometric bias for the faintest
sources. This was done by placing artificial sources of known brightness 
in representative SSDF mosaics (i.e., a pair of coextensive 3.6 and 4.5\,$\mu$m `tiles' of size $2\times1$\,deg$^2$; 
Ashby et al. 2013b), which were observed to the nominal survey depth. Then, we performed a detailed comparison between the resulting photometric measurements and the input brightnesses, as described below.  

\subsection{Simulation procedures}\label{s_simproc}
We began by generating point spread function (PSF) images to represent the artificial sources.  
This is consistent with high-resolution {\it HST}/WFC3 observations showing that the vast majority 
of galaxies having magnitudes in our range of interest are point sources at IRAC 
resolution (Fig.\,25 of Ashby et al. 2013a). We first identified 18 point sources in the SSDF science mosaics and verified by visual inspection that they did not contain any artifacts or anomalies. These point sources were then scaled and median-stacked at their centroid positions. The resulting PSF images
were constructed with 41 by 41 0\farcs6 pixels to match the spatial resolution
of the SSDF science mosaics. The FWHMs of these images were found to be
1\farcs69 and 1\farcs85 in the 3.6 and 4.5\,$\mu$m bands, respectively. These values are close to those measured for IRAC in single exposures, i.e., 1\farcs62 and 1\farcs77. 

\begin{figure}
\includegraphics[trim=0mm 0mm 0mm 0mm,clip=True,width=\columnwidth]{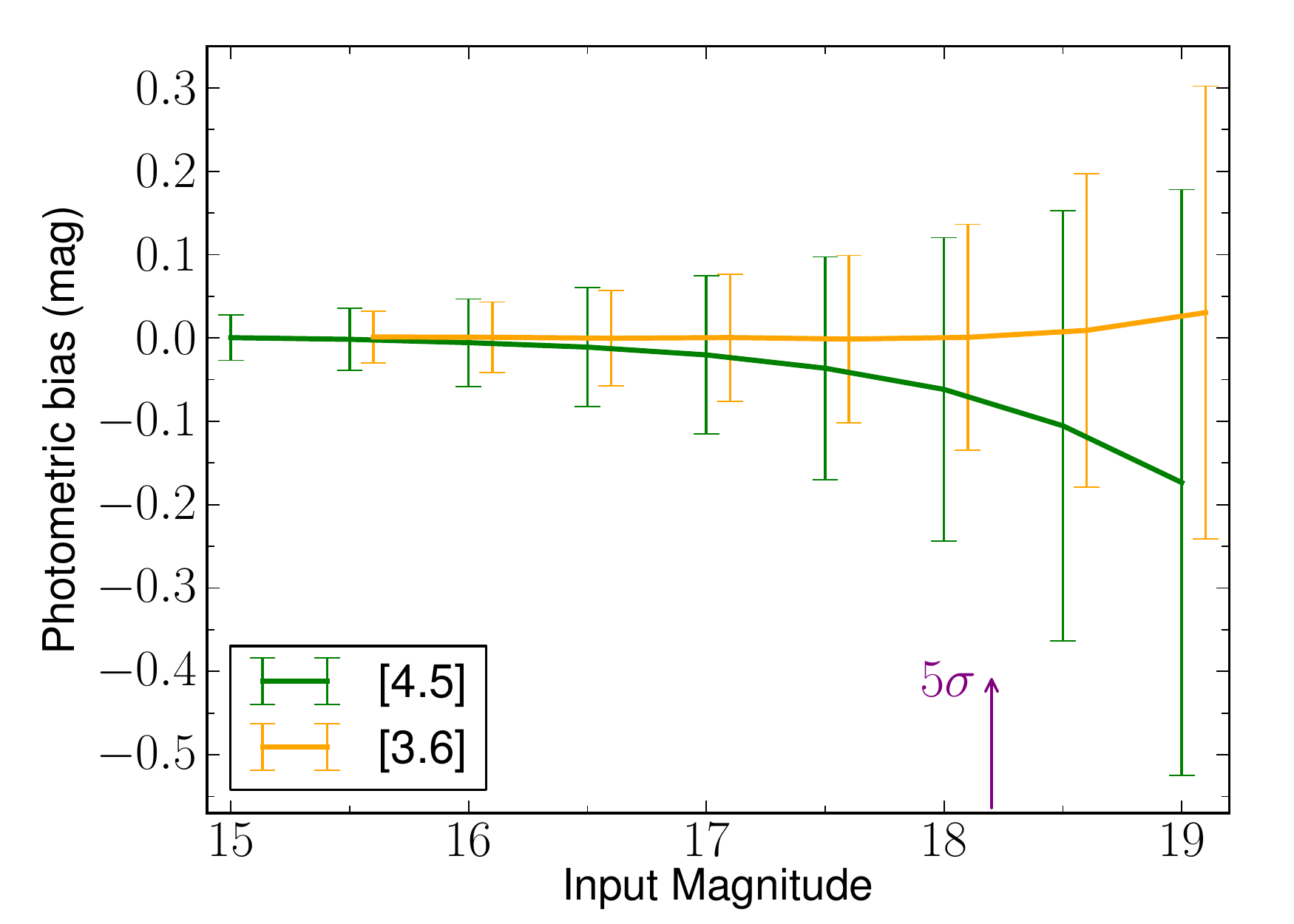}
\caption{\label{aper_corr} Photometric bias in the 3.6 and 4.5\,$\mu$m SSDF mosaics. The photometric bias was measured as the difference between the input magnitude and the output aperture-corrected magnitude, as derived from the simulations. The solid lines represent the median bias and error bars are one standard deviation. The arrow marks the $5\sigma$ sensitivity level in [4.5] fluxes. The bias trend in [4.5] indicates an artificial brightening of sources towards faint input magnitudes, mainly caused by the contamination of flux from nearby objects and noise peaks. On the other hand, [3.6] extracted magnitudes become slightly fainter towards faint input magnitudes, which is due to positional shifts in the [4.5] selection aperture with respect to the real centroid of the source. The aperture-corrected magnitudes in the SSDF catalog are combined with the photometric bias to obtain the final photometry (see text).   }
\end{figure}

\begin{figure}
\includegraphics[trim=0mm 0mm 0mm 0mm,clip=True,width=\columnwidth]{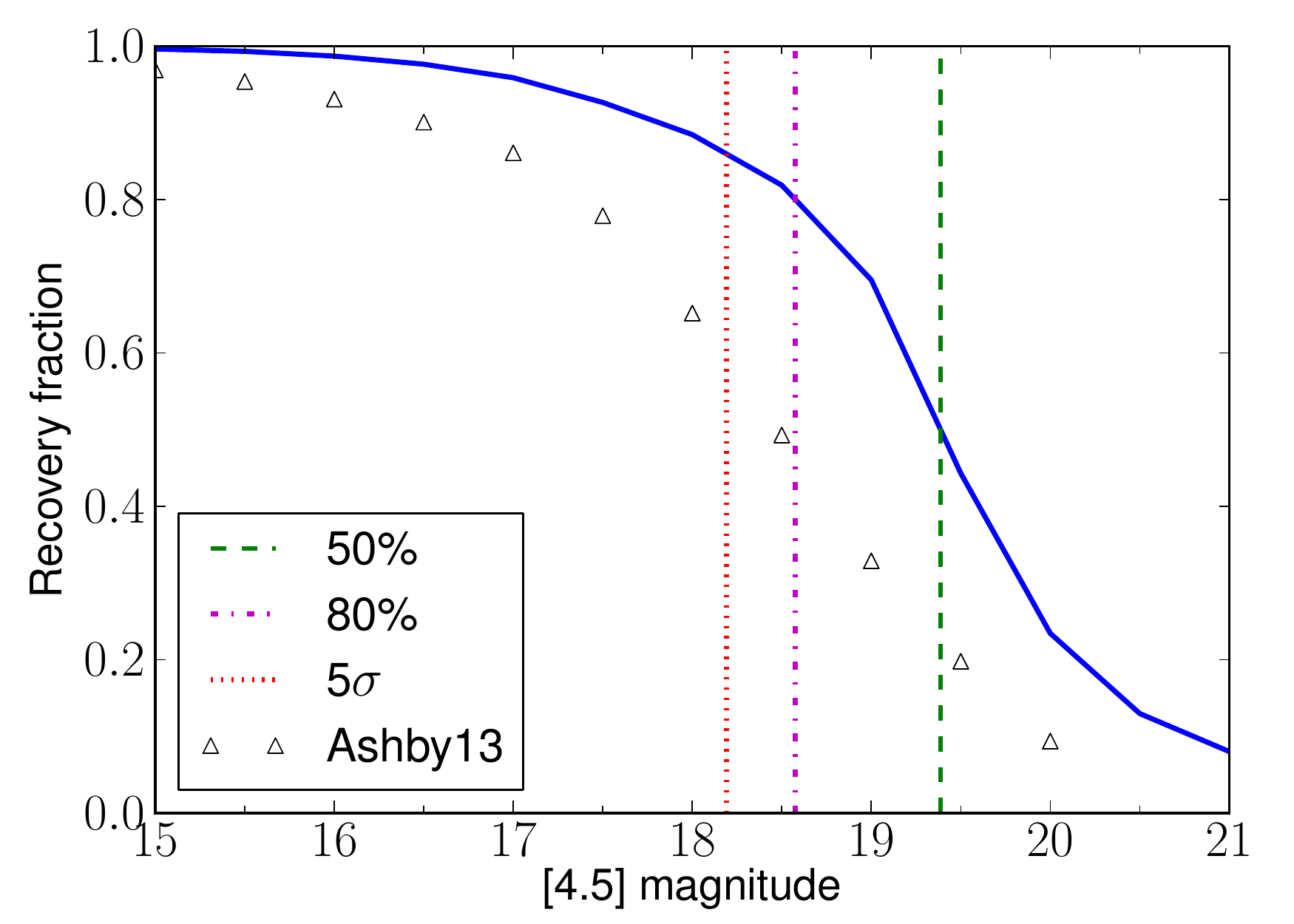}
\caption{\label{completeness} Recovered fraction of simulated sources as a function of input magnitude in [4.5]. The dash-dotted line at 18.19 represents the 5$\sigma$ level of photometric sensitivity. The dashed and dotted lines mark the points where the recovery fraction reaches 80\% at 18.58 and 50\% at 19.39, respectively. Open triangles represent the recovery fraction derived in \citet{ashby13b}. We measure higher completeness than seen by \citet{ashby13b} because we employ a more complicated source identification procedure, which boosts the detections in complex cases (e.g., blends, positional shifts) that otherwise would be rejected. }
\end{figure}

The PSF images were then placed in the science mosaic. They were placed at random positions,
but at a minimum distance from each other. This minimum distance varies linearly with the magnitude of the simulated source from 18\asec\ to 6\asec\ through $[4.5]=15-21$. In addition, simulated sources were not allowed to fall within regions contaminated by stars brighter than $K_s=12$\,mag. The size of the exclusion regions around these stars was determined following the method described in Section \ref{s_datasets}.

Outside the masked regions, 1500 simulated point sources having [4.5]=15\,mag were added to 
random locations of the 4.5\,$\mu$m science mosaic. This number of simulated sources is low enough to avoid crowding and alterations in the photometric background. An equal number of sources were also 
put at identical locations in the 3.6\,$\mu$m science mosaic. The 3.6\,$\mu$m sources were 
set to have colors $[3.6]-[4.5]=0.7$, appropriate for the sample selection described in Section \ref{ss_subsamples}. The modified science mosaics were then processed with SExtractor. This was also done on the original, unmodified science mosaic. Identical SExtractor parameter settings were used in all instances, following those presented in \citet{ashby13b}. The process was repeated until a total of 80,000 [4.5]=15\,mag
sources were detected and photometered. The simulations were then carried out in the same
manner for input magnitudes in the range $[4.5]=15.5-21$ mag with steps of 0.5\,mag.
We used the resulting pairs of SExtractor catalogs to determine our detection completeness 
and photometric bias. Specifically, we retrieved all cataloged sources found within 
6\asec\ of the position of each simulated source in both the original and 
modified mosaics. This search radius was set empirically to encompass all possible shifts in the measured centroids of sources due to the distortion caused by the simulated source. Sources in the two catalogs were judged to match when they differed by less than 50\% 
in flux and were separated by less than half the PSF FWHM (0.9\asec). This left a number of non-matched sources from the original and modified mosaics: $N_{\mathrm{orig}}$ and $N_{\mathrm{mod}}$, respectively. Then, a detection of the simulated source was determined if one of the following cases applied: 

\begin{description}
\item[A:] $N_{\mathrm{orig}}=0$, $N_{\mathrm{mod}}=1$. This is the most typical case, where only one source in the modified mosaic could not be matched to another in the original mosaic and was therefore identified as the simulated source. 
\item[B:] $N_{\mathrm{orig}}=0$, $N_{\mathrm{mod}}>1$. As in \textbf{A}, all sources in the original mosaic were uniquely identified in the modified mosaic. However, there were a few sources in the latter with no counterpart. This happened because the simulated source was erroneously recovered by SExtractor as multiple sources. We identified the brightest one of these with the simulated source.
\item[C:] $N_{\mathrm{orig}}>0$, $N_{\mathrm{mod}}>N_{\mathrm{orig}}$. The local photometric influence of the simulated source caused a change in the position and flux of several sources in the modified mosaic. As a result, not all sources in the original mosaic were found a match in the modified mosaic. This left at least two candidates in the modified mosaic to represent the simulated source. Consequently, one of them would correspond to an unmatched source in the original mosaic. To simplify the identification process, we restricted the selection of unmatched sources based on the input location of the simulated source: we considered the closest one and the closest two from the original and modified mosaics, respectively. If the source from the original mosaic was brighter than the simulated one, then the former ought to have shifted its position less than the latter. Thus, the source in the original mosaic was matched to the source in the modified mosaic that lay closest to it. Otherwise, the simulated source may have not shifted significantly due to the presence of the source in the original mosaic. In this case, the simulated source was identified with the nearest source in the modified mosaic.
\item[D:] $N_{\mathrm{mod}}>0$, $N_{\mathrm{orig}}\geq N_{\mathrm{mod}}$. In this situation, either a blend occurred or the simulated source distorted the local background in such a way that some sources in the original mosaic were not recovered in the modified mosaic. In the latter case, the candidate to represent the simulated source was the one found closest to it. A detection was judged if the candidate was the result of a blend between the simulated source and one or more in the original mosaic, provided that the simulated source dominated the total flux. This was confirmed when the flux ratio between the candidate and the simulated source was less than 2.

\end{description}

A direct product of these photometric simulations is the relation between input and 4\asec-aperture recovered magnitudes. For bright sources, this quantity should match the aperture correction, which represents the flux loss due to the finite aperture size. The 4.5\,$\mu$m-selected catalog from \citet{ashby13b} includes aperture corrections of $(\Delta _\mathrm{[3.6]}, \Delta_\mathrm{[4.5]})=(0.33,0.33)$, derived from PSF growth curves. The values returned by our simulations for $[4.5]=15$\,mag sources are $(\Delta _\mathrm{[3.6]}, \Delta_\mathrm{[4.5]})=(0.32,0.36)$. For consistency with the rest of our procedures, we undo the corrections applied in \citet{ashby13b} and use the values derived here.

\subsection{Photometric bias}\label{s_apercorr}
Due to the broad PSF, IRAC photometry suffered from a non-negligible level of source confusion. This caused the aperture flux to be contaminated by neighboring sources and photometric background noise. In general, these effects were more significant for fainter sources. Thus it was necessary to measure the average photometric bias as a function of magnitude, and use it to correct the full SSDF source catalog. For this purpose we employed the results from the simulations described in Appendix \ref{s_simproc}. \\
\indent We calculated the photometric bias as the median difference between the input magnitude of the simulated sources and the recovered aperture-corrected magnitude, for those sources that were detected. The photometric biases for [4.5] and [3.6] (sources were selected in 4.5\,$\mu$m) are shown in Figure \ref{aper_corr}. This bias became progressively larger in [4.5] toward fainter magnitudes, in the sense that those sources had greater excess of flux due to contamination. An important contribution to the contamination in faint sources came from the background noise fluctuations. Sources falling on top of noise peaks became brighter, while those overlapping with noise troughs could avoid detection. Therefore, the net effect was an overestimation of the [4.5] flux in point sources, which became increasingly important toward the faint end. \\
\indent In the case of the 3.6\,$\mu$m photometry, the photometric bias followed the opposite trend than seen at 4.5\,$\mu$m. This can be understood as being driven by the 4.5\,$\mu$m selection. As mentioned in Appendix \ref{s_simproc}, a faint source was likely to be measured in the 4.5\,$\mu$m mosaic at a shifted location from its true center, which was caused by the addition of a nearby background flux spike in the 4.5\,$\mu$m mosaic and boosting the extracted flux value. However, within the same aperture in the 3.6\,$\mu$m mosaic, that flux spike was not present. We have verified these effects by visual inspection of the modified mosaics in the simulations. On average, the 3.6\,$\mu$m measurement did not add extra flux from the background and still lost input flux due to the aperture shift. This induced a photometric bias that slightly underestimated the 3.6\,$\mu$m aperture fluxes.    \\
\indent The detection algorithm in Appendix \ref{s_simproc} already required some photometric bias correction to compare the measured and input fluxes in items \textbf{C} and \textbf{D}. Therefore, we ran a first pass of the simulation to compute a pre-correction, which was then used in the second pass to obtain the final results. In the first pass we only considered the photometry of recovered sources via \textbf{A} and \textbf{B}, whose detection was independent on the photometry itself. In the second pass, we ran the simulation using the full detection algorithm, where we used the pre-corrections to perform comparison with input fluxes. In this algorithm, we did not use 4\asec\ apertures. Instead, 3\asec\ and 5\asec\ corrected [4.5] fluxes were used for \textbf{C} and \textbf{D}, respectively. The pre-corrections were computed in these apertures. We chose 3\asec\ because it was the aperture with the lowest photometric scatter, and 5\asec\ due to the larger sizes that source blends generally had.  

\subsection{Completeness}\label{s_completeness}
With the results from the mock source simulations described in Appendix \ref{s_simproc}, we can compute the detection fraction as a function of input 4.5\,$\mu$m magnitude. This photometric completeness is shown in Figure \ref{completeness}, reaching 80\% at 18.58 and 50\% at 19.39. We obtain significantly higher values than \citet{ashby13b}, which is due to the different procedures used in the detection algorithm. The procedure used in this work considers a larger number of cases where a source may be detected. This includes the positional shifts $>1\asec$ in the measured photometry and the flux variations $>0.5$\,mag due to source confusion (see items \textbf{B} - \textbf{D} in Appendix \ref{s_simproc}).\\
\indent The scatter in the extracted 4.5\,$\mu$m fluxes allows us to determine the photometric sensitivity. The 5$\sigma$ limit is [4.5]=18.19, very similar to the level of 18.2 found in \citet{ashby13b}.

\section{COSMOS vs EGS as reference samples} \label{a_EvsC}
All the procedures presented so far have used COSMOS data as the reference sample, but we have also performed the same calculations with EGS. The comparison between these sets of results provides a sense of the systematic uncertainty associated with the choice of the reference survey. The redshift distributions, number densities and stellar masses present some differences, but we show below that these variations do not qualitatively alter our results. \\
\indent First, we present a brief description of some of the relevant aspects of the EGS and COSMOS datasets:

\begin{itemize}
\item \textbf{EGS}: \citet{barro11a} select sources in IRAC with 4\asec\ apertures and reach S/N = 5$\sigma$ at $\sim 21$ magnitude. The survey covers 0.48 deg$^2$ and photometric redshifts are provided with an accuracy of $\delta z/(1+z)=0.034$. The IRAC photometry in EGS is almost 3 magnitudes deeper than SSDF, reaching a higher source completness throughout the range of magnitudes considered in this work ($[4.5]<18.6$). Likewise, the higher depth in EGS makes it robust against the photometric bias that affects the SSDF (see Appendix \ref{s_apercorr}). \citet{barro11a} apply aperture corrections of $(\Delta _\mathrm{[3.6]}, \Delta_\mathrm{[4.5]})=(0.32,0.36)$ derived from PSF growth curves. These values are exactly the same as our photometric corrections in the bright limit (see Figure \ref{aper_corr}).

\item \textbf{COSMOS}: \citet{muzzin13a} explain that images from optical+NIR bands are PSF-matched and source selection is done in the $K_s$ band with 2\farcs1 color apertures. The $K_s$ band images are then used as high resolution templates in a fitting procedure to deblend confused IRAC sources. The 4.5 $\mu$m photometry reaches S/N = 5$\sigma$ at $\sim 20$, which is 2 magnitudes deeper than the SSDF and therefore completeness in COSMOS is not a concern. The survey region covers 1.62 deg$^2$ and the photometric redshifts are accurate to $\delta z/(1+z)=0.013$. The aperture corrections are very different from the schemes used in EGS and by us in the SSDF. From PSF growth curves, they derive the correction to the $K_s$ AUTO flux. The ratio between this corrected AUTO flux and the 2\farcs1 flux in $K_s$ is the used as the aperture correction for all other bands. 
\end{itemize}

\begin{figure}
\includegraphics[trim=0mm 0mm 5mm 0mm,clip=True,width=\columnwidth]{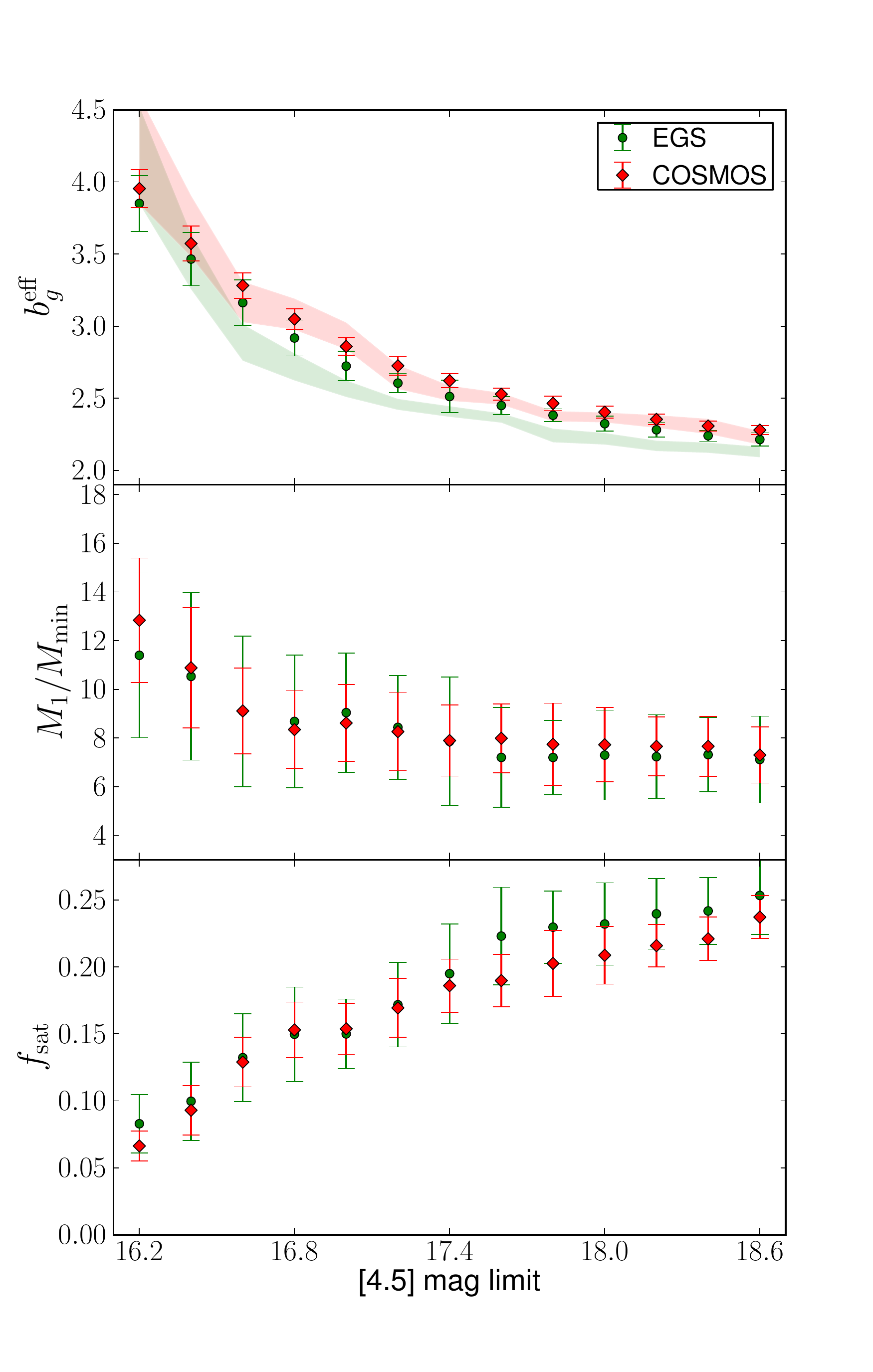}
\caption{\label{f_combined2} Results from the HOD fits to the SSDF data using COSMOS and EGS as reference catalogs. Each point denotes a sample defined by a limiting apparent magnitude threshold. In the top panel, the shaded regions represent the $\pm 1\sigma$ interval of direct large-scale bias fits. These are consistent with the HOD bias. There is an excellent agreement between both datasets in all panels.   }
\end{figure}

In COSMOS, the IRAC aperture-flux corrections factors are $\sim$50\% lower on average than in EGS, taking into account the different aperture sizes. In other words, for the same population of galaxies, EGS measures a higher IRAC apparent flux than COSMOS. Surprisingly, this has a very little impact in the stellar masses for a given luminosity, where the difference is just an offset of $\sim0.07$ dex between these catalogs.  \\
\indent In Figure \ref{f_combined2} we compare the HOD-derived quantities $b_g^\mathrm{eff}$, $M_1/M_\mathrm{min}$ and $f_\mathrm{sat}$ for EGS and COSMOS. This figure demonstrates the agreement between these datasets. Note that the redshift distributions used to model the ACFs are generated from these reference catalogs, as described in Section \ref{ss_phi}. Regarding the SHMR, the peak using COSMOS was found at $\mathrm{log}M_\mathrm{peak}=12.44\pm0.08$, whereas it is $12.35\pm0.10$ for EGS. These measurements are mutually consistent given their uncertainties. In addition, the slopes of the relation are also practically the same.

\section{Integral constraint} \label{s_ic}

\begin{figure}
\includegraphics[trim=0mm 0mm 10mm 0mm,clip=True,width=\columnwidth]{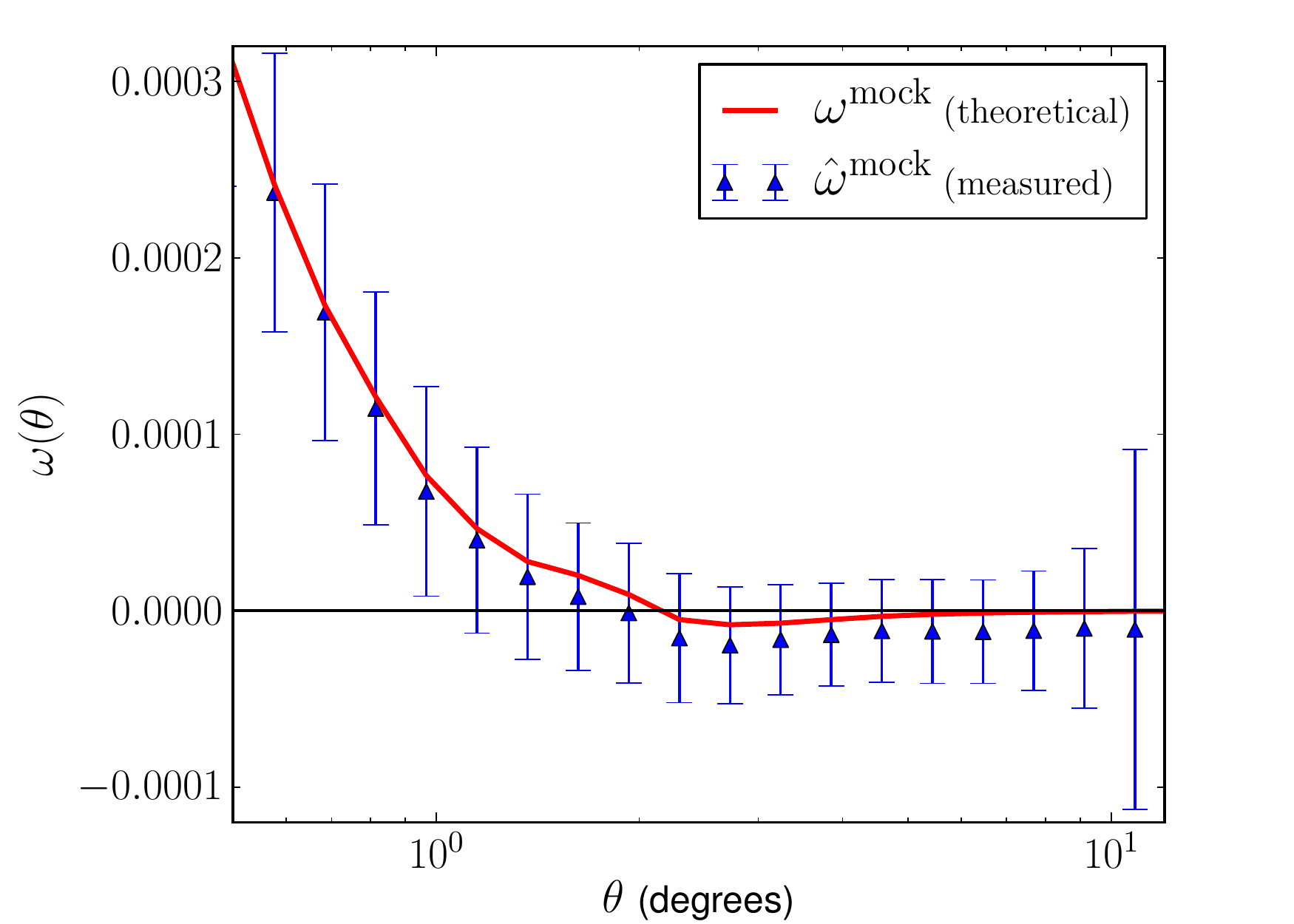}
\caption{\label{mocks} \textbf{Red}: Theoretical angular correlation function of dark matter following a redshift distribution as in Figure \ref{dndz1D_set}. \textbf{Blue}: Measured angular correlation function of mock galaxy fields within the SSDF survey region. Their parent distribution follows the same statistics as the theoretical curve. Error bars represent one standard deviation. The suppression in the measured curve is due to the normalization of the mean galaxy density within a survey region that is a fraction of the total sky.   }
\end{figure}

The estimators of the angular correlation function (such as the one in Equation \ref{ham1}) suffer from a well-known systematic supression due to the finite size of the survey, called the \emph{integral constraint} \citep{peebles80}. By construction, the estimator requires the probability to be normalized over the survey area. This means that:
\begin{equation}
\int \limits_\mathrm{survey} \hat{\omega}(\theta)d\Omega = 0. \label{ic}
\end{equation}
However, the true $\omega(\theta)$ is normalized with the entire sky, so that
\begin{equation}
\oint \limits_\mathrm{sky} \omega(\theta)d\Omega = 0.
\end{equation}
Eq.(\ref{ic}) shows that $\hat{\omega}(\theta)$ will be different from $\omega(\theta)$ whenever the survey is a fraction of the sky. In order to calculate the correction $\hat{\omega}(\theta) \rightarrow \omega(\theta)$ for the SSDF ACFs, we run simulations with mock realizations of the galaxy field in the survey region. These are generated with some known $\omega(\theta)$, which is then compared to the measured estimator $\hat{\omega}(\theta)$. We adopt the power spectrum of dark matter $P(k,z=0)$ and a redshift selection function equal to that of our main galaxy sample, $\phi^\mathrm{mock}(z)=\phi_\mathrm{cut}^{12}(z)/\int \phi_\mathrm{cut}^{12}(z^\prime)dz^\prime $ (see Figure \ref{dndz1D_set}). Following \citet{tegmark02}, the angular power spectrum is computed as

\begin{equation}
C_l^\mathrm{mock} = \frac{2}{\pi}\int \limits_0^\infty dk \, k^2 \, P(k) \left[ \int \limits_0^\infty dz\, \phi^\mathrm{mock}(z)\,G(z)\,j_l(\chi(z)k) \right]^2,
\end{equation}
where $G(z)$ is the growth factor, $\chi(z)$ the radial comoving distance and $j_l$ is the spherical bessel function. We use the routine SYNFAST in the HEALPix \footnote{http://healpix.sf.net} distribution \citep{gorski05} to produce 1000 different sky realizations drawn from this angular power spectrum. These overdensity maps are cropped to the SSDF mask and renormalized within that region. A pixelized version of the estimator in Equation \ref{ham1} (see Equation 18 in Scranton et al. 2002) is then employed to calculate the angular correlation function. The theoretical expression of this statistic takes the form of 

\begin{equation}
\omega^\mathrm{mock}(\theta) = \sum \limits_l \frac{(2l+1)}{4\pi}C_l^\mathrm{mock} \, P_l(\mathrm{cos}\theta), \label{w_theor}
\end{equation}
where $P_l$ are the Legendre polynomials. Figure \ref{mocks} shows this theoretical expectation along with the measured statistic from the mock simulations. We define the difference between both curves as $\Delta(\theta)$. In general, this quantity should scale with the overall bias of the ACF being corrected. Therefore, we correct the galaxy ACFs as $\omega^\mathrm{gal}(\theta) = \hat{\omega}^\mathrm{gal}(\theta) + \gamma \, \Delta(\theta)$. Here, $\gamma$ is a relative bias factor determined by the quotient $\gamma = \hat{\omega}^\mathrm{gal}/\omega^\mathrm{mock}$ at $\theta=0.5$, where $\Delta \simeq0$. For our largest galaxy sample, the large-scale correction is $\sim 10^{-4}$. In all samples, the correction is considerably smaller than the errors at all scales and it does not play a significant role in the results presented in this study.

\section{The halo model} \label{a_halo}
We use the halo mass function from \citet{tinker10}, which considers spherically collapsed haloes with an average density 200 times greater than the critical density of the Universe. It takes the form

\begin{equation}
\frac{dn(M,z)}{dM}=\frac{\overline{\rho}_m}{M}f(\nu)\,d\nu \label{e_mf}
\end{equation}
where $\overline{\rho}_m$ is the comoving average matter density of the Universe. The function $f(\nu)$ is empirically determined by simulations in \citet{tinker10} and parametrized with the variable

\begin{equation}
\nu(M,z)=\frac{\delta_c(z)}{\sigma(M,z)}.
\end{equation}
Here, $\delta_c$ is the critical density for halo collapse \citep{press74} for which we adopt the redshift evolution from \citet{weinberg03}

\begin{equation}
\delta_c(z)=\frac{3}{20}(12\pi)^{2/3}(1+0.131\,\mathrm{log}\Omega_m(z)),
\end{equation}
with the universal fraction of matter evolving as
\begin{equation}
\Omega_m(z)= \left[ 1+ \frac{\Omega_\Lambda}{\Omega_{m0}(1+z)^3}\right]^{-1}.   
\end{equation}
The rms of the matter density field inside spheres of $R=(3M/4\pi \overline{\rho}_m)^{1/3}$ is 

\begin{equation}
\sigma^2(M,z)=G^2(z) \int\limits_0^\infty dk \frac{k^2 P_\mathrm{lin}(k)}{2\pi^2}W^2(kR)
\end{equation}
where $P_\mathrm{lin}$ is the linear matter power spectrum today, $W(x)=(3/x^3)(\mathrm{sin}x - x\,\mathrm{cos}x)$ and the growth factor is \citep{linder05,weinberg12}

\begin{equation}
G(z)= \mathrm{exp}\left[ -\int\limits_0^z \frac{dz^{\prime}}{1+z^{\prime}} \Omega_m(z^{\prime})^{-0.55}  \right].
\end{equation}
With the mass function we can write the predicted total number density of galaxies as
\begin{equation}
n_g = \int\limits_{M_\mathrm{low}}^{M_\mathrm{high}}dM \frac{dn}{dM}(M)  N(M) \label{ng}
\end{equation}
where the integration limits are set hereafter by $M_\mathrm{low}=10^8 M_\odot$ and $M_\mathrm{high}=10^{16} M_\odot$. The NFW halo density profile is

\begin{equation}
\rho_h(M,r) = \frac{\rho_s}{(r/r_s)(1+r/r_s)^2}.  
\end{equation}
Here, $r_s = r_{200}/c$, where $r_{200} =\left[3M/(4\pi\,200 \overline{\rho}_m) \right]^{1/3} $ and the concentration parameter is given by \citet{duffy08}

\begin{equation}
c(M,z) = A(M/M_\mathrm{pivot})^B (1+z)^C, 
\end{equation}
with $A=6.71$, $B=-.091$, $C=-0.44$ and $M_\mathrm{pivot}=2.86\times10^{12}M_\odot$. We have also tried other concentration models from the literature \citep{bullock01,gao08} and explored variations in the normalization. We find the ACFs to be relatively insensitive to these changes within the angular scales probed by our data. The central density $\rho_s$ can be determined through

\begin{equation}
M = \int\limits_0^{r_{200}}dr4\pi r^2 \rho_h(M,r) 
\end{equation}
so that

\begin{equation}
\rho_s = \frac{ 200 \overline{\rho}_m c^3}{3 \left[\mathrm{ln}(1+c)-c/(1+c)\right]}.
\end{equation}
For the large-scale halo bias, we adopt the prescription from \citet{sheth01}

\begin{eqnarray}
b_h(M,z)=b_h(\nu)&=& 1 +\frac{1}{\delta_c \sqrt{a}}\bigg[  \sqrt{a}(a\nu^2)+\sqrt{a}b(a\nu^2)^{1-c}  \nonumber
\\ &&- \frac{(a\nu^2)^{c}}{(a\nu^2)^{c} + b(1-c)(1-c/2)}       \bigg]
\end{eqnarray}
with the updated parameters from \citet{tinker05} $a=0.707$, $b=0.35$, $c=0.8$. Under the halo definition we use (spherical-overdensity, Tinker et al. 2008), haloes are allowed to overlap as long as the center of one halo is not contained within the radius of another halo. The full scale-dependent bias is given by \citet{tinker12}
\begin{equation}
b_h(M,z,r) = b_h(M,z) \frac{[1+1.17 \xi_{m}(R^\ast,z)]^{1.49}}{[1+0.69 \xi_{m}(R^\ast,z)]^{2.09}}
\end{equation}
with
\begin{equation}
\label{e.scale_bias}
  R^\ast= \left\{ \begin{array}{ll}
      r\,  & {\rm if\ \ } r>=2R_{\rm halo} \\ \\
      2R_{\mathrm{halo}}\, & {\rm if\ \ } r<2R_{\rm halo} \\
        \end{array}
        \right.
\end{equation}
which sets a constant bias in the regime where haloes overlap. The non-linear matter power spectrum is obtained with the  
CAMB package \citep{lewis00}. It transforms to the correlation function as
\begin{equation}
\xi(r) = \frac{1}{2\pi^2}\int\limits_0^\infty dk\, k^2 P(k) \frac{\mathrm{sin}kr}{kr}   \label{xi2p}.
\end{equation}
Since the virialized regime of satellites within haloes will be different from the large-scale interaction between central galaxies, it is convenient to express the spatial correlation function as a sum of two terms:

\begin{equation}
\xi_{g}(r) = 1+ \xi^{1h}_{g}(r)+\xi^{2h}_{g}(r).    \label{1_2halo}
\end{equation}
The one-halo term is highly non-linear and dominates at scales smaller than the average halo size (i.e., virial radius), while the 2-halo term becomes more important at large-scales. Furthermore, the one-halo term can be separated into central-satellite and satellite-satellite parts. In the former, the correlation follows the form of a NFW density weighted by spherical volume, since by construction satellites are distributed according to that profile from the central galaxy. In the latter, the satellite-satellite correlation follows the form of a NFW profile (still representing the distribution of satellites from the central galaxy) convolved with itself. In the case of the 2-halo term, the correlation traces the convolution between $\xi_{m}$ and density profiles of different halos. Given the many convolutions, it is better to work in Fourier space, where all these become simple products. Thus, Equation (\ref{1_2halo}) can be rewritten as

\begin{equation}
P_{g}(k) = \left[ P^{cs}_{g}(k)+ P^{ss}_{g}(k) \right]^{1h} +P^{2h}_{g}(k).    \label{b2}
\end{equation}
The explicit form of the 1-halo terms is

\begin{equation}
P_{g}^{cs}(k,z) = \frac{2}{n_g^2} \int\limits_{M_\mathrm{low}}^{M_\mathrm{high}} dM N_s(M) N_c(M) \frac{dn}{dM}(M,z) u(k,M,z),\label{1h1}
\end{equation}
\begin{equation}
P_{g}^{ss}(k,z) =\frac{1}{n_g^2} \int\limits_{M_\mathrm{low}}^{M_\mathrm{high}} dM N_s(M) N_c(M) \frac{dn}{dM}(M,z) u^2(k,M,z),\label{1h2}
\end{equation}
where $u$ is the Fourier transform of a NFW profile \citep{cooray02}. For the 2-halo term, we must not consider overlapping haloes if one of their radii contains the center of the other. This is done by adopting halo-exclusion \citep{zheng04}, where we set the minimum separation allowed for 2 haloes to $d=\mathrm{max}(R_\mathrm{halo1},R_\mathrm{halo2})$ \citep{leauthaud11}. This implies that measuring halo correlations at distances smaller that $r$, we can integrate all possible pairs where the individual radii are bound to an upper limit $R_\mathrm{lim}=r$. The 2-halo term thus reads

\begin{eqnarray}
P_{g}^{2h}(k,r,z) =P_{m}(k,z) \bigg[ \frac{1}{{n_g^\prime(r)}^2}\int\limits_{M_\mathrm{low}}^{M_\mathrm{lim}(r)} dM N(M)\cdot \nonumber \\ \frac{dn}{dM}(M,z) b_h(M,r,z) u^2(k,M,z)\bigg]^2,
\end{eqnarray}
where the scale-dependent halo bias is introduced and $M_\mathrm{lim}(r)=M(r=r_{200})$ enforces halo-exclusion. This integration limit restricts the average density of the galaxies considered \citep{tinker05}:

\begin{equation}
n_g^\prime(r) = \int\limits_{M_\mathrm{low}}^{M_\mathrm{lim}(r)}dM \frac{dn}{dM}(M)  N(M),
\end{equation}
compared to the total $n_g$ in Equation (\ref{ng}). After Fourier transforming $P_{g}^{2h}$ into ${\xi_{g}^{2h}}^\prime$, the probability function needs to be supressed to account for the missing galaxies in $n_g^\prime$ as

\begin{equation}
1+{\xi_{g}^{2h}}(r,z)=\left(\frac{n_g^\prime(r)}{n_g}  \right)^2 [ 1+{\xi_{g}^{2h}}^\prime(r,z)]
\end{equation}
Adding the one halo terms from eqs. (\ref{1h1}, \ref{1h2})  completes the HOD description of our model.

\section{Low-redshift bump} \label{nosc}

\begin{figure}
\includegraphics[trim=0mm 0mm 5mm 0mm,clip=True,width=\columnwidth]{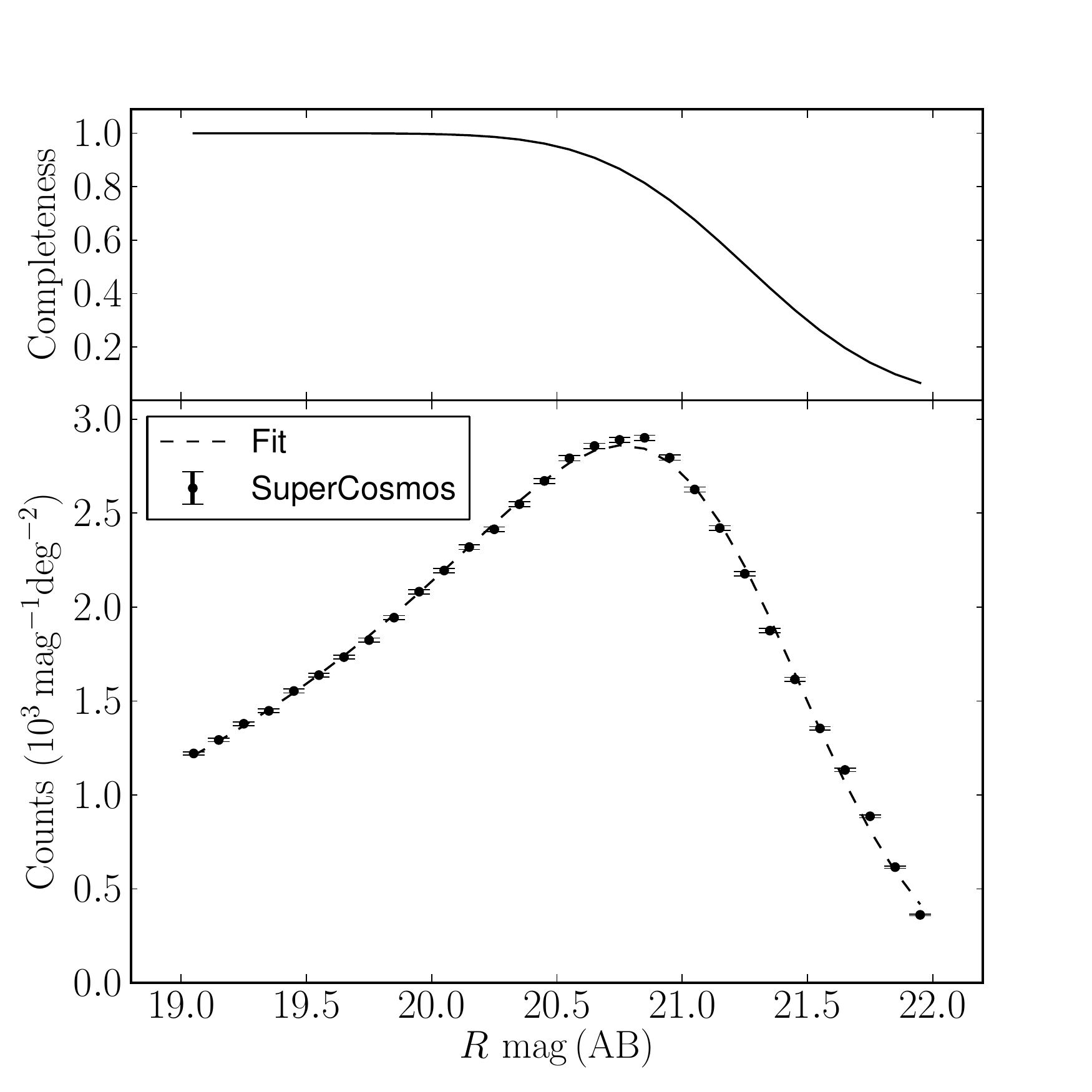}
\caption{\label{scLF} Distribution of $R$-band magnitudes from Super Cosmos sources in the SSDF region (points). We fit the data with a function consisting of a power law times an error function (dashed line). The best-fit error function is displayed in the top panel and represents the $R$-band photometric completeness of the Super Cosmos sample.    }
\end{figure}

Low-$z$ galaxies have an important contribution to the redshift distribution in our brightest galaxy sample, $[4.5]<16.2$, as shown in Figure \ref{dndz1D_set}. In this Section, we use the optical Super Cosmos survey data \citep{hambly01} to match and remove these sources from the SSDF catalog and evaluate how this changes the HOD results from Section 5. This is intended to serve as a consistency check for the methods used so far to model the low-redshift galaxy clustering. \\
\indent Super Cosmos (hereafter SC) is a full sky survey produced from digitized photographic plates in the $B$, $R$ and $I$ bands, with a typical depth of $R(AB)\sim21$ mag. We retrieved from the Super Cosmos Science Archive\footnote{surveys.roe.ac.uk/ssa/} all sources in the SSDF footprint with detection in $R$ and at least one other band. We chose $R$ as the main optical band because, in combination with our infrared cuts, it is particularly effective in selecting $z<1$ galaxies at $R(\mathrm{AB})\lesssim 22.5$ \citep{papovich08}. We removed from the SSDF catalog all the sources in the SC sample that matched within a search radius of $1\asec$. The SSDF clustering computation and modeling followed the same procedures described through Sections 3-6, except for a modification of equation 2. This equation describes the SSDF redshift distribution as a sum of contributions from the individual galaxies in the control sample, given a particular [3.6] and [4.5] selection. The modification consists of including a weight factor to each individual contribution based on the galaxy's $R$-band magnitude. This weight, $\mathcal{W}(R)\in \left[ 0,1 \right]$, should represent the probability of a galaxy in the SSDF to be undetected in SC. We defined this probability as $\mathcal{W}(R)=1-\mathcal{U}(R)$, where $\mathcal{U}(R)$ is the $R$-band completeness in SC. We estimated the completeness directly from the distribution of $R$ magnitudes from the SC catalog, shown in the lower panel of Figure \ref{scLF}. This distribution, $\mathcal{D}(R)$, was assumed to obey a power-law (Schechter 1976) that is supressed at the faint end by the completeness function:

\begin{equation}
\mathcal{D}(R) = d_0 R^{d_1} \, \mathcal{U}(R). \label{e_scLF}
\end{equation}
Here, $d_0$ and $d_1$ are the power-law coefficients and  
\begin{equation}
\mathcal{U}(R) = \frac{1}{2}\left(1-\mathrm{Erf}\left(s\left(R-R_0\right)\right)\right),
\end{equation}
\begin{equation}
\mathrm{Erf}(y) = \frac{2}{\sqrt{\pi}}\int_0^y e^{-t^2}dt,
\end{equation}
where $R_0$ and $s$ are parameters that control the shape of the error function. We fit $\mathcal{D}(R;d_0,d_1,R_0,s)$ to the SC data. This is shown in Figure \ref{scLF}, where the upper panel represents the corresponding $\mathcal{U}(R)$ component. 

\begin{figure}
\includegraphics[trim=5mm 0mm 15mm 0mm,clip=True,width=\columnwidth]{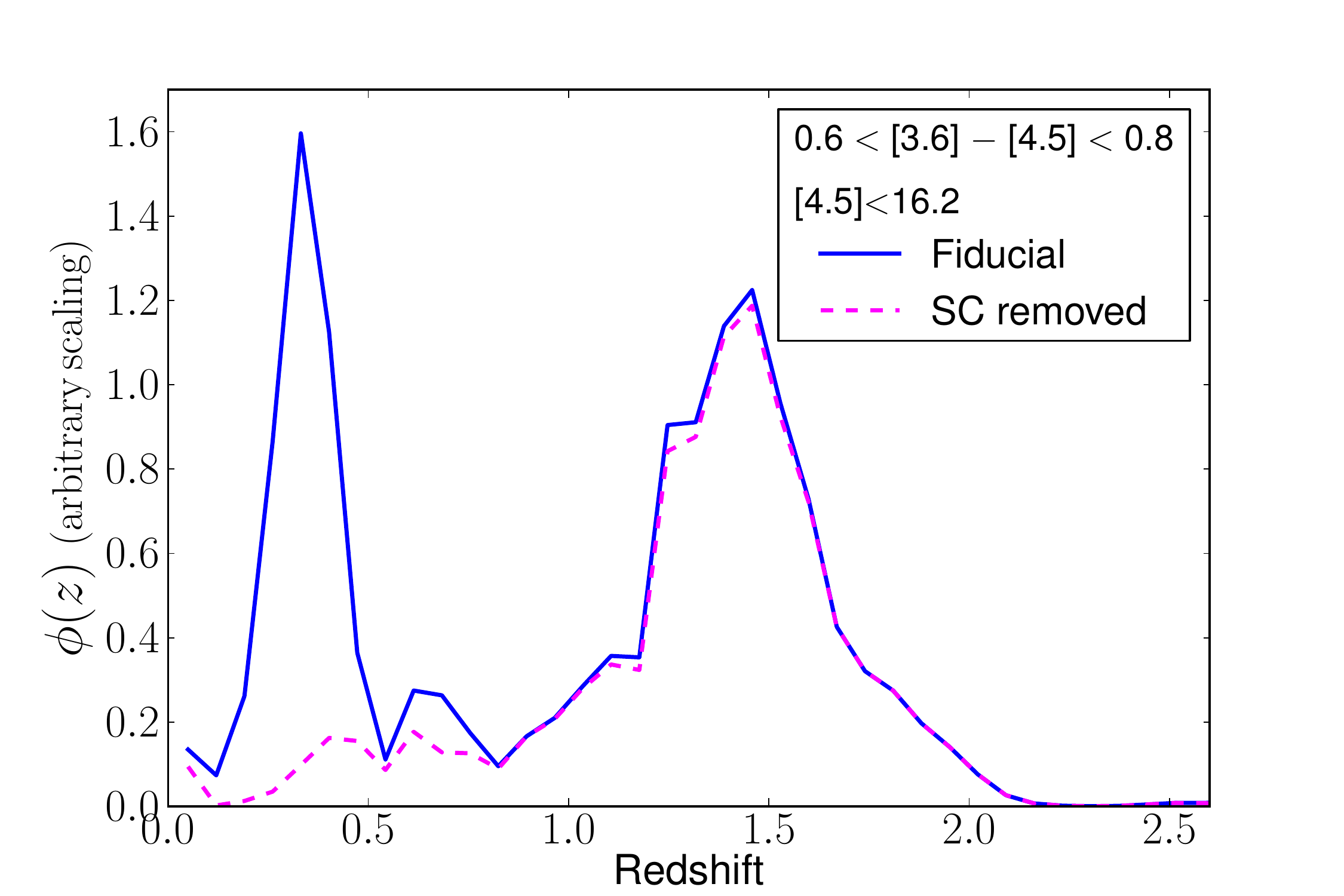}
\caption{\label{dndz_nosc} Redshift distribution of our brightest galaxy sample obtained from the COSMOS-based control sample. The solid line represents the fiducial distribution as derived in Section \ref{s_ctrlsamples}, which includes sources selected only with IRAC data. The dashed line shows the simulated effect of removing all sources with $R$-band detections in Super Cosmos. This optical selection is very effective at suppressing the low-redshift bump.    }
\end{figure}

The application of $\mathcal{W}(R)$ to the galaxies in the control sample produces a strong suppresion of the low-redshift contribution. This can be seen in Figure \ref{dndz_nosc} for the $[4.5]<16.2$ subsample. Compared to the fiducial distribution derived in Section \ref{ss_subsamples}, the removal of SC sources eliminates almost completely the bump at $z\sim 0.3$. The comoving number density at $z=1.5$ (Equation \ref{ng_obs}) increases only $7\%$. The HOD fit suffers a decrease in $M_\mathrm{min}$ of $0.02$ dex and an increase in $M_1$ of $0.05$ dex. For subsamples with thresholds fainter than $[4.5]=16.2$, these variations become even smaller. Overall, the impact of removing SC data on the results presented in this paper is negligible. This provides solid support to the modeling of the low-redshift galaxy clustering described in Sections 3-5. \\
\indent We have not used the analysis from this Section to derive the main results of the paper because there are potential systematic effects in the SC catalog that we have not thoroughly inspected. For instance, the SC photometry suffers from differential coverage depth across the field, which could cause an artificial contribution to the clustering. In addition, we do not have a complete statistical description of the large photometric $R$-band errors present in SC. An improved treatment in this analysis would entail the use of such errors to deconvolve the SC distribution of $R$ magnitudes, in order to be consistent with the high photometric quality of the control sample.

\section{No prior on number density} \label{a_tests}
\begin{figure}
\includegraphics[trim=0mm 0mm 5mm 0mm,clip=True,width=\columnwidth]{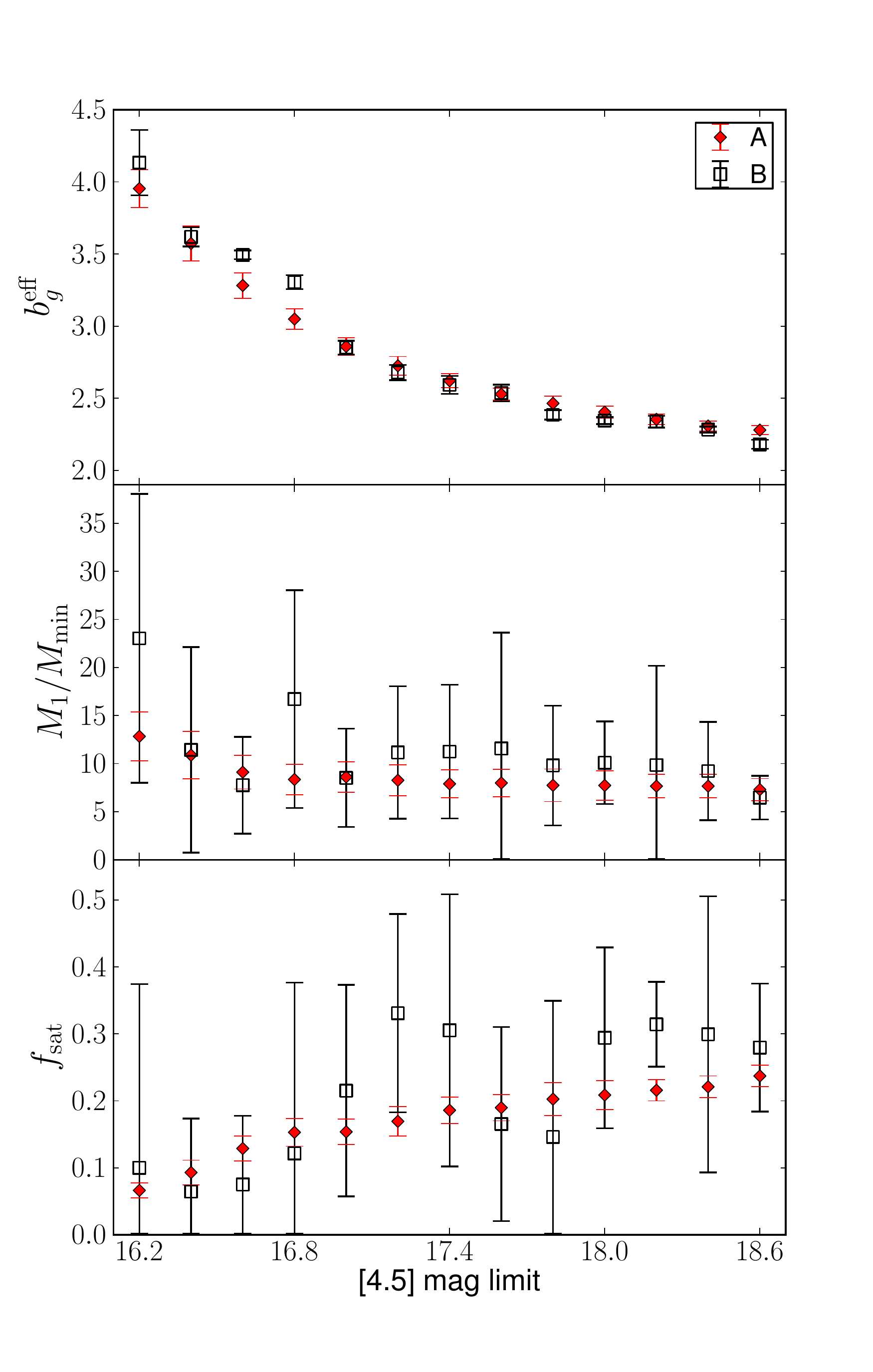}
\caption{\label{1vs5param} Results from the HOD fits to the SSDF data using the COSMOS reference catalog. Models A and B refer to fits using 1 and 5 free parameters, respectively (see text). The galaxy number density is fixed in A to the observed number counts, but it is left as a free parameter in B. Both models are fully consistent, which implies that the observed clustering can reproduce the correct number density and that the values for the fixed parameters in A are reasonable. We choose A as our fiducial model since it yields a more monotonic behavior between the derived parameters and the sample luminosity, as expected.   }
\end{figure}

In the fitting procedure we have fixed all basic HOD parameters except $M_1^\prime$, and used galaxy number densities obtained through a combination of the SSDF observed number counts and the control sample. Here, we check what happens if we leave all those parameters to vary freely and discard any prior information on the number density. Thus, $n_g$ becomes a derived quantity trough Equation \ref{ng}. We still need to use the normalized redshift distributions from the control sample, however. For the sake of clarity in this section, we will call the fiducial fit of this paper \emph{model A} (1-parameter fit, $n_g$ fixed), and the unconstrained one \emph{model B} (5-parameter fit, $n_g$ as derived quantity).  \\
\indent The goodness of fit, $\chi^2_\nu$, remains on average the same between models \emph{A} and \emph{B}, which points to the latter not really being statistically favored. The $b_g$, $M_1/M_\mathrm{min}$ and $f_\mathrm{sat}$ relations also do not change appreciably, as shown in Figure \ref{1vs5param}. This implies that: (1) the observed clustering is able to correctly reproduce the measured galaxy number density, which is a free parameter in \emph{B}, and (2) the values of the fixed parameters in \emph{A} are reasonable choices. On average, the changes in the derived parameters and $n_g$ are $\lesssim20$\%, which supports the validity of the halo occupation model. We choose \emph{A} as the fiducial model because the relation between the derived parameters and sample luminosity is more monotonic than in \emph{B}. The latter shows roughly the same average trends but with a stronger level of stochasticity. In essence, there would be no substantial information gain by adopting \emph{B} instead of \emph{A}. In addition, \emph{A} includes the observed number density, which places an important constraint on the HOD model. The inferred densities from Equation \ref{ng_obs} do contain some uncertainty since they are partially derived from the control sample, but we believe that using them produces a more physically consistent HOD model. \\
\indent When performing a fit of the SHMR with results from \emph{B}, the errors become large enough to be consistent at about 1-$\sigma$ level with \emph{A}. For example, in the M13 parametrization, for \emph{A} we obtained $\mathrm{log}M_\mathrm{peak}=12.44\pm0.08$, $\beta=1.64\pm0.09$ and $\gamma=0.60\pm0.02$, while for \emph{B} these become $\mathrm{log}M_\mathrm{peak}=12.43\pm0.09$, $\beta=2.14\pm0.98$ and $\gamma=0.50\pm0.11$. We obtain similar values when using EGS as the reference catalog (see Appendix \ref{a_EvsC}).

\end{appendix}



\end{document}